# The Federal Reserve's Response to the Global Financial Crisis and Its Long-Term Impact: An Interrupted Time-Series Natural Experimental Analysis[*]

Arnaud Cedric KAMKOUM[**]

May 20, 2023

## Abstract

This paper examines the monetary policies the Federal Reserve implemented in response to the Global Financial Crisis. More specifically, it analyzes the Federal Reserve's quantitative easing (QE) programs, liquidity facilities, and forward guidance operations conducted from 2007 to 2018. The essay's detailed examination of these policies culminates in an interrupted time-series (ITS) analysis of the long-term causal effects of the QE programs on U.S. inflation and real GDP. The results of this formal design-based natural experimental approach show that the QE operations positively affected U.S. real GDP but did not significantly impact U.S. inflation. Specifically, it is found that, for the 2011Q2-2018Q4 post-QE period, real GDP per capita in the U.S. increased by an average of 231 dollars per quarter relative to how it would have changed had the QE programs not been conducted. Moreover, the results show that, in 2018Q4, ten years after the beginning of the QE programs, real GDP per capita in the U.S. was 14% higher relative to what it would have been during that quarter had there not been the QE programs. These findings contradict Williamson's (2017) informal natural experimental evidence and confirm the conclusions of VARs and new Keynesian DSGE models that the Federal Reserve's QE policies positively affected U.S. real GDP. The results suggest that the current U.S. and worldwide high inflation rates are likely not because of the QE programs implemented in response to the financial crisis that accompanied the COVID-19 pandemic. They are likely due to the unprecedentedly large fiscal stimulus packages used, the peculiar nature of the financial downturn itself, the negative supply shocks from the war in Ukraine, or a combination of these factors. To the best of my knowledge, this paper is the first study to measure the macroeconomic effects of QE using a design-based natural experimental approach.

[*] This paper was part of my master's thesis written at Bursa Uludag University. The thesis is freely accessible from https://econpapers.repec.org/RePEc:osf:thesis:d7pvg, https://doi.org/10.31237/osf.io/d7pvg, and https://acikerisim.uludag.edu.tr/jspui/handle/11452/30308. I thank my supervisor, Professor Görkem BAHTİYAR, for his invaluable guidance and support. I also thank the other two members of my thesis defense committee, Professors Hasan BAKIR and Selda GÖRKEY, for their important questions, comments, and suggestions.

[**] I have a Master of Arts in economics (with *High Honors*), a Bachelor of Science in mathematics (with *High Honors*), and a Bachelor of Arts in economics (with *High Honors*). I am a two-time recipient of the Hewlett Foundation MITx DEDP MicroMasters Scholarships and a two-time holder of the Turkish Government Scholarships. I can be contacted at kace.excellence@gmail.com and https://www.researchgate.net/profile/Arnaud-Cedric-Kamkoum. ORCID: https://orcid.org/0000-0002-8632-2312.

# 1. INTRODUCTION

The financial crisis that started in the U.S. at the end of 2007 and later spread to other countries was the most severe economic and financial disaster since the Great Depression. The crisis began in the U.S. housing market in August 2007, rapidly extended to other sectors of the U.S. economy, and became global following the collapse of various U.S.-based international financial institutions. To counter the negative effects of the crisis, the Federal Reserve (the central bank of the United States) and other central banks conducted monetary policies that are widely considered unconventional. This paper examines the monetary policies the Federal Reserve implemented in response to the crisis. More specifically, it analyzes the Federal Reserve's quantitative easing (QE) programs, liquidity facilities, and forward guidance operations implemented from 2007 to 2018. The essay's detailed description of these policies is concluded by examining the long-term causal effects of the QE programs on U.S. inflation and real GDP using the *formal design-based natural experimental method* of interrupted time-series (ITS) analysis. The empirical investigation was motivated by a desire to verify Williamson's (2017) *informal natural experimental evidence* that contradicts the results of vector autoregressions (VARs) and new Keynesian dynamic stochastic general equilibrium (DSGE) models that the QE programs positively affected U.S. real GDP.

The paper contributes to the literature in three significant ways:
1. *It provides a clear and practical classification of domestic balance sheet policies*. It does this by classifying expansionary and contractionary domestic balance sheet policies into two mutually exclusive categories each: *quantitative easing* (*QE*) and *quasi-quantitative easing* (*qQE*); and *quantitative tightening* (*QT*) and *quasi-quantitative tightening* (*qQT*), respectively. Where QE is defined as a central bank outright and large-scale purchase of domestic public or private assets that increases the quantity of bank reserves. qQE is defined as a central bank large-scale expansionary domestic operation conducted to influence the economy beyond the policy rate that is either (i) non-outright and may increase the quantity of bank reserves temporarily or (ii) outright but does not increase the quantity of bank reserves. QT is defined as a central bank outright and large-scale sale of domestic public or private assets that decreases the quantity of bank reserves. And qQT refers to a central bank large-scale contractionary domestic operation conducted to influence the economy beyond the policy rate that is either (i) non-outright and may decrease the quantity of bank reserves temporarily or (ii) outright but does not decrease the quantity of bank reserves. Existing studies on this subject mostly classify domestic balance sheet policies into overlapping categories, rendering their distinction difficult and confusing. My classification of these policies into mutually exclusive categories helps overcome the difficulty and confusion.

2. *It provides a comprehensive introduction to the natural experimental method of interrupted time-series (ITS) analysis.* An interrupted time series is a special type of time series in which the series shows an interruption in its values and can be used to estimate the impact of an intervention. The name *interrupted time-series analysis* (*ITS analysis* or *ITSA*) is used to describe the statistical methods and models of studies that use interrupted time series. Although ITS analysis is a widely used natural experimental approach in the evaluation of health-care and education policies and has been employed in various study areas of economics, its use in estimating the effects of monetary policy on inflation and real output is extremely limited. I am optimistic that the overview of the ITS analysis method presented in this essay will help increase its use in macroeconomics.

3. *It is one of the few studies that examine the macroeconomic effects of monetary policy using a formal design-based natural experimental approach. And, to the best of my knowledge, it is the first paper to use a formal design-based natural experimental method to measure the macroeconomic effects of QE.* Because of the difficulty in using design-based methods in monetary economics, very few papers have employed such techniques to analyze the effects of monetary policy on inflation and real GDP. Indeed, the majority of current studies that measure the macroeconomic effects of monetary policy use either VARs or new Keynesian DSGE models. Thus, this paper enriches the literature with empirical evidence from a distinctive methodological approach.

The rest of the paper is organized as follows. **Section 2** presents an introduction to monetary policy and describes how it was implemented by the Federal Reserve before the Global Financial Crisis. Specifically, the section (i) examines the frameworks, classifications, effectiveness, transmission mechanisms, and measurement of monetary policy; (ii) describes how open market operations, reserve requirements, and discount window lending were used as policy tools by the Federal Reserve before the crisis to achieve its target for the federal funds rate; and (iii) presents the ongoing debate about the expansionary nature of the Federal Reserve's monetary policies during the 2002-2006 period. It is in this section that I categorize domestic balance sheet policies into QE, qQE, QT, and qQT. **Section 3** analyzes the Federal Reserve's monetary policies conducted in response to the crisis. The section provides a detailed description of the various QE and qQE operations the Federal Reserve implemented between 2007 and 2018. It also examines the policies conducted during the normalization process that consisted of normalizing the balance sheet on the one hand and the federal funds rate on the other hand. Furthermore, the section assesses the difference between the Federal Reserve's response to the crisis and that of the Bank of Canada. **Section 4** provides a comprehensive overview of the ITS analysis method and uses this technique to examine the causal effects of the Federal Reserve's QE programs on U.S. inflation and real GDP. In the overview presented, I discuss the various ways in which an ITS study can be designed, describe the main stages involved in the conduction of an ITS study, and explain the difference between ITS analysis and regression discontinuity in time (RDiT). Finally, **Section 5** concludes.

# 2. MONETARY POLICY AND ITS IMPLEMENTATION BY THE FEDERAL RESERVE BEFORE THE GLOBAL FINANCIAL CRISIS

## 2.1. Introduction

Monetary policy refers to decisions taken by an economy's monetary authority. The monetary authority is generally a central bank, and the economy it oversees may be that of a country or group of countries. Decisions the central bank takes include choosing one or more policy goals and determining how the goal(s) will be achieved.[1] Possible policy goals the central bank may choose are low unemployment, stable prices, stable domestic interest rates, stable foreign exchange rates, and stable financial markets.

Since the policy goal(s) is(are) achieved over a considerably long period, the central bank usually has to choose one or more medium-term targets and one or more short-term instruments through which it can reach the goal(s). Medium-term policy targets and short-term policy instruments may be measures of the quantity as well as the price of money. For medium-term policy targets, such measures include monetary aggregates and medium-term nominal interest rates. And for short-term policy instruments, they include reserve aggregates and short-term nominal interest rates.[2] In addition to these measures, credit aggregates (aggregates of loans to nonfinancial sectors), foreign exchange rates, and the amount of foreign reserves may also be chosen as medium-term targets or short-term instruments.[3]

---

[1] A central bank's policy goals may also be set by the government.

[2] Nominal interest rates represent measures of the price of money. Monetary aggregates represent measures of the quantity of money like M1, M2, M3, and M4. Reserve aggregates include measures like the monetary base and the amount of total reserves.

[3] To qualify as a good short-term policy instrument, a variable must be easy to control and measure and its effect on the policy goal(s) must be strong and predictable.

Once the medium-term target(s) and short-term instrument(s) have been chosen, the central bank must decide which policy tools will be used to control the behavior of its policy instrument(s). Potential policy tools include open market operations, discount window lending, reserve requirements, interest on reserves, lending to specific credit markets, large-scale purchases and sales of domestic assets, foreign exchange interventions, and policy announcements.

The stages of the monetary policy decision-making process described above are summarized in **Figure 1** below.

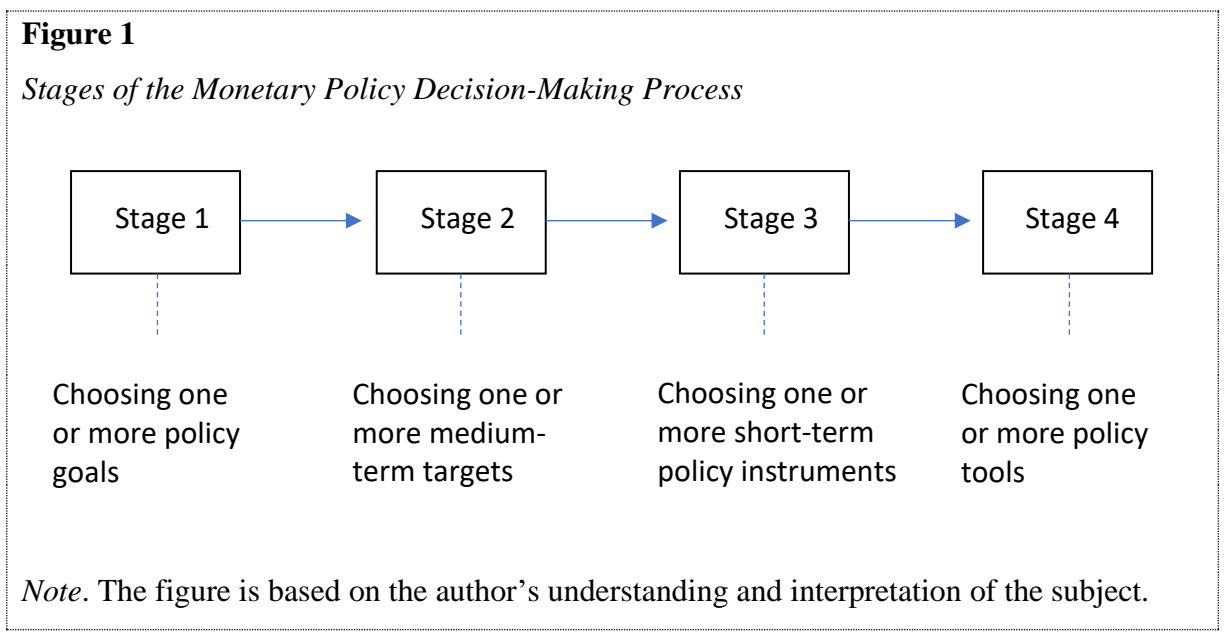

**Figure 1**

*Stages of the Monetary Policy Decision-Making Process*

| Stage 1 | Stage 2 | Stage 3 | Stage 4 |

Choosing one or more policy goals | Choosing one or more medium-term targets | Choosing one or more short-term policy instruments | Choosing one or more policy tools

*Note*. The figure is based on the author's understanding and interpretation of the subject.

The rest of this section is structured as follows. **Section 2.2** describes the frameworks, classifications, effectiveness, and measurement of monetary policy. And **Section 2.3** examines the Federal Reserve's monetary policies before the Global Financial Crisis.

## 2.2. Frameworks, Classifications, Effectiveness, Transmission Mechanisms, and Measurement of Monetary Policy

### *2.2.1. How Monetary Policy Is Implemented: Monetary Policy Frameworks*

During the monetary policy decision-making process described above, the central bank's choice about the relative weights to be given to each of its policy goals is what mainly determines its *monetary policy framework*—i.e., how it will actually implement the decisions taken (how the conduct of monetary policy will take place). A monetary policy framework (strategy or regime) may generally be based on a *hierarchical* or *dual mandate* (L. H. Meyer, 2001b; Mishkin, 2019). A hierarchical mandate is one in which price stability is the primary goal of monetary policy, and other goals are pursued provided price stability is maintained (L. H. Meyer, 2001b; Mishkin, 2019). On the other hand, a dual mandate is one in which price stability and full employment are two coequal primary goals of monetary policy (L. H. Meyer, 2001b; Mishkin, 2019). In other words, while the hierarchical mandate gives more weight only to price stability, the dual mandate gives more (and equal) weight to both price stability and full employment relative to other monetary policy goals.

When a monetary policy framework is based on the hierarchical mandate and has an explicit numerical target (objective) for the inflation rate, it is known as *inflation targeting* (L. H. Meyer, 2001b; Mishkin, 2019). Inflation targeting started during the 1990s as a result of the macroeconomic consensus that began in the 1980s. It is currently the dominant monetary policy strategy worldwide: being employed by major central banks such as the Reserve Bank of New Zealand, the Bank of Canada, the Bank of England, the Swedish central bank (Sveriges Riksbank), the Reserve Bank of Australia, the Bank of Spain, among others (Mishkin, 2019). Other characteristics of inflation targeting include high transparency in the policymaking process and central bank accountability regarding the achievement of the inflation target (Mishkin, 2019).

*Price-level targeting* is another monetary policy framework based on the hierarchical mandate. Here, instead of the *growth rate* of prices (inflation), a predetermined path for the nominal *level* of prices is targeted (Kahn, 2009; Mester, 2018a; see also Vestin, 2006; Walsh, 2017). Price-level targeting is comparable to inflation targeting in the sense that the slope of the predetermined path is equal to the target growth rate of prices (target inflation) of an inflation targeting strategy (Kahn, 2009). However, contrary to inflation targeting, where a shift of inflation from target permanently moves the economy's price path to a different level, in price-level targeting, the central bank responds to the shift with an equal and opposite change that restores the economy's initial price-path level (Kahn, 2009). The Swedish central bank is the only central bank that has employed this framework in the past (from 1931 to 1937) (Kahn, 2009).

There is no particular name for a monetary policy strategy that is based on the dual mandate. The central bank of the United States (the Federal Reserve System) is the only major central bank that uses this framework.

Although real-world monetary policy strategies are based on the hierarchical and dual mandates as described above, there are other types of frameworks that are mostly theoretical (have not yet been implemented by any major central banks). Examples include *nominal GDP targeting*, where the central bank would conduct policy by targeting a specific level of nominal GDP (Mester, 2018a), and *average inflation targeting*, where the central bank would aim to stabilize average inflation measured over several years (Nessen & Vestin, 2005).

### 2.2.2. Classifications of Monetary Policy

A particular monetary policy may be classified in numerous ways. It may be classified based on its effect on the price and quantity of money in the economy, its interaction with the business cycle, the reaction or nonreaction of the central bank to current economic conditions, the commitment or noncommitment of the central bank to past announcements, or the policy instrument used by the central bank. These various classifications are described below.

*2.2.2.1. Based on the Effect of Monetary Policy on the Price and Quantity of Money*

Based on its effect on the price and quantity of money in the economy, a specific monetary policy may be classified as either *expansionary* or *contractionary*. An expansionary, easy, or loose monetary policy is a monetary policy that decreases the price or increases the quantity of money in the economy. In contrast, a contractionary or tight monetary policy is a monetary policy that increases the price or decreases the quantity of money in the economy.

All other things being equal, when a central bank decreases its lending rates, decreases its policy rate target, increases the number of its lending facilities, increases its asset purchases, decreases its asset sales, or relaxes reserve requirements, it is easing monetary policy. And all other things being equal, when it increases its lending rates, increases its policy rate target, decreases the number of its lending facilities, decreases its asset purchases, increases its asset sales, or makes reserve requirements stricter, it is tightening monetary policy.

Since central bank announcements or communications to the public about the likely future course of monetary policy—commonly called *forward guidance*—also affect the price and quantity of money, they are part of expansionary and contractionary monetary policies. Indeed, when the central bank provides forward guidance, it intends to affect current household and business expenditure decisions (thus, the price and quantity of money) by influencing household and business expectations about the future course of monetary policy (Federal Reserve System, 2015c). In other words, just like an actual central bank lending, asset purchase, or asset sale, a central bank announcement may qualify as an expansionary or contractionary monetary policy. A central bank announcement that leads to a decrease in the price of money or an increase in the quantity of money qualifies as an expansionary monetary policy. And a central bank announcement that leads to an increase in the price of money or to a decrease in the quantity of money qualifies as a contractionary monetary policy.

Central bank announcements can further be subdivided into two, namely, *Delphic* and *Odyssean* forward guidance (Campbell et al., 2012); or into three, namely, *open-ended*, *calendar-based*, and *state-contingent* forward guidance (Borio & Zabai, 2016). Delphic forward guidance occurs when the central bank publicly states a forecast of macroeconomic performance and likely or intended monetary policy actions without committing to a particular course of action (Campbell et al., 2012). On the other hand, with Odyssean forward guidance, the central bank publicly commits itself (Campbell et al., 2012). While calendar-based forward guidance is guidance about a specific period, state-contingent forward guidance is guidance dependent on economic conditions, and open-ended forward guidance is mere qualitative information concerning the future path of monetary policy that contains no precise dates or quantitative economic data (Borio & Zabai, 2016).

*2.2.2.2. Based on the Reaction of the Central Bank to Current Economic Conditions*

Based on the reaction or nonreaction of the central bank to current economic conditions, a monetary policy can be qualified as either *activist* or *nonactivist*. While activist monetary policies refer to monetary policies implemented to respond to current economic conditions, nonactivist monetary policies are those conducted regardless of the current state of the economy (Dornbusch et al., 2018; Mishkin, 2019).

*2.2.2.3. Based on the Interaction of Monetary Policy With the Business Cycle*

With respect to its interaction with the business cycle, a monetary policy can be *countercyclical*, *cyclical*, or *acyclical*. Countercyclical or anticyclical monetary policies are activist monetary policies that are expansionary when the economy is contracting and contractionary when the economy is expanding. In contrast, cyclical monetary policies are activist monetary policies that are expansionary when the economy is expanding and contractionary when the economy is contracting. Acyclical monetary policies are neither cyclical nor countercyclical.

*2.2.2.4. Based on the Commitment of the Central Bank to Past Announcements*

Concerning the classification of monetary policies according to the central bank's commitment to past announcements or communications, a monetary policy can be characterized as *rule-based* or *discretionary*. Rule-based monetary policies are policies implemented according to preannounced rules describing how monetary policy variables will be determined in the future (Dornbusch et al., 2018). With such policies, the central bank commits itself to a specific course of policy actions. The commitment anchors inflation expectations and helps keep inflation at an appropriate level. Rule-based monetary policies can be nonactivist, like the constant-money-growth-rate rule and McCallum rule; or activist, like the Taylor rule (Mishkin, 2019). On the contrary, discretionary monetary policies are policies whereby the central bank does not commit itself to a particular course of policy actions. A monetary authority that implements discretionary monetary policies may determine the values of its policy variables at any point in time as it sees fit at that specific moment (independently of any past statements or announcements).

*2.2.2.5. Based on the Policy Instrument Used by the Central Bank*

Depending on the type of policy instrument used by the central bank, a monetary policy can be classified as an *interest rate policy* or a *balance sheet policy* (Borio & Disyatat, 2009, 2010; Borio & Zabai, 2016, 2018). An interest rate policy is a monetary policy through which the central bank influences the economy by setting or closely controlling a short-term nominal interest rate and guiding expectations about the future of this rate (Borio & Disyatat, 2009, 2010; Borio & Zabai, 2016, 2018). This short-term nominal interest rate is what is called the policy rate. On the other hand, a balance sheet policy is a policy whereby the central bank influences the economy beyond the short-term nominal rate, by modifying the size or composition of its balance sheet and guiding expectations about their future (Borio & Disyatat, 2009, 2010; Borio & Zabai, 2016, 2018).

Since the same amount of bank reserves can coexist with different levels of the policy rate and the same policy rate can coexist with different amounts of bank reserves, the central bank can conduct an interest rate policy at any size of its balance sheet and without engaging in balance sheet policies; and can conduct a balance sheet policy at any level of the policy rate without affecting the rate (Borio & Zabai, 2016, 2018). However, central banks mostly resort to balance sheet policies when the policy rate is too low (near zero). This is generally the case during severe economic disruptions.

Interest rate and balance sheet policies can further be subdivided into two types each. The two types of interest rate policies include (i) *negative interest rate policy*, whereby the central bank's target for the policy rate is set or announced below zero, and (ii) *non-negative interest rate policy*, whereby the target for the policy rate is never set or announced below zero. On the other hand, the two types of balance sheet policies are (i) *domestic balance sheet policies*, which involve central bank lending to domestic entities, purchases and sales of domestic assets, and announcements about these, and (ii) *non-domestic balance sheet policies* (or foreign exchange policies), which involve central bank lending to foreign entities, purchases and sales of foreign assets, and announcements about these. Although it is straightforward to group balance sheet policies this way, classifying and defining domestic balance sheet policies can be particularly challenging. Indeed, different authors categorize and define domestic balance sheet policies very differently (Bank of Canada, 2009; Borio & Disyatat, 2009, 2010; Borio & Zabai, 2016, 2018; Klyuev et al., 2009; Stone et al., 2011). This paper's categorization of domestic balance sheet policies is discussed in the paragraphs ahead.

Before moving to this paper's categorization of domestic balance sheet policies, it is important to mention that monetary policies that are not directly related to the adjustment of a positive policy rate are generally characterized as *unconventional*, *nontraditional*, or *nonstandard*. Since many central banks' policy rates decreased and reached or passed zero during the Global Financial Crisis, most monetary policies implemented during and after this period are widely considered unconventional. In other words, negative interest rate policy and balance sheet policies implemented in response to the Global Financial Crisis are commonly termed unconventional monetary policies.

*Categorization of Domestic Balance Sheet Policies*

Of course, a given domestic balance sheet policy can be unambiguously characterized as either expansionary or contractionary. What divide authors and cause confusion are the classification and description of expansionary domestic balance sheet policies.[4] Building upon information from Buiter (2008), the Bank of Canada (2009), Benford et al. (2009), Borio and Disyatat (2009, 2010), Farmer (2012, 2013), Kuroda (2013, 2017), Borio and Zabai (2016, 2018), the Federal Reserve System (2016), Williamson (2017), Mishkin (2019), and the Bank of England (2021), I classify expansionary domestic balance sheet policies into two *mutually exclusive* categories, namely, *quantitative easing* (*QE*) and *quasi-quantitative easing* (*qQE*). I define QE as a central bank outright and large-scale purchase of domestic public or private assets that increases the quantity of bank reserves. And I define qQE as a central bank large-scale expansionary domestic operation conducted to influence the economy beyond the policy rate that is either (i) non-outright and may increase the quantity of bank reserves temporarily or (ii) outright but does not increase the quantity of bank reserves.[5] In other words, as opposed to QE, which increases the quantity of bank reserves for an extended period, qQE either does not increase the quantity of bank reserves or increases it only for a limited time.

*Non-outright qQE* operations include large-scale temporary (self-reversing) operations such as large-scale term purchases (repurchase agreements) and term loans. They increase the quantity of bank reserves (thus, of central bank liabilities) temporarily when the central bank finances them by crediting accounts of depository institutions it holds. They do not increase the quantity of bank reserves (despite increasing the quantity of central bank liabilities) when the central bank finances them with money from specially created nonbank funds or facilities. In contrast, *outright qQE* operations include large-scale permanent (non-self-reversing) asset purchases that are funded through central bank securities issuances, matched reductions in the quantity of other assets, or increases in government deposit liabilities.[6] Hence, outright qQE operations either leave both the quantity of bank reserves and the total amount of central bank liabilities unchanged (when funded through matched reductions in the quantity of other assets) or leave the quantity of bank reserves unchanged while increasing the total amount of central bank liabilities (when funded through central bank securities issuances or increases in government deposit liabilities).

---

[4] More specifically, the sources of division among authors are the classification and description of expansionary domestic balance sheet policies involving actual central bank lending and asset purchases. Although central bank announcements (forward guidance) about domestic balance sheet policies that lead to decreases in the price of money or increases in the quantity of money also qualify as expansionary domestic balance sheet policies, the debate is about actual central bank lending and asset purchases, ignoring announcements about these lending and purchases.

[5] Here, *outright* means permanent or non-self-reversing, as opposed to temporary or self-reversing. An outright operation (purchase or sale) is one that is permanent (non-self-reversing). A *non-outright* operation (purchase or sale) is one that is temporary (self-reversing). *Quasi-* means having some similarities to, but being different from. That is, quasi- means *almost*.

Since QE and qQE are balance sheet policies, the central bank implements them with the objective of influencing the economy beyond the policy rate, especially when the rate is too low. As such, they differ from ordinary open market operations and discount window lending.

While QE is mainly conducted with the purpose of reducing long-term interest rates in the whole economy, qQE is primarily implemented with the intention of ameliorating liquidity and risk conditions in particular credit markets. Moreover, although each of QE and qQE may change both the quantity and composition (quality) of central bank assets, QE mainly changes the quantity while qQE primarily changes the composition. This change in composition may involve a shift toward riskier, less liquid, or longer-maturity assets.

Corresponding definitions for *quantitative tightening* (*QT*) and *quasi-quantitative tightening* (*qQT*) can be easily derived from the above definitions of QE and qQE. That is, QT refers to a central bank outright and large-scale sale of domestic public or private assets that decreases the quantity of bank reserves. And qQT refers to a central bank large-scale contractionary domestic operation conducted to influence the economy beyond the policy rate that is either (i) non-outright and may decrease the quantity of bank reserves temporarily or (ii) outright but does not decrease the quantity of bank reserves.

---

[6] Since a *sterilized* domestic balance sheet operation is a permanent (non-self-reversing) operation whose effect on the quantity of bank reserves has been reversed by a different permanent operation, it is part of outright qQE operations. In other words, it is important to distinguish between a balance sheet operation that has been *self-reversed* and one that has been *reversed by a different operation*. The former is a non-outright qQE operation while the latter is an outright qQE operation.

In summary, I qualify a central bank operation as *quantitative* if it is performed at large scale, is outright, and changes the quantity of bank reserves. In other words, a quantitative operation is one that *significantly and permanently* changes the *quantity of bank reserves*. A quantitative easing (QE) operation significantly and permanently increases the quantity of bank reserves. And a quantitative tightening (QT) operation significantly and permanently decreases the quantity of bank reserves. A *quasi-quantitative* operation is *almost* quantitative in that it is performed at large scale and changes (temporarily or permanently) the *quantity of central bank liabilities*. However, it is *not really* quantitative because either (i) it is non-outright, like in the case of non-outright qQE and non-outright qQT operations, or (ii) it does not change the quantity of bank reserves (despite changing the quantity of central bank liabilities), like in the case of outright qQE and outright qQT operations. A "non-quantitative" operation, should such terminology be given a thought by the reader (its usage is not advised), is one conducted at small scale. Thus, it is synonymous with a traditional (conventional) central bank operation. That is, a "non-quantitative" operation is one implemented in the course of a non-negative interest rate policy. Therefore, although not advised, the reader may consider ordinary open market operations and discount window lending (which are operations conducted at small scale) as "non-quantitative easing" policies.

Existing studies on this subject mostly classify domestic balance sheet policies into overlapping categories (Bank of Canada, 2009; Borio & Disyatat, 2009, 2010; Borio & Zabai, 2016, 2018; Klyuev et al., 2009), rendering their distinction difficult and confusing. The above classification of these policies into mutually exclusive categories helps overcome the difficulty and confusion.

*2.2.3. Theoretical Views on Monetary Policy's Role in the Economy and its Effectiveness Against Output and Asset Price Fluctuations (A Literature Review of Theoretical Work on the Effectiveness of Monetary Policy)*

Views on what monetary policy can or cannot do have significantly evolved over time. From Adam Smith and David Ricardo through John Maynard Keynes and Milton Friedman to present-day authors, different generations of economists have offered diverse and sometimes contradictory opinions on the function of monetary policy in the economy. Although viewpoints about its role have considerably converged over time, the theme of discussion has expanded from one mainly concentrated on its impact on output fluctuations to one that also focuses on its importance regarding asset-price fluctuations. This subsection briefly describes the main theoretical views about monetary policy's function and effectiveness, starting from the period before the Great Depression of 1929 to that before the Great Recession of 2008. (A literature review of theoretical work on measuring the effects of monetary policy is presented in **Section 2.2.4**, and a literature review of empirical work specific to quantitative easing is given in **Section 4.2**.)

Before the Great Depression of 1929, the dominant economic ideas were those of the *classical* and *neoclassical* doctrines. Prominent economists who represented these schools of thought include Adam Smith, David Ricardo, Jean-Baptiste Say, John Stuart Mill, Thomas Robert Malthus, Irving Fisher, Francis Edgeworth, Alfred Marshall, Leon Walras, Stanley Jevons, Vilfredo Pareto, and Arthur Pigou. The main views of these doctrines that are relevant to monetary policy are as follows (Mankiw, 2019; Mishkin, 2019; Sandelin et al., 2014; Vaggi & Groenewegen, 2006):[7]

1. *Wages and prices are flexible (in both the short and long runs).* Consequently, the economy is always at its full-employment equilibrium level, and there is no need for government intervention in the form of monetary or fiscal policy.[8] Wage and price flexibility imply that if aggregate output deviates from its full-employment equilibrium level, wages and prices will adjust automatically, quickly, and completely to restore equilibrium. Thus, any unemployment in the economy is either frictional or voluntary since,[9] in the aggregate, people willing to work at the market wage level will always be able to find jobs.[10]

---

[7] These views varied to some extent among the aforementioned economists. For instance, contrary to the other classical-neoclassical economists, Malthus accepted that effective demand could be insufficient or excessive (Keynes, 1936/2018).

2. *Money is neutral (in both the short and long runs).* That is, a change in the money supply does not affect real variables; it only influences nominal variables such as the price level and its growth rate (inflation).[11]

3. *Money demand is only for transactions and does not depend on the interest rate*.[12] In other words, apart from being a medium of exchange, money has no additional value in the eyes of economic agents.

---

[8] The economy is also assumed to be perfectly competitive. Markets in the economy operate under perfect competition whereby in a given market, there are countless and homogeneous participants, no barriers to entry and exit, and perfect information.

[9] *Frictional unemployment* refers to the unemployment caused by the movement of workers in and out of jobs in normal times (Dornbusch et al., 2018). *Voluntary unemployment* is a situation whereby at a given market wage level, some people choose not to work—hence, they are willingly unemployed (Blanchard, 2021). Frictional and voluntary unemployment are consistent with the full-employment equilibrium (Keynes, 1936/2018).

[10] The idea that the economy is always at full employment is equivalent to *Say's law* (Keynes, 1936/2018), which states that supply creates its own demand. Say's law refers to aggregate supply and demand and implies that there cannot be a general glut in the economy since aggregate demand always adapts to aggregate supply (Keynes, 1936/2018; Vaggi & Groenewegen, 2006). The law may also be stated as follows: the sum of all revenues in an economy is equal to the economy's total output (Vaggi & Groenewegen, 2006).

[11] *Monetary neutrality* follows from the *classical dichotomy*—the separation of real and nominal variables in an economic model—and provides an explanation for the *classical quantity theory of money*. In the identity $MV = PT$ (known as the *quantity equation* or *Fisher's equation of exchange*, where $M$ denotes the quantity of money, $V$ the velocity of circulation of money, $P$ the price level, and $T$ the number of transactions during a given period), if $V$ and $T$ are assumed to be constant, we get the classical quantity theory of money (Mankiw, 2019; Mishkin, 2019). The theory states that changes in the quantity of money lead to proportional changes in the price level (Mankiw, 2019; Mishkin, 2019). The quantity equation may also be represented as $MV = PY$, where $Y$ denotes aggregate real output.

[12] Moreover, the interest rate is determined at the *goods market equilibrium* (thus, is real), represents the price of current consumption (or equivalently, the reward for saving), and is flexible. Interest rate flexibility implies that if aggregate saving exceeds aggregate investment or vice versa, the interest rate will automatically adjust until aggregate saving equals aggregate investment.

Differently stated, the first of the three classical-neoclassical views listed above means that the market instantly corrects short-term deviations of aggregate output from its potential level without a need for government intervention in the form of monetary or fiscal policy. Indeed, government intervention may actually harm the economy by hindering the performance of the wage and price mechanisms. Moreover, taken together, the second and third views mean that monetary policy is ineffective. That is, a change in the money supply has no effect on the interest rate, aggregate real output, and employment.

When the Great Depression occurred, the classical-neoclassical diagnosis was that higher-than-normal real wages were the cause of the crisis. It was believed that with wage and price flexibility, a real-wage deflation would bring the downturn to an end (De Vroey, 2016; McDonald, 2022). However, despite the real-wage deflation, the crisis persisted for several years (De Vroey, 2016; McDonald, 2022). This failure of the classical-neoclassical doctrines to account for and provide a solution to the prolonged downturn made several economists doubt the soundness of their theories. Among these economists was John Maynard Keynes. In his seminal (and quite controversial) 1936 book titled *The General Theory of Employment, Interest, and Money* (mostly called the *General Theory*), Keynes criticized the classical-neoclassical theories for their impracticality and incompetence in tackling unemployment. Keynes's (1936/2018) views in the *General Theory* that differ from the classical-neoclassical thoughts and are relevant to monetary policy are as follows:

1. *Wages and prices are usually not flexible*. Although workers as a whole may not resist a *general* reduction in the level of *real* wages in the aggregate economy (resulting from an increase in the general price level, with nominal wages left unchanged), individual or groups of workers will resist a *relative* reduction in the level of their *nominal* wages that occurs comparatively to that of other workers.[13]

---

[13] Also, as opposed to what classical-neoclassical theories assert, a general reduction (increase) in the level of real wages does not lead to a fall (rise) in the supply of labor (Keynes, 1936/2018).

Moreover, the supply of labor is usually above that which is employed.[14] That is, aggregate employment (and output) is generally below its full-employment equilibrium level, and, hence, *involuntary unemployment* does exist.[15] Indeed, the situation where aggregate employment (and thus, output) is at its full-employment equilibrium level is a special case that rarely occurs.[16]

2. *Money is not neutral*. Although indirectly, the money supply (quantity of money) affects both nominal and real variables.[17] It affects real variables such as aggregate real income, output, and employment through its influence on the interest rate and inducement to invest.[18] Nevertheless, this indirect effect is limited compared to the impact of government spending on aggregate demand.

---

[14] "For more labour than is at present employed is usually available at the existing money-wage [nominal wage], even though the price of wage-goods [consumption goods] is rising and, consequently, the real wage falling" (Keynes, 1936/2018, p. 10).

[15] According to Keynes (1936/2018), people are *involuntarily unemployed* "if, in the event of a small rise in the price of wage-goods [consumption goods] relatively to the money-wage [nominal wage], both the aggregate supply of labour willing to work for the current money-wage and the aggregate demand for it at that wage would be greater than the existing volume of employment" (Keynes, 1936/2018, pp. 14–15). In addition, to him, there is only one level of aggregate employment (and output) consistent with equilibrium. This level of employment can be less than or equal to, but cannot be greater than full employment (Keynes, 1936/2018).

[16] "The effective demand associated with full employment is a special case, only realised when the propensity to consume and the inducement to invest stand in a particular relationship to one another. This particular relationship, which corresponds to the assumptions of the classical theory, is in a sense an optimum relationship. But it can only exist when, by accident or design, current investment provides an amount of demand just equal to the excess of the aggregate supply price of the output resulting from full employment over what the community will choose to spend on consumption when it is fully employed" (Keynes, 1936/2018, p. 25). "(…) the characteristics of the special case assumed by the classical theory happen not to be those of the economic society in which we actually live, with the result that its teaching is misleading and disastrous if we attempt to apply it to the facts of experience" (Keynes, 1936/2018, p. 3).

[17] "Money, and the quantity of money, are not direct influences at this stage of the proceedings [of the analysis of the price level as resulting from the equilibrium between the demand for and supply of products]. They have done their work at an earlier stage of the analysis. The quantity of money determines the supply of liquid resources, and hence the rate of interest, and in conjunction with other factors (particularly that of confidence) the inducement to invest, which in turn fixes the equilibrium level of incomes, output and employment and (at each stage in conjunction with other factors) the price-level as a whole through the influences of supply and demand thus established" (Keynes, 1936/2018, pp. xviii–xix).

[18] Contrary to what some people believe, Keynes neither said nor implied that monetary policy (money supply) is ineffective in the determination of aggregate output and the business cycle (Keynes, 1936/2018, introduction by Krugman). He only pointed out that monetary policy would be less effective if the interest rate were very low (since an increase in the money supply wouldn't easily decrease the interest rate further)(Keynes, 1936/2018, introduction by Krugman).

3. *Money demand is not only for transactions but also for precautionary and speculative purposes*. In other words, money demand is also influenced by the rate of interest.

The first of the three points above indicates that, in contrast to the classical-neoclassical economists who believed that nominal wages are flexible and the existing level of aggregate employment is always equal to the full-employment level, Keynes considered that nominal wages are not usually flexible and the economy is generally below its full-employment equilibrium level. Thus, according to Keynes, involuntary unemployment does exist. To him, *effective demand* (and hence, aggregate output) is generally below the full-employment equilibrium level, and, therefore, the government must intervene to push it to this level because it cannot adjust automatically.[19] The second and third points imply that, to Keynes, the interest rate is determined at the equilibrium between money demand and money supply, not at the equilibrium between aggregate saving and aggregate investment as argued by the classical-neoclassical theories (see **Footnote 12**).[20] Taken together, the three points mean that, to Keynes, when effective demand (thus, aggregate output) is insufficient, the government must intervene to push it to a level consistent with full employment; however, this intervention should be in the form of fiscal policy since money supply has a limited effect on aggregate demand.[21]

---

[19] Keynes defined *effective demand* as the aggregate demand that equals aggregate supply (Keynes, 1936/2018). That is, effective demand is the equilibrium aggregate demand.

[20] The interest rate considered here is the nominal interest rate, not the real interest rate as in the classical-neoclassical doctrines.

[21] "If, however, we are tempted to assert that money is the drink which stimulates the system to activity, we must remind ourselves that there may be several slips between the cup and the lip. For whilst an increase in the quantity of money may be expected, *cet. par.*, to reduce the rate of interest, this will not happen if the liquidity-preferences of the public are increasing more than the quantity of money; and whilst a decline in the rate of interest may be expected, *cet. par.*, to increase the volume of investment, this will not happen if the schedule of the marginal efficiency of capital is falling more rapidly than the rate of interest; and whilst an increase in the volume of investment may be expected, *cet. par.*, to increase employment, this may not happen if the propensity to consume is falling off" (Keynes, 1936/2018, pp. 151–152).

Soon after Keynes's criticism of the classical-neoclassical theories in the *General Theory* as outlined above, many economists started supporting his economic thoughts. Like Keynes, his early followers heavily pushed for active government stabilization of aggregate demand through public spending. Although their interpretations of Keynes's ideas significantly varied among themselves and from the original theories of Keynes, it is possible to label them as one group: the *early Keynesians*. This way, early Keynesians would include disciples of the *Keynesian synthesis* (like Don Patinkin, Robert Clower, Axel Leijonhufvud, and Edmond Malinvaud), *neo-Keynesian synthesis* (like John Hicks, Paul Samuelson, Robert Solow, Franco Modigliani, William Baumol, and James Tobin), and *Cambridge Keynesianism* (like Roy Harrod, Joan Robinson, Richard Kahn, and Nicholas Kaldor) (Sandelin et al., 2014). Most early Keynesians believed that monetary policy is ineffective and hardly matters to changes in the business cycle and aggregate output (Mishkin, 2019; Sims, 2011). This belief is based on their assessment that the relationship between nominal interest rate and aggregate investment (which they thought to be the only channel by which monetary policy can influence aggregate demand) is very weak (Mishkin, 2019). The culmination of early Keynesian ideas was probably the conclusion from their analysis of the *Phillips curve* that there exists a long-run (stable) trade-off relationship between inflation and unemployment (or equivalently, between inflation and output growth) (Papademos, 2003; Sandelin et al., 2014).

In the 1960s, a group of economists led by Milton Friedman—known as *monetarists* or Chicago school economists—refuted the early Keynesian ideas and adopted the view that changes in the money supply are the primary cause of output and business cycle fluctuations (Mankiw, 2019; Mishkin, 2019; Sims, 2011).[22] That is, contrary to early Keynesians (and classical-neoclassical economists), monetarists believed that monetary policy is effective and does matter to changes in aggregate demand and output (Mishkin, 2019; Sims, 2011). The *monetarism* appellation came from their contention that stabilization policies should be limited to stabilizing inflation expectations at low levels (Sandelin et al., 2014). Furthermore, monetarists disagreed with the early Keynesian thought that a weak or nonexistent association between nominal interest rate and aggregate investment is synonymous with a weak or nonexistent relationship between money supply and aggregate investment (Mishkin, 2019). To them, there could be a strong relationship between money supply and aggregate investment resulting from a strong relation between *real* interest rate and aggregate investment even if the association between *nominal* interest rate and aggregate investment is weak (Mishkin, 2019). That is, real and nominal interest rates could affect aggregate investment very differently. Moreover, monetarists believed that the relation between interest rate and aggregate investment is not the only channel by which monetary policy can influence aggregate demand (Mishkin, 2019). To them, although not fully and clearly identifiable, several other channels exist, and monetary policy can still be effective even if there is no relationship between interest rates and aggregate investment.

---

[22] Therefore, the monetarist view implies that a stable money supply leads to a stable economy (Mankiw, 2019).

Also, independent studies by Milton Friedman and Edmund Phelps in 1968 on the theory of the *natural rate of unemployment* (or equivalently, the *natural level of output*) led to the monetarist view that movements of aggregate output from its full-employment equilibrium level are due to the *adaptive expectations* of economic agents, not because of changes in aggregate demand as claimed by Keynes and the early Keynesians (Papademos, 2003; Sandelin et al., 2014). According to Friedman and Phelps, the adaptive expectations of economic agents would mislead them into false beliefs about the level of real wages. However, these false beliefs would only be temporary, and agents would quickly correct them in such a way that aggregate output shifts from full-employment levels would only occur in the short term, not in the long term (Papademos, 2003). Moreover, Friedman integrated adaptive expectations into the examination of the Phillips curve and concluded that the long-run inflation-unemployment trade-off relationship suggested by early Keynesians is not valid (Sandelin et al., 2014). According to him, in the long run, the Phillips curve is a vertical line at the natural rate of unemployment; and slops downward only in the short term (because of money illusion) (Sandelin et al., 2014). Coincidentally, these findings came at a time when stagflation—the simultaneous existence of high inflation and high unemployment in the economy—had increased suspicions about the early Keynesian long-term inflation-output trade-off suggestion.

Like monetarists, many other groups of economists rejected the ideas of Keynes and early Keynesians. One such group was known as the *new classical* school of thought. New classical economists were led by Robert Lucas (1976) and his critique that the parameters of early Keynesian macroeconomic models are not structural (stable) since they depend on the expectations of economic agents regarding the implementation of monetary policy. That is, according to Lucas (1976), early Keynesian models are not useful in forecasting the effects of alternative economic policies. His critique is commonly called the *Lucas critique of econometric policy evaluation*, or simply the *Lucas critique*. Shortly after the findings of Friedman and Phelps, new classical economists further developed the Phillips curve analysis by integrating it with *rational expectations* and the classical-neoclassical theory of wage-price flexibility. Their conclusion supported the monetarist interpretation that the long-run inflation-unemployment trade-off relationship suggested by early Keynesians is not valid (Sandelin et al., 2014). However, contrary to monetarists, they contended that the trade-off does not exist even in the short run (Sandelin et al., 2014). Therefore, according to them, it is neither possible nor necessary to stabilize output fluctuations using monetary policy (Papademos, 2003).[23] In addition to Lucas, other influential new classical economists include Thomas Sargent, Edward Prescott, and Finn Kydland (Sandelin et al., 2014).[24]

In response to the criticisms of the new classicals about Keynesian macroeconomics, a new group of economists who endeavored to defend Keynesianism was quickly formed. They later came to be called the *new Keynesians*. Major figures of this doctrine include George Akerlof and Joseph Stiglitz (Sandelin et al., 2014). Although new Keynesians accept the new classical concept of rational expectations, they still maintain the Keynesian view of inflexible (rigid) nominal prices and wages (Sandelin et al., 2014). As such, they believe that if aggregate output deviates from its full-employment equilibrium level, wages and prices will not adjust automatically to restore equilibrium. For this reason, they affirm that monetary policy should indeed be employed to stabilize aggregate output fluctuations.

---

[23] Although Lucas contended that *monetary non-neutrality* is possible in situations where economic agents confuse unforeseeable monetary shocks with changes in relative prices, most new classical economists who came after him—i.e., proponents of the *real business cycle* (*RBC*) side of the new classical school—assumed that money is completely neutral (Sandelin et al., 2014). That is, according to them, monetary policy cannot be used to affect unemployment (aggregate output) (Sandelin et al., 2014). Consequently, RBC models explain business cycle fluctuations using technological shocks and changes in tastes (preferences), not monetary policy shocks (Sandelin et al., 2014).

[24] Edward Prescott and Finn Kydland are part of the RBC side of the new classical school (Sandelin et al., 2014).

In addition to the above diverging views among different schools of thought about the role of monetary policy and its effectiveness against aggregate output fluctuations, in the 1990s, a new debate surfaced over monetary policy's importance with respect to asset-price fluctuations. The contention was about whether monetary policy should be employed to (i) pop a potential asset-price bubble before it fully forms, so as to prevent the consequences of the bursting of a larger bubble, or (ii) control the fall in asset prices that happens when a bubble fully forms and bursts, so as to stabilize aggregate output and inflation (Mishkin, 2019). In other words, the disagreement was about whether the central bank should respond *before* or *after* a bubble has fully formed and burst. At the time of this debate, the then-chairman of the Federal Reserve System—Alan Greenspan—argued against popping asset-price bubbles (Mishkin, 2019). Greenspan's view that the central bank should respond only after the bubble has fully formed and burst came to be known as the *Greenspan doctrine*, and was widely approved among central bankers (Mishkin, 2019). Greenspan's position was based on the argument that not only is it practically impossible to identify a bubble, but even if it could be identified, popping it by raising interest rates would have dangerous consequences on the aggregate economy (Mishkin, 2019). On the other hand, economists who supported the view that the central bank should pop an asset-price bubble before it fully forms contended that waiting to respond after the bubble has fully formed and burst would be costlier to the economy because stabilizing output and inflation after such an event would be extremely hard. This debate about the role of monetary policy concerning asset-price bubbles is commonly called the *leaning against asset-price bubbles* versus *cleaning up after the bubbles burst debate*, or simply the *lean versus clean debate* (Mishkin, 2019).

## 2.2.4. Theoretical Views on Measuring the Effects of Monetary Policy: Monetary Policy Transmission Mechanisms (A Literature Review of Theoretical Work on Measuring the Effects of Monetary Policy)

The ways in which monetary policy affects aggregate demand or output are called *monetary transmission mechanisms*. Because of their contrasting views on the role and effectiveness of monetary policy, different schools of thought propose strikingly distinct transmission mechanisms of monetary policy. While monetarists believe in a direct transmission mechanism, Keynesians (both early and new) believe in an indirect one. Consequently, while monetarists measure the effectiveness of monetary policy using *nonstructural models*, Keynesians measure it using *structural models* (Mishkin, 2019; Sims, 2011; see also Diebold, 1998; Pescatori & Zaman, 2011). Classical, neoclassical, and new classical economists do not believe in any of the transmission mechanisms since, as described earlier, they argue that monetary policy does not affect real variables.[25]

In the paragraphs that follow, after presenting an overview of structural and nonstructural models, I compare the Keynesian structural modeling approach to the monetarist nonstructural modeling methodology and examine the various Keynesian monetary transmission channels.

### 2.2.4.1. An Overview of Structural and Nonstructural Models

Generally defined, a structural model is one that is built or specified based on a particular economic theory (Diebold, 1998; Mishkin, 2019; see also Heckman, 2005, 2008; Pescatori & Zaman, 2011). It makes explicit assumptions regarding economic agents' behaviors and the economy, and clearly distinguishes endogenous and exogenous variables. It may be (i) a large-scale model built using systems of equations that represent decision rules (called a *simultaneous equations model* or *structural equations model*, *SEM*), like in the case of early

---

[25] See **Footnote 23**.

Keynesian macroeconomic models; or (ii) a model built on microeconomic foundations, like in the case of new Keynesian *dynamic stochastic general equilibrium* (*DSGE*) models (Diebold, 1998; Heckman, 2005, 2008; Mishkin, 2019; Pescatori & Zaman, 2011). Thus, the definition includes both models that are immune to the Lucas critique (new Keynesian models) and those that violate the critique (early Keynesian models). Henceforth, unless otherwise stated, the Keynesian appellation will refer to both early and new Keynesians.

The SEM methodology consists of identifying and estimating economic time series using multiple equations that represent decision rules from Keynesian theory. It was initiated by Trygve Haavelmo, who built upon Jan Tinbergen's proposition that economic theory should be tested with statistical models (Sims, 2011; see also Heckman, 2005, 2008). Because they were not based on rational expectations and lacked microeconomic foundations, SEMs were criticized and characterized as being only semi-structural (Diebold, 1998; Pescatori & Zaman, 2011). New Keynesian DSGE models, which are considered fully structural because they are based on rational expectations and have microeconomic foundations, were developed as a response to these criticisms (Diebold, 1998; Pescatori & Zaman, 2011). Both new Keynesian DSGE models and early Keynesian SEMs measure the effects of monetary policy by clearly specifying the various channels of the monetary transmission mechanism. Model identification is based on theory, and estimation or prediction is made by simulating exogenous shocks to selected variables of the model.

Generally defined, a nonstructural model is one that is built based on little or no economic theory (Diebold, 1998; Pescatori & Zaman, 2011; see also Heckman, 2005, 2008; Mishkin, 2019). It examines the effect of one variable on another by directly analyzing the relationship between the two variables, and imposes few or no conditions on the variables (Diebold, 1998; Pescatori & Zaman, 2011; see also Heckman, 2005, 2008; Mishkin, 2019). This definition includes (i) models based on historical narratives, like the *monetarist models* used by Milton Friedman and Anna Schwartz (Mishkin, 2019); (ii) univariate and multivariate time-series models, such as *autoregressive* (*AR*), *moving average* (*MA*), *autoregressive moving average* (*ARMA*), *autoregressive integrated moving average* (*ARIMA*), cointegration, error correction, and *vector autoregressive* (*VAR*) models (Diebold, 1998; Pescatori & Zaman, 2011); and (iii) models based on *randomized controlled trials* (*RCT*s) and *natural experiments* (Monnet & Velde, 2020; see also Heckman, 2005, 2008).[26]

The monetarist methodology of measuring the effects of monetary policy consists of identifying historical exogenous variations in the money supply (or its growth rate) and linking those variations to changes in aggregate output (or its growth rate) through single-equation regressions. Early studies that used this methodology were done in the 1960s. Examples include Friedman and Schwartz (1963a, 1963b). More recent examples include Romer and Romer (1989) and Friedman (2005). Although less frequently than in the 1960s and the two decades that followed, the monetarist method of modeling the effects of monetary policy is still used nowadays.

VAR models are arguably the most widely used nonstructural models for studying the relationship between money supply and aggregate output. A VAR model is an *n*-equation, *n*-variable system that describes how each variable depends on its past values, the past values of the other *n*-1 variables, and some exogenous shocks (Committee for the Nobel Prize in Economic Sciences in Memory of Alfred Nobel, 2011; see also Diebold, 1998; Stock & Watson, 2019). In other words, in a VAR model, all variables are endogenous (Diebold, 1998). The VAR modeling methodology establishes causation (causality) based on the notion of *Granger causality*. The idea originated from Granger (1969) and was advocated in macroeconomics by Sims (1972) (see also Christiano, 2012; Diebold, 1998; Sims, 1980).

---

[26] According to Sims (2002), the *original meaning* of a structural or nonstructural model in econometrics is different from the description provided above. He says that a model is structural in its original meaning in econometrics if it can be used to forecast the impact of a change in the economy, which may be due to a policy intervention or a natural occurrence. In other words, in that original sense, models that have explicit equations or variables representing changes in the economy are structural, and models that do not have any explicit equations or variables representing changes in the economy are nonstructural (Sims, 2002). This implies that models described here as nonstructural (such as ARIMA and VAR models) may be considered structural in the original econometric sense if they have explicit equations or variables corresponding to a change in the economy. It is based on this reasoning that Sims qualifies some VARs as structural—the so-called *structural vector autoregressions* or *SVAR*s—even though they are not actually structural in the sense of DSGE models and SEMs.

Variable *X* Granger-causes variable *Y* if past values of *X* improve our ability to predict values of *Y* (Granger, 1969; Sims, 1972; see also Diebold, 1998; Cromwell et al., 1994). Although a VAR model measures the effect of money supply on aggregate output by directly examining the relationship between the two variables (similar to a monetarist model), the VAR modeling methodology differs from the monetarist approach since VAR exogenous variations in money supply are theoretical (hypothetical) shocks, not observed historical changes. Moreover, as opposed to monetarist models, since VAR models treat all variables as endogenous, changes in the business cycle or aggregate output can be accepted as leading (preceding or causing) variations in the money supply, not only the other way round.

RCTs (real experiments) cannot be employed to examine the causal impact of monetary policy on economic outcomes since central banks cannot randomize monetary policy. However, a well-designed natural experiment can produce results that approximate those of an RCT. A natural experiment, also called a quasi-experiment, is a study that exploits a situation where units are exposed to an observable and almost random variation in treatment due to a natural occurrence, policy intervention, or institutional change (Committee for the Nobel Prize in Economic Sciences in Memory of Alfred Nobel, 2021; Fuchs-Schündeln & Hassan, 2016; Stock & Watson, 2019; see also Angrist & Krueger, 2001; B. D. Meyer, 1995; Monnet & Velde, 2020; Shadish et al., 2002). The experiment is natural or quasi-, as opposed to randomized or real, in the sense that the variation in treatment is not deliberately introduced or manipulated by a researcher (Fuchs-Schündeln & Hassan, 2016; B. D. Meyer, 1995; Shadish et al., 2002). Although the variation in treatment may consist of more than two categories, most studies consider only two: a category in which there is a change (units that experience the change are called the treatment group) and another in which there is little or no change (units that experience little or no change are called the comparison or control group).

Research designs used in natural experiments include difference in differences, regression discontinuity, instrumental variables, and interrupted time series, among others.[27] These methods are mostly employed in applied microeconomics, but their use in macroeconomics is growing. Fuchs-Schündeln and Hassan (2016) survey some of the macroeconomics papers that use them. Examples of monetary policy papers that use natural experiments include Williamson (2017) and Richardson and Troost (2009).[28] Because of the difficulty of employing design-based methods in monetary economics, a major contribution of this paper is its use of a detailed natural experimental approach to measure the causal effects of monetary policy on inflation and real GDP.

Although monetarist studies, which determine the impact of monetary policy by identifying historical exogenous variations in money supply, may be characterized as "historical natural experiments" (Cantoni & Yuchtman, 2021; Sims, 2010; see also Friedman, 2005), they are not natural experiments in the sense of the definition provided above. This is because counterfactuals of monetarist studies are specified only implicitly by the studies' identifying assumptions, not explicitly by the existence of observable control groups. The definition of a natural experiment provided above corresponds to its meaning as described by proponents of the movement Angrist and Pischke (2010) call the *Credibility Revolution in Empirical Economics*. In other words, here, a natural experiment is a *design-based* observational study.[29]

---

[27] Some of these methods are used in both nonstructural and structural modeling approaches. For instance, instrumental variables are also used in SEMs (Angrist & Krueger, 2001).

[28] Although Williamson's (2017) study may be considered an *informal* natural experiment since it uses a control and treatment group, it does not qualify as a *formal* design-based natural experiment because it does not involve statistical analyses.

[29] Therefore, it can be said that there are two types of natural experiments: *design-based natural experiments* and *historical natural experiments*. The design-based natural experimental approach is mainly used in applied microeconomics and consists of methods such as difference in differences, regression discontinuity, instrumental variables, and interrupted time series, among others. On the other hand, the historical natural experimental approach is primarily used in macroeconomics and is the hallmark of the monetarist school.

In distinguishing among different types of models, a crucial factor to consider is whether a model's approach to causal inference is based on the concept of *controlled variation* or that of *prediction*. In a model whose approach to causal inference is based on the concept of controlled variation, for variables *X* and *Y* of the model, there is said to be a causal effect of *X* on *Y* if manipulating *X* while holding all other variables constant produces a change in *Y* (Heckman, 2005, 2008). On the other hand, in a model whose approach to causal inference is based on the concept of prediction, there is said to be a causal effect of variable *X* on variable *Y* if past values of *X* improve our ability to predict values of *Y* (Holland, 1986). While SEMs, RCTs, and natural experiments make causal inference based on controlled variation, Granger-causality models like VARs make it based on prediction (Heckman, 2008; see also Diebold, 1998; Holland, 1986).

Causal inference based on controlled variation is the primary way causation (causality) is established in economics and dates back to John Stuart Mill and Alfred Marshall (Heckman, 2005, 2008). It is represented in almost every economics writing through the hallmark Latin phrase *ceteris paribus*, or its English translation *all other things being equal* (*all other things held constant*). Some authors argue that causal inference based on prediction, as described by Granger-causality models like VARs, does not represent causation in the sense of cause and effect (Cromwell et al., 1994; Holland, 1986; Shadish et al., 2002).

*2.2.4.2. The Keynesian Structural Versus Monetarist Nonstructural Modeling Approaches*
Although it is quite simplistic, **Figure 2** below compares the Keynesian structural and monetarist nonstructural modeling approaches to measuring the effects of monetary policy on the economy. As the figure indicates, the Keynesian structural modeling approach describes the monetary transmission mechanism by showing that a variation in the money supply (*M*) affects aggregate demand or output (*Y*) through a clearly specified channel. On the other hand, the monetarist nonstructural (unrestricted reduced-form) modeling approach assumes that a specific channel of monetary transmission cannot be clearly identified and, therefore, proposes that the impact of the money supply on aggregate demand or output should be measured by directly analyzing correlations between the two variables (Mishkin, 2019; Sims, 2011). In other words, the monetarist nonstructural modeling approach considers the monetary transmission mechanism to be a complex process whose internal workings cannot be clearly understood or explained—hence, it is comparable to a *black box*.

**Figure 2**

*The Keynesian Structural and Monetarist Nonstructural Modeling Approaches to Measuring the Effects of Monetary Policy*

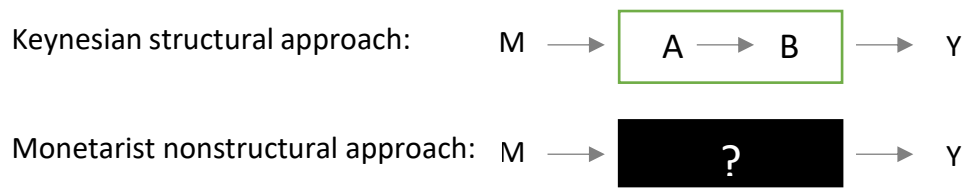

*Note*. $M$ = money supply; $Y$ = aggregate demand or output; A→B = clearly specified channel of the monetary transmission mechanism showing the relationship between two given economic variables *A* and *B* (such possible channels are discussed in **Section 2.2.4.3** below); ? = unknown channel linking the money supply to aggregate demand or output. Reprinted from Kamkoum (2022; see also 2023b), CC BY 4.0. Kamkoum (2022) adapted with permission from Mishkin (2019), copyright 2019 by F. S. Mishkin.

While it is impossible to say which of the two approaches is better, each has some clear advantages over the other. Advantages of the Keynesian structural modeling approach over the monetarist nonstructural modeling approach include its capability to give a detailed interpretation of the economy, forecast how a specific monetary policy modification can impact aggregate demand, and estimate how institutional changes can alter the relation between money supply and aggregate demand (Mishkin, 2019). However, because this approach relies on our knowledge of the correct structure of the model (that is, our understanding of all the transmission channels), it is only as good as the structural model on which it depends (Mishkin, 2019). A slight misspecification of the model is enough to generate inaccurate results. On the other hand, advantages of the monetarist nonstructural modeling approach over the Keynesian structural modeling method include its non-reliance on preconceived restrictions on the monetary transmission mechanism and its ability to gauge the full impact of a variation in the money supply on aggregate demand or output (Mishkin, 2019). Nevertheless, the nonstructural method may be misleading since correlation does not always imply causation.

*2.2.4.3. Keynesian Monetary Transmission Channels*

The Keynesian monetary transmission channels mentioned earlier can be classified into three main types: (i) *traditional Keynesian interest rate channels*, (ii) *other asset price channels*, and (iii) *credit channels* (Ireland, 2006; Mishkin, 2019). These channels represent the different ways through which both traditional and nontraditional monetary policies affect aggregate demand (Federal Reserve System, 2016). In other words, interest rate and balance sheet policies influence the economy through the same transmission channels (Federal Reserve System, 2016). Details of these channels are discussed in the following subsections.

*2.2.4.3.1. Traditional Keynesian Interest Rate Channels of the Monetary Transmission Mechanism*

According to the traditional Keynesian interest rate channels, a central bank policy that induces an increase in the short-term nominal interest rate also causes a rise in the short-term real interest rate since prices are inflexible (Ireland, 2006; Mishkin, 2019). Then, following the expectations hypothesis of the term structure, the rise in the short-term real interest rate causes the long-term real interest rate to increase. In turn, the increase in the long-term real interest rate raises the real cost of borrowing for businesses and households. This rise in the real cost of borrowing leads to a decrease in business investment and household consumption of durable goods. Finally, the decrease in investment and consumption causes aggregate demand (or output) and employment to fall. Therefore, in **Figure 2** above, according to the traditional Keynesian interest rate channels, *A* represents real interest rates and *B* represents investment and consumption.

*2.2.4.3.2. Other Asset Price Channels of the Monetary Transmission Mechanism*

Traditional Keynesian interest rate channels describe how a variation in the money supply affects aggregate demand by influencing *bond prices* (Mishkin, 2019). Other asset price channels analyze the effect of a change in the money supply on aggregate demand through the money supply's influence on *other asset prices*, such as currency prices (foreign exchange rates) and stock prices (Mishkin, 2019).

According to the *foreign exchange rates channels*, when domestic real interest rates fall, deposits denominated in the domestic currency become less attractive than those denominated in foreign currencies (Mishkin, 2019). This leads to a domestic currency depreciation since domestic currency deposits lose value relative to their foreign currency counterparts. The domestic currency depreciation makes goods produced domestically less expensive than those produced abroad. In the end, there is an increase in net exports, causing domestic aggregate demand and employment to rise. Thus, in **Figure 2** above, according to the foreign exchange rates channels, *A* represents exchange rates and *B* represents net exports.

In contrast, according to the *stock prices channels*, a rise in the short-term nominal interest rate raises stock prices, which leads to a decrease in business investment and household financial wealth. The decrease in investment and financial wealth causes aggregate consumption and output to fall (Federal Reserve System, 2016; Ireland, 2006). Therefore, in **Figure 2**, according to the stock prices channels, *A* represents stock prices and *B* represents investment and consumption.

*2.2.4.3.3. Credit Channels of the Monetary Transmission Mechanism*

Credit channels of the monetary transmission mechanism describe how a variation in the money supply affects aggregate demand by influencing the total debt of the nonfinancial sectors; that is, by altering the aggregate debt households, the government, and nonfinancial firms (businesses) obtain from financial institutions (Dornbusch et al., 2018). The theory about credit channels is commonly referred to as the *credit view* of the monetary transmission mechanism (Dornbusch et al., 2018; Mishkin, 2019). The credit view advances that an increase in adverse selection and moral hazard problems decreases lending and aggregate demand. There are two major credit channels by which monetary policy influences aggregate demand: the *bank lending channel* and *balance sheet channel* (Ireland, 2006, Mishkin, 2019). They are discussed below.

The bank-lending-channel view is based on the special function of banks as intermediaries that solve adverse selection and moral hazard problems in the financial system. This view indicates that a central bank operation that causes a contraction in the supply of bank reserves and a contraction in bank deposits forces banks that are dependent on deposits to reduce lending. The reduction in bank lending causes firms and households that are dependent on bank loans to reduce investment and consumption, respectively. Then, the decrease in business investment and household consumption causes aggregate demand and employment to fall (Ireland, 2006, Mishkin, 2019). In other words, a central bank policy that limits banks' ability to perform their role as financial intermediaries increases adverse selection and moral hazard problems and leads to a decline in aggregate investment, consumption, and demand. Therefore, in **Figure 2** above, according to the bank lending channel, *A* represents bank reserves and deposits while *B* represents investment and consumption.

The balance-sheet-channel view describes how changes in firms' and households' net worth (the difference between assets and liabilities) affect their ability and desire to borrow. There are two main balance sheet channels: the *cash flow channel* and *unanticipated price level channel* (Mishkin, 2019). Cash flow represents the difference between cash receipts and cash expenditures. According to the cash flow channel, a central bank policy that induces a decrease in nominal interest rates improves business and household balance sheets by increasing their cash flow (Mishkin, 2019). This results in less adverse selection and moral hazard in the economy, leading to increased lending, investment, consumption, and aggregate demand. Thus, in **Figure 2**, according to the cash flow channel, *A* represents business and household cash flow while *B* represents investment and consumption. According to the unanticipated price level channel, a monetary policy that induces an unanticipated increase in the general price level increases real net worth (by decreasing the real value of liabilities at a constant real value of assets), which then results in less adverse selection and moral hazard in the economy, leading to an increase in aggregate investment, consumption, and demand (Mishkin, 2019). Therefore, in **Figure 2**, according to the unanticipated price level channel, *A* represents business and household real net worth while *B* represents investment and consumption.

**2.3. The Federal Reserve's Monetary Policies Before the Global Financial Crisis: Interest Rate Policy With Limited Reserves**

The central bank of the United States, the Federal Reserve System (mostly called the Federal Reserve or the Fed), conducts monetary policy to achieve the three goals of maximum employment, stable prices, and moderate long-term interest rates mandated by the U.S. Congress (Federal Reserve System, 2016). The goal of moderate long-term interest rates is generally not considered an independent objective since the Federal Reserve can reach it by achieving price stability (L. H. Meyer, 2001b). Also, the maximum employment goal is usually interpreted as full employment (maximum sustainable employment), which refers to the highest level of employment that can be maintained in the economy without exercising an upward pressure on inflation (L. H. Meyer, 2001a, 2001b). For the above two reasons, the Congress-mandated three goals are generally viewed as two goals and commonly called the Federal Reserve's dual mandate of price stability and full employment (L. H. Meyer, 2001b).

The aforementioned two goals of price stability and full employment are given equal weights by the Federal Reserve when conducting monetary policy (Federal Reserve System, 2016). Because of this and the fact that an inflation-targeting monetary authority must give more weight to price stability than to any other policy goals, the Federal Reserve does not officially consider itself an inflation-targeting central bank (Bernanke, 2003; Jahan, 2018). However, it is arguably a flexible inflation targeter since it has adopted important elements of the inflation-targeting framework, such as high transparency in the policymaking process and proactiveness in averting inflationary pressures (Bernanke, 2003; Mester, 2018a, 2018b; L. H. Meyer, 2001b).

The Federal Reserve's goal of price stability is achieved when U.S. households and businesses do not consider expectations of changes in the average level of prices in their economic decisions (Greenspan, 1994). In other words, this goal is met when inflation is stable at or near a suitable target and is believed likely to remain so. In January 2012, the Federal Reserve announced an annual inflation rate of 2 percent as an appropriate target for the U.S. long-run inflation rate (Federal Reserve System, 2016). It was the first time the Federal Reserve made its inflation target explicit. An explicit target makes monetary policy more transparent to the public and anchors inflation expectations (Mester, 2018a). The U.S. inflation rate is mainly measured according to the personal consumption expenditures (PCE) price index. The Federal Reserve's actions can directly affect inflation since the average level of prices is primarily determined by monetary policy in the long run. When inflation is low and stable, the economy functions more effectively. Because non-monetary factors like changes in population size, population composition, types of jobs, and skill levels also affect the level of employment, the Federal Reserve does not have a fixed target for its goal of full employment (Federal Reserve System, 2016).

Before the Global Financial Crisis, the Federal Reserve conducted an interest rate monetary policy. It generally did so by choosing a non-negative target level or range for a short-term nominal interest rate called the *federal funds rate* and using several policy tools to reach this target. The federal funds rate is the rate at which depository institutions (banks) lend reserve balances to each other overnight—that is, it is the rate at the federal funds market.[30] The Federal Reserve's effective lower bound for the federal funds rate is the 0.00-0.25 range. Before the crisis, with *limited reserves* in the banking system, the Federal Reserve mainly used three policy tools to achieve its target for the federal funds rate: *open market operations*, *reserve requirements*, and *discount window lending* (Federal Reserve System, 2016). After describing these interest rate monetary policy tools, I finish the section by presenting the ongoing debate about the expansionary nature of the Federal Reserve's monetary policies immediately before the Global Financial Crisis (during the 2002-2006 period).

---

[30] Unless specifically stated, *federal funds rate* generally refers to the *effective federal funds rate* (the rate at the federal funds market—that is, the rate at which banks lend reserve balances to each other overnight), not the *target federal funds rate* set and announced to the public by the Federal Reserve during Federal Open Market—FOMC—meetings.

*2.3.1. Open Market Operations*

Open market operations consist of small-scale temporary and permanent purchases and sales of securities issued or guaranteed by the U.S. Treasury or U.S. government agencies (Federal Reserve System, 2016). Their necessity is determined by the target federal funds rate set by the Federal Open Market Committee (FOMC) (the policy-making entity of the Federal Reserve). When the FOMC establishes that an open market operation is needed to achieve or maintain the target federal funds rate, it instructs the Open Market Desk (the policy-implementing entity of the Federal Reserve, often called the Desk) at the Federal Reserve Bank of New York to conduct the operation (Federal Reserve System, 2016). They are called *open market* operations because the securities purchases and sales are made through competitive auctions between the Desk and all eligible primary dealers, not through direct interactions between the Desk and the U.S. Treasury or U.S. government agencies (Federal Reserve System, 2016; Hopper, 2019).[31]

Open market operations influence the level of the federal funds rate by directly affecting the quantity of reserves in the banking system (Federal Reserve System, 2016; Ihrig & Wolla, 2020; Wolla, 2019). In other words, although the Federal Reserve does not directly set the level of the federal funds rate, it easily affects it by setting a target for it and altering the quantity of bank reserves through open market operations. While a Federal Reserve open market purchase increases the quantity of bank reserves and exercises downward pressure on the federal funds rate, a sale decreases bank reserves and puts upward pressure on the federal funds rate (Federal Reserve System, 2016).

---

[31] *Primary dealers* are Federal-Reserve-designated securities dealers active in the market for U.S. government securities and operate out of private banking institutions (Federal Reserve System, 2016; Mishkin, 2019).

Before the Global Financial Crisis, the Federal Reserve affected the federal funds rate primarily through open market operations. Each business day, it would determine and conduct the number of open market operations necessary to achieve and maintain the target federal funds rate (Federal Reserve System, 2016). However, as discussed in **Section 3** of this paper, this process of affecting the federal funds rate through open market operations is valid only when the quantity of reserves in the banking system is *limited*. Since the quantity of bank reserves largely increased and became *ample* during the Global Financial Crisis, as discussed in **Section 3**, the Federal Reserve resorted to a different process to affect the federal funds rate during and after that period.

*2.3.2. Discount Window Lending*

Discount window lending describes direct borrowing by depository institutions from a Federal Reserve Bank. There are three categories of discount window lending: *primary credit*, *secondary credit*, and *seasonal credit* (Federal Reserve System, 2016). Primary credit is lending accessible to depository institutions that meet financial conditions set by the Federal Reserve Bank from which they are borrowing (Federal Reserve System, 2016). Secondary credit is lending accessible to depository institutions that are eligible for discount window loans but that do not meet the requirements for primary credit (Federal Reserve System, 2016). Finally, seasonal credit is lending accessible to depository institutions that encounter significant seasonal fluctuations in their loans and deposits (Federal Reserve System, 2016). Although rates on these three types of discount window loans differ, the name *discount rate* is commonly used to refer to the rate on the primary credit since it is the Federal Reserve's principal discount window program (Federal Reserve System, 2016).

The Federal Reserve uses discount window lending as a policy tool by setting the interest rates of discount window loans above the target for the federal funds rate (Federal Reserve

System, 2016). These higher rates for discount window loans function as an upper limit on the federal funds rate and help keep it at or close to the central bank's target. **Figure 3** below shows the discount window primary credit rate and effective federal funds rate for comparison.

**Figure 3**

*The Discount Window Primary Credit Rate and Effective Federal Funds Rate Over the Years*

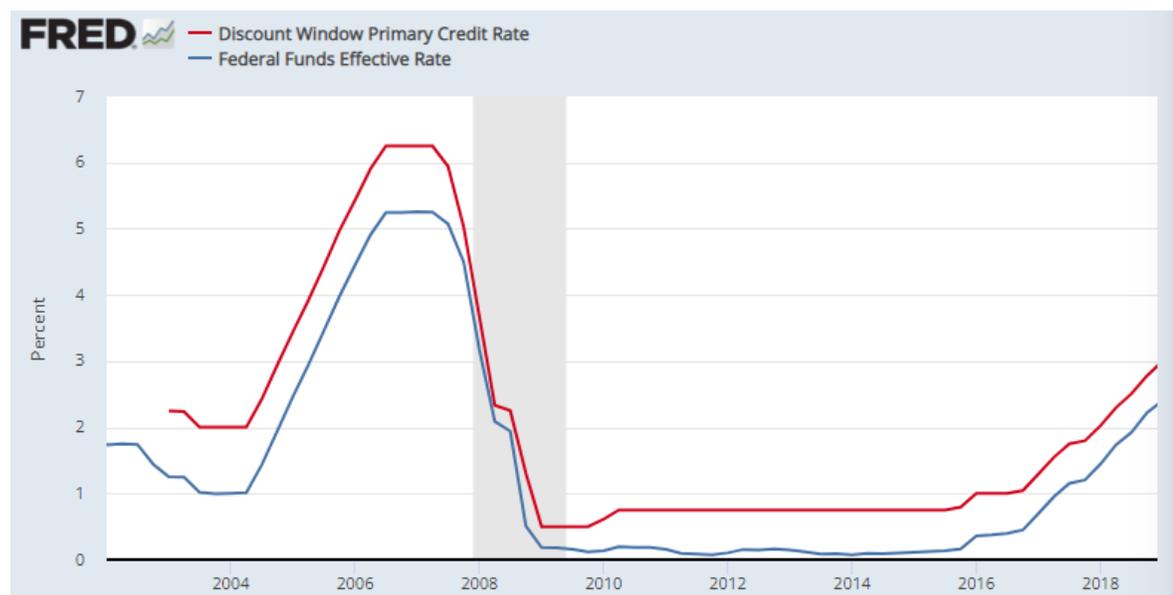

*Note*. The shaded area represents the December 2007-June 2009 U.S. recession. From Federal Reserve System/FRED (2022a, 2022b), in the public domain.

Since discount window loans are generally given at an interest rate above the federal funds rate, depository institutions mostly borrow directly from a Federal Reserve bank when their overnight funding cannot be met in the federal funds market (when they cannot borrow from other depository institutions) (Federal Reserve System, 2016). Also, depository institutions generally borrow from the discount window when they are in an emergency because such borrowing carries a stigma: it indicates that a borrowing institution may be in trouble and desperate for funds.

*2.3.3. Reserve Requirements*

Reserve requirements represent the percentage of deposits that depository institutions are required to hold as reserves. The Federal Reserve has not changed the reserve requirement ratios since 1992 (Ihrig et al., 2015). However, as shown in **Table 1** below, the limits for the deposit liabilities have frequently changed.

**Table 1**

*Low Reserve Tranche Amounts and Exemption Amounts for Depository Institutions*

| Effective date (beginning of maintenance period) | Low reserve tranche amount (millions of U.S. dollars) | Exemption amount (millions of U.S. dollars) |
|---|---|---|
| December 28, 2000 | 42.8 | 5.5 |
| December 27, 2001 | 41.3 | 5.7 |
| December 26, 2002 | 42.1 | 6.0 |
| December 25, 2003 | 45.4 | 6.6 |
| December 23, 2004 | 47.6 | 7.0 |
| December 22, 2005 | 48.3 | 7.8 |
| December 21, 2006 | 45.8 | 8.5 |
| December 20, 2007 | 43.9 | 9.3 |
| January 1, 2009 | 44.4 | 10.3 |
| December 31, 2009 | 55.2 | 10.7 |
| December 30, 2010 | 58.8 | 10.7 |
| December 29, 2011 | 71.0 | 11.5 |
| December 27, 2012 | 79.5 | 12.4 |
| January 23, 2014 | 89.0 | 13.3 |
| January 22, 2015 | 103.6 | 14.5 |
| January 21, 2016 | 110.2 | 15.2 |
| January 19, 2017 | 115.1 | 15.5 |
| January 18, 2018 | 122.3 | 16.0 |
| January 17, 2019 | 124.2 | 16.3 |

*Note*. From Federal Reserve System (2022f), in the public domain.

The reserve requirement ratios are as follows: 0% for depository institutions whose deposit liabilities are less than or equal to the exemption amount, 3% for those whose deposit liabilities are above the exemption amount but less than or equal to the low reserve tranche amount, and 10% for those whose deposit liabilities are above the low reserve tranche amount (Federal Reserve System, 2022f). Since October 1, 2008, the Federal Reserve has been paying interest on the required and excess reserves of depository institutions (this is discussed in **Section 3** under policies implemented by the Federal Reserve in response to the Global Financial Crisis).

### *2.3.4. Taylor's Critique and Bernanke's Response: The Debate About the Expansionary Nature of the Federal Reserve's Monetary Policies Immediately Before the Global Financial Crisis (During the 2002-2006 Period)*

There is an ongoing debate about whether or not the Federal Reserve's monetary policies during the 2002-2006 period (immediately before the Global Financial Crisis) were too expansionary. Although there are several economists on each side of the debate, views of those on the side that argues that the Federal Reserve's policies during this period were too expansionary can be summarized by *Taylor's critique*, and views of those that do not agree with this argument can be summarized by *Bernanke's response* to the criticism. The critique and its response are presented below.

According to Taylor (2009), the Federal Reserve's monetary policies during the 2002-2006 period were too expansionary and were the primary cause of the Global Financial Crisis. Using **Figure 4** below, he argues that from 2002 to 2006, had the Federal Reserve followed a monetary policy approach similar to the one it had implemented during the previous 20 years, the federal funds rate would have been higher than observed, there would have been no housing boom and bust, and, consequently, there would have been no financial crisis.

As Panel A of **Figure 4** shows, Taylor believes that during the 2002-2006 period, the federal funds rate was consistently below what it would have been if the Federal Reserve had followed the Taylor rule. To him, as shown in Panel B, these unusually low levels of the federal funds rate were the primary cause of the boom and bust in the U.S. housing starts during that period. Because the Global Financial Crisis followed the boom and bust, Taylor concludes that the low levels of the federal funds rate were the leading cause of the crisis.

**Figure 4**

A: *Federal Funds Rate (%), Actual and Taylor-Rule Counterfactual*

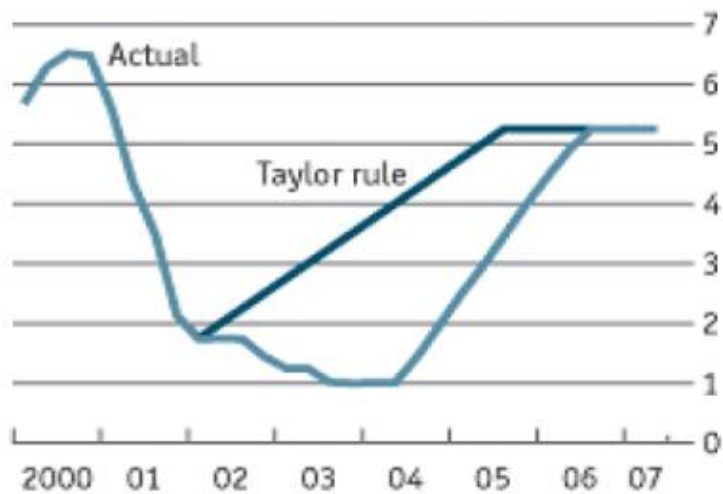

B: *The Boom-Bust in Housing Starts Compared with the Counterfactual*

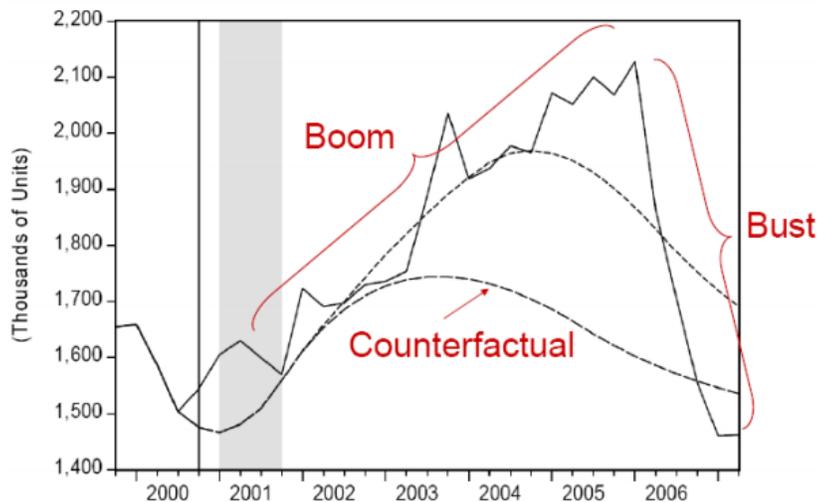

*Note*. In Panel A, the line labeled *Actual* shows the actual federal funds rate during the 2000-2006 period. The one labeled *Taylor rule* represents what the federal funds rate would have been if the Federal Reserve had followed the Taylor rule during that period. In Panel B, the jagged line shows actual housing starts during the 2000-2006 period. And the line labeled *counterfactual* represents what housing starts would have been if the Federal Reserve had followed the Taylor rule during that period. Reprinted from Kamkoum (2022; see also 2023b), CC BY 4.0. Kamkoum (2022) reprinted with permission from Taylor (2009), copyright 2009 by J. B. Taylor.

However, according to Bernanke (2015), Taylor's criticism of the Federal Reserve's monetary policies during this period is not valid. Using **Figure 5** below, Bernanke argues that with a modified Taylor rule in which core PCE inflation is used instead of the GDP deflator and an output gap coefficient of 1 is used instead of 0.5, for the period from 2002 to 2006, the actual federal funds rate is no longer below the Taylor-rule federal funds rate as stated by Taylor. Thus, Bernanke believes that Taylor's critique is invalid.

**Figure 5**

*Federal Funds Rate (%), Actual and Modified Taylor-Rule Counterfactual*

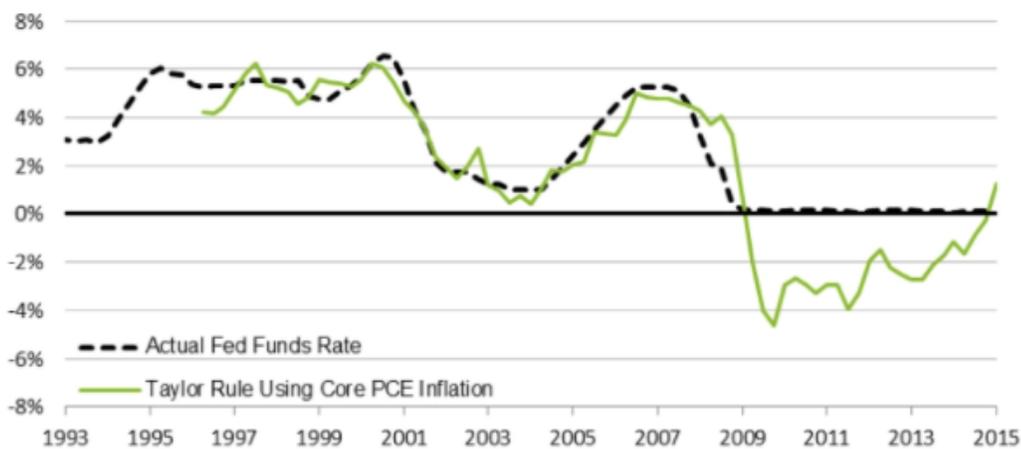

*Note*. Reprinted from Kamkoum (2022; see also 2023b), CC BY 4.0. Kamkoum (2022) reprinted with permission from Bernanke (2015), copyright 2015 by The Brookings Institution.

# 3. THE GLOBAL FINANCIAL CRISIS AND THE FEDERAL RESERVE'S RESPONSE

## 3.1. Introduction

The financial crisis that started in the U.S. at the end of 2007 and quickly spread to other countries was the most severe economic and financial disaster since the Great Depression (Mishkin, 2019; Terrones et al., 2009). The crisis began in the U.S. housing market in August 2007, rapidly extended to other sectors of the U.S. economy, and became global following the collapse of various U.S.-based international financial institutions. To counter the negative effects of the crisis, the central bank of the United States, the Federal Reserve, and other central banks conducted policies that are commonly considered novel and unusual. This section of the paper describes the policies the Federal Reserve implemented during and after the crisis. The section is organized as follows. **Section 3.2** describes the various policies the Federal Reserve conducted in response to the crisis, **Section 3.3** presents the policies the Federal Reserve implemented during policy normalization, and **Section 3.4** briefly examines the difference between the Federal Reserve's response and that of the Bank of Canada.

## 3.2. The Federal Reserve's Response to the Global Financial Crisis

Before the Global Financial Crisis, as explained in **Section 2**, the Federal Reserve implemented an interest rate monetary policy whereby it influenced the economy by affecting the federal funds rate through its use of three policy tools: open market operations, discount window lending, and reserve requirements. However, as the following paragraphs describe, in response to the Global Financial Crisis, the Federal Reserve mainly implemented a balance sheet monetary policy whereby it influenced the economy beyond the federal funds rate through its use of quantitative easing (QE) and quasi-quantitative easing (qQE) policies or policy tools.

**Figure 6** below shows that when the financial crisis broke, the Federal Reserve responded by aggressively decreasing its target for the federal funds rate (thus, conducting open market purchases more actively) and cutting the administered discount rate. These active measures resulted in corresponding drops in the market-determined federal funds rate or effective federal funds rate.[32]

**Figure 6**

*The Federal Funds Rate and Discount Rate During the Global Financial Crisis*

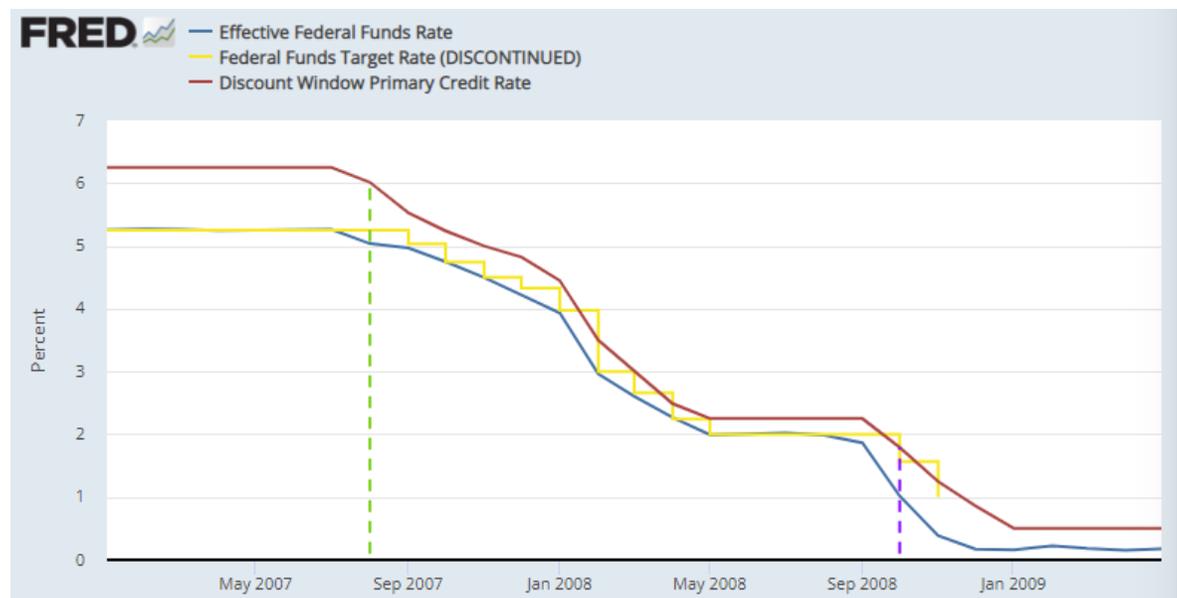

*Note*. The dashed green and purple lines represent August 2007 and October 2008, respectively. From Federal Reserve System/FRED (2022a, 2022b, 2022c), in the public domain.

As the figure shows, from August 2007 to October 2008, the effective federal funds rate and discount window primary credit rate both decreased by more than four percentage points (the effective federal funds rate dropped from about 5.25% to 1%, and the discount rate decreased from 6% to about 1.75%).

---

[32] As explained in **Footnote 30**, *federal funds rate* generally refers to the *effective federal funds rate* (the actual rate at the federal funds market), not the *target federal funds rate* set and announced by the Federal Reserve during FOMC meetings.

During the same period of August 2007 to October 2008, because reductions in the federal funds rate and discount rate were not enough to combat liquidity issues in the economy, the Federal Reserve also actively conducted qQE operations. Moreover, during that period, to control the public's expectations regarding monetary policy, the Federal Reserve also actively provided forward guidance about the federal funds rate.

This rapid response of the Federal Reserve to the financial crisis using a policy combination of near-zero federal funds rate, near-zero discount rate, forward guidance, and qQE operations was, to a great extent, unusual. Nevertheless, as the collapse of several US-based international financial institutions and especially Lehman Brothers during that same period proved, even this unusual policy mix was still insufficient to stop the crisis. Because of this and the fact that the crisis had taken a worse turn following Lehman Brothers' collapse (on September 15, 2008), the Federal Reserve announced in November 2008 that it was going to start conducting QE operations the following month.

### *3.2.1. Quasi-Quantitative Easing (qQE) Operations*

As defined in **Section 2**, a quasi-quantitative easing (qQE) operation, program, or policy (or simply quasi-quantitative easing) refers to a central bank large-scale expansionary domestic operation conducted to influence the economy beyond the policy rate that is either (i) non-outright and may increase the quantity of bank reserves temporarily or (ii) outright but does not increase the quantity of bank reserves. Based on this definition, the Federal Reserve's liquidity programs, reinvestment program, and operation twist (maturity extension program) conducted in response to the Global Financial Crisis all qualify as qQE operations. Together with their corresponding announcements or forward guidance, these various qQE operations are described below.

*3.2.1.1. Liquidity Programs*

In December 2007, due to the reluctance of banks to borrow from the discount window despite near-zero lending rates, the Federal Reserve was forced to create a program called the *Term Auction Facility* (*TAF*), through which loans were auctioned to financially sound depository institutions (Federal Reserve System, 2007; Mishkin & White, 2014; Weinberg, 2015).[33] Financially sound depository institutions allowed to participate in TAF auctions were those eligible for the discount window primary credit. Depository institutions were less reluctant to borrow through the TAF than through the discount window because the loans were provided through a market mechanism and could be received only several days after the auction (Federal Reserve System, 2016). The first TAF auction was worth $20 billion and was scheduled for Monday, December 17, with settlement on Thursday, December 20. The final TAF auction was conducted on March 8, 2010 (Federal Reserve System, 2015b, 2021; Weinberg, 2015). At its peak, outstanding credit provided through the TAF program reached $493 billion (Weinberg, 2015).

Apart from the discount window and Term Auction Facility loans it provided to depository institutions, the Federal Reserve also provided loans to non-depository financial institutions during the crisis, although it had for decades only lent to depository institutions (Federal Reserve System, 2016; Moe, 2012). These loans were given to primary dealers through the *Term Securities Lending Facility* (*TSLF*) and *Primary Dealer Credit Facility* (*PDCF*) programs (Federal Reserve System, 2016; Mishkin & White, 2014; Weinberg, 2015).

The TSLF and PDCF programs were established by the Federal Reserve in March 2008 to address liquidity shortages faced by primary dealers. Under the TSLF program, the Federal Reserve loaned relatively liquid Treasury securities to primary dealers for 28 days, taking less liquid securities as collateral (Federal Reserve System, 2008g; Weinberg, 2015). The primary purpose of the TSLF was to promote liquidity in the financing markets for Treasury and other collateral. Like the TAF, TSLF securities were loaned through auctions. The first TSLF auction was conducted on March 27, 2008, and the last on February 1, 2010 (Federal Reserve System, 2020b; Weinberg, 2015). The program peaked at $236 billion in outstanding credit on September 26, 2008 (Weinberg, 2015). On the other hand, under the PDCF, the Federal Reserve provided overnight loans to primary dealers, just like it does to depository institutions through the discount window (Bernanke, 2012; Federal Reserve System, 2008e). The PDCF ended on the same date as the TSLF, on February 1, 2010 (Federal Reserve Bank of New York, 2022b; Federal Reserve System, 2020b; Weinberg, 2015).

---

[33] "Actions taken by the Federal Reserve include the establishment of a temporary Term Auction Facility (approved by the Board of Governors of the Federal Reserve System) and the establishment of foreign exchange swap lines with the European Central Bank and the Swiss National Bank (approved by the Federal Open Market Committee)" (Federal Reserve System, 2007).

In addition to providing liquidity to depository institutions through the TAF and to primary dealers via the TSLF and PDCF, after the collapse of Lehman Brothers on September 15, 2008, the Federal Reserve started giving loans directly to borrowers and investors in specific key credit markets. These loans were provided through the following programs: the *Asset Backed Commercial Paper Money Market Mutual Fund Liquidity Facility* (*AMLF*), *Commercial Paper Funding Facility* (*CPFF*), *Money Market Investor Funding Facility* (*MMIFF*), and *Term Asset Backed Securities Loan Facility* (*TALF*) (Bernanke 2012; Mishkin & White, 2014; Weinberg, 2015).

The AMLF was announced on September 19, 2008 (Federal Reserve System, 2008f; Weinberg, 2015). Its purpose was to fund purchases of asset-backed commercial paper (ABCP) that money market mutual funds (MMMFs) wanted to sell. Through the AMLF, U.S. depository institutions, U.S. bank holding companies, and U.S. branches and agencies of foreign banks took non-recourse loans from the Federal Reserve (Weinberg, 2015). The first AMLF operation took place on September 22, 2008 (Federal Reserve System, 2020a). Just like the TSLF and PDCF, the AMLF was closed on February 1, 2010 (Federal Reserve System, 2020a, 2020c; Weinberg, 2015). During the period the program was implemented, it reached a peak of $152 billion in outstanding loans (Weinberg, 2015).

The CPFF was announced on October 7, 2008 (Federal Reserve System, 2008a). The program began on October 27, 2008, and closed on February 1, 2010 (Federal Reserve System, 2008c, 2020c). Under the CPFF, eligible U.S. issuers of commercial paper accessed liquidity by directly selling three-month unsecured and asset-backed commercial paper to a limited liability company (LLC) called CPFF LLC, created specifically for that purpose (Bernanke, 2012; Federal Reserve System, 2008a, 2020c).

The MMIFF was announced on October 21, 2008 (Federal Reserve System, 2008c). Its purpose was to complement the CPFF and AMLF by supporting lending to U.S. money market investors (Federal Reserve System, 2008c, 2020c; Weinberg, 2015). The three programs were designed to enhance credit availability by improving liquidity in short-term debt markets (Federal Reserve System, 2008c). The MMIFF expired on October 30, 2009, with no loans having been made under it (Federal Reserve System, 2020c; Weinberg, 2015).

The Federal Reserve announced the TALF on November 25, 2008 (Federal Reserve System, 2008d). The program was launched on March 03, 2009 (Federal Reserve System, 2009b). It was established to provide liquidity to households and small businesses with eligible asset-backed securities (ABS) (Federal Reserve System, 2008d, 2009b, 2020c; Weinberg, 2015). Eligible ABS were U.S. dollar-denominated ABS collateralized by student loans, auto loans, credit card loans, and loans guaranteed by the Small Business Administration (SBA). The TALF was terminated on June 30, 2010, for new loans collateralized by newly issued commercial mortgage-backed securities (CMBS), and on March 31, 2010, for new loans backed by all other types of collateral (Federal Reserve System, 2020c; Weinberg, 2015). During the period the TALF was implemented, loans provided under it reached a peak of $48 billion (Weinberg, 2015).

As **Figure 7** below shows, at its peak, outstanding credit provided by the Federal Reserve through its various liquidity programs surpassed $1.2 trillion (Federal Reserve System, 2016).[34]

**Figure 7**

*The Federal Reserve's Total Credit Outstanding During and After the Global Financial Crisis*

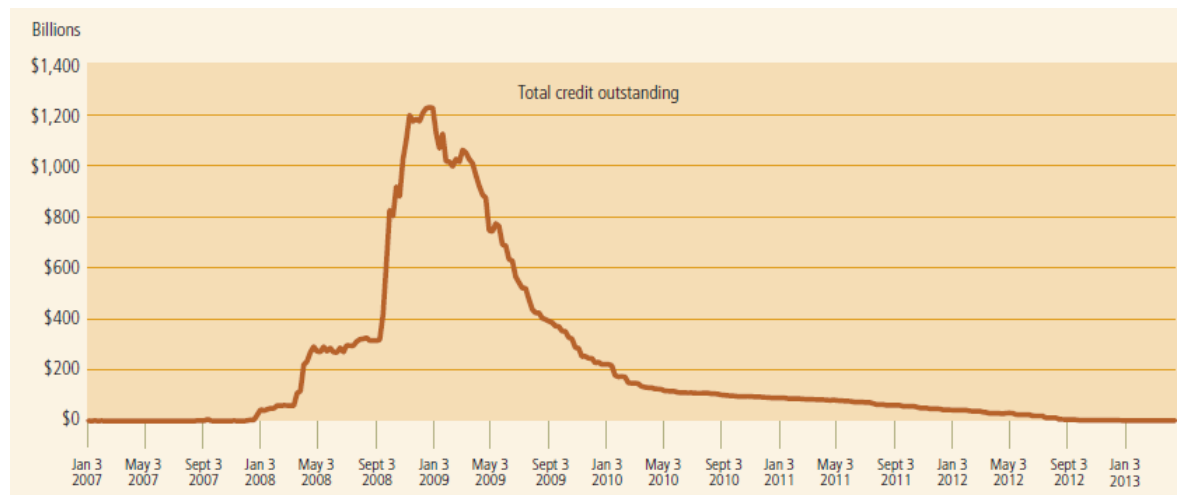

*Note*. From Federal Reserve System (2016), in the public domain.

The *Bagehot rule* states that during a financial crisis, a central bank should lend quickly and freely at high interest rates to any well-collateralized financial institution (Mishkin & White, 2014). According to Mishkin and White (2014), the Federal Reserve violated the Bagehot rule during the Global Financial Crisis because the loans it provided through its liquidity programs were not at high interest rates.

---

[34] In addition to its domestic liquidity programs, the Federal Reserve created a program called the Central Bank Liquidity Swaps program to solve the dollar liquidity problems of other central banks (Federal Reserve System, 2016).

*3.2.1.2. Reinvestment Program*

On August 10, 2010, the Federal Reserve announced it would begin maintaining its securities holdings at a constant level by reinvesting principal payments from agency debt (direct obligations of the government-sponsored enterprises Fannie Mae, Freddie Mac, and the Federal Home Loan Banks) and agency mortgage-backed securities (agency MBS for short, which are mortgage-backed securities guaranteed by Fannie Mae, Freddie Mac, and Ginnie Mae) in long-term Treasury securities (Federal Reserve Bank of New York, 2010b; Federal Reserve System, 2010a).[35] The long-term Treasury securities purchases were expected to begin on August 17, 2010 (Federal Reserve Bank of New York, 2010b). The program did not affect the existing policy of reinvesting principal payments from maturing Treasury securities in Treasury securities (Federal Reserve Bank of New York, 2010a). Through the reinvestments, the Federal Reserve kept the nominal size of domestic securities held in its System Open Market Account (SOMA) constant, thereby continuing to own a large quantity of long-term securities and maintaining downward pressure on long-term interest rates. The constant size of the securities holdings was $2.054 trillion, the August 4, 2010, value.

On September 21, 2011, the Federal Reserve announced it would change its reinvestment policy by reinvesting principal payments from agency debt and agency MBS in agency MBS instead of long-term Treasury securities (Federal Reserve Bank of New York, 2022a, 2022c; Federal Reserve System, 2011a). The shift in policy was intended to support liquidity conditions in mortgage markets. On December 16, 2015, the Federal Reserve announced that it expected to continue the reinvestment program long after it began increasing the federal funds rate target (Federal Reserve Bank of New York, 2022a; Federal Reserve

---

[35] "To help support the economic recovery in a context of price stability, the Committee will keep constant the Federal Reserve's holdings of securities at their current level by reinvesting principal payments from agency debt and agency mortgage-backed securities in longer-term Treasury securities. The Committee will continue to roll over the Federal Reserve's holdings of Treasury securities as they mature" (Federal Reserve System, 2010a).

System, 2015a, 2016).[36] On June 14, 2017, the Federal Reserve further announced that it would begin a balance sheet normalization process in which it would gradually reduce its Treasury and agency mortgage-backed securities holdings by decreasing reinvestments of principal payments from these securities (Federal Reserve Bank of New York, 2022a; Federal Reserve System, 2017a, 2017b).[37] This balance sheet normalization program of decreasing reinvestments of principal payments from Treasury and agency mortgage-backed securities was finally initiated in October 2017 (Federal Reserve Bank of New York, 2022a; Federal Reserve System, 2017c). The normalization process was still underway when the Federal Reserve began implementing new QE programs in response to the COVID-19 pandemic in March 2020.

*3.2.1.3. Maturity Extension Program (Operation Twist)*

When the Federal Reserve stated on September 21, 2011, that it would change its reinvestment policy by reinvesting principal payments from agency debt and agency MBS in agency MBS instead of long-term Treasury securities, it also announced that it would begin a maturity extension program in which it would purchase $400 billion of Treasury securities with remaining maturities of 6 to 30 years and sell an equal amount of Treasury securities with remaining maturities of 3 years or less by the end of June 2012 (Federal Reserve Bank of New York, 2022a; Federal Reserve System, 2011a). Therefore, through the maturity extension program, the Federal Reserve intended to replace shorter-term Treasury securities in its portfolio with long-term Treasury securities. By doing so, it aimed to reduce the supply of long-term Treasury securities in the corresponding financial market, thereby putting downward pressure on long-term interest rates. Although the sales of shorter-term

---

[36] "The Committee is maintaining its existing policy of reinvesting principal payments from its holdings of agency debt and agency mortgage-backed securities in agency mortgage-backed securities and of rolling over maturing Treasury securities at auction, and it anticipates doing so until normalization of the level of the federal funds rate is well under way" (Federal Reserve System, 2015a).

[37] "The Committee currently expects to begin implementing a balance sheet normalization program this year, provided that the economy evolves broadly as anticipated. This program, which would gradually reduce the Federal Reserve's securities holdings by decreasing reinvestment of principal payments from those securities, is described in the accompanying addendum to the Committee's Policy Normalization Principles and Plans" (Federal Reserve System, 2017b).

Treasury securities would put upward pressure on short-term Treasury rates, the pressure was not expected to be significant since the federal funds rate was at the zero lower bound.

Although the maturity extension program was intended to be terminated by the end of June 2012, the Federal Reserve announced on June 20, 2012, that it would extend it through the end of 2012 due to weak economic conditions (Federal Reserve Bank of New York, 2022a; Federal Reserve System, 2011a, 2012a, 2013c, 2019).[38] The Federal Reserve stated that this continuation of the program would increase the program's total amount by $267 billion (Federal Reserve Bank of New York, 2012; Federal Reserve System, 2013c). When the program ended in December 2012, the Federal Reserve had sold or redeemed $667 billion of Treasury securities with maturities of 3 years or less and used the proceeds to buy an equivalent amount of Treasury securities with maturities of 6 to 30 years (Federal Reserve Bank of New York, 2022a; Federal Reserve System, 2013c, 2016).

**Table 2** below summarizes the Federal Reserve's various qQE operations (programs) conducted in response to the Global Financial Crisis.

---

[38] "The Committee also decided to continue through the end of the year its program to extend the average maturity of its holdings of securities. Specifically, the Committee intends to purchase Treasury securities with remaining maturities of 6 years to 30 years at the current pace and to sell or redeem an equal amount of Treasury securities with remaining maturities of approximately 3 years or less. This continuation of the maturity extension program should put downward pressure on longer-term interest rates and help to make broader financial conditions more accommodative" (Federal Reserve System, 2012a).

**Table 2**

*The Federal Reserve's Quasi-Quantitative Easing (qQE) Programs Conducted in Response to the Global Financial Crisis*

| Quasi-Quantitative Easing (qQE) Program | | Announcement Date | Period Active | Description |
|---|---|---|---|---|
| Liquidity Programs | Term Auction Facility (TAF) | December 12, 2007 | December 17, 2007 - March 8, 2010 | It was created to address the reluctance of depository institutions (banks) to borrow from the discount window despite near-zero lending rates. Loans were auctioned to financially sound depository institutions eligible for the discount window primary credit. At its peak, outstanding credit provided through the program reached $493 billion. |
| | Term Securities Lending Facility (TSLF) | March 11, 2008 | March 27, 2008 - February 1, 2010 | It was established to address liquidity shortages faced by primary dealers. Under the program, the Federal Reserve loaned relatively liquid Treasury securities to primary dealers for 28 days, taking less liquid securities as collateral. Like in the TAF program, securities were loaned through auctions. The program peaked at $236 billion in outstanding credit. |
| | Primary Dealer Credit Facility (PDCF) | March 16, 2008 | March 17, 2008 - February 1, 2010 | It was set up to address liquidity shortages faced by primary dealers. Under the program, the Federal Reserve provided overnight loans to primary dealers, just like it does to depository institutions through the discount window. |
| | Asset Backed Commercial Paper Money Market Mutual Fund Liquidity Facility (AMLF) | September 19, 2008 | September 22, 2008 - February 1, 2010 | It was designed to increase credit availability to money market mutual funds. Through the program, U.S. depository institutions, U.S. bank holding companies, and U.S. branches and agencies of foreign banks took non-recourse loans from the Federal Reserve. During the period the program was implemented, it reached a peak of $152 billion in outstanding loans. |
| | Commercial Paper Funding Facility (CPFF) | October 7, 2008 | October 27, 2008 - February 1, 2010 | It was initiated to increase credit availability to U.S. issuers of commercial paper. Under the program, eligible U.S. issuers of commercial paper accessed liquidity by directly selling three-month unsecured and asset-backed commercial paper to a limited liability company (LLC) called CPFF LLC, created specifically for that purpose. |
| | Money Market Investor Funding Facility (MMIFF) | October 21, 2008 | Expired on October 30, 2009 without ever being active | It was created to complement the CPFF and AMLF by increasing credit availability to U.S. money market investors. The program expired on October 30, 2009, with no loans having been made under it. |
| | Term Asset Backed Securities Loan Facility (TALF) | November 25, 2008 | March 03, 2009 - June 30, 2010 | It was established to provide liquidity to households and small businesses with eligible asset-backed securities (ABS). Eligible ABS were U.S. dollar-denominated ABS collateralized by student loans, auto loans, credit card loans, and loans guaranteed by the Small Business Administration (SBA). During the period the program was conducted, it reached a peak of $48 billion. |
| Reinvestment Program | | August 10, 2010 | August 17, 2010 - March 2020 | It was created to maintain the Federal Reserve's securities holdings at a constant level. Between August 17, 2010, and September 20, 2011, the Federal Reserve achieved this goal by reinvesting principal payments from agency debt and agency mortgage-backed securities (agency MBS) in long-term Treasury securities. From September 21, 2011, onward, the Federal Reserve accomplished the objective by reinvesting principal payments from agency debt and agency MBS in agency MBS instead of long-term Treasury securities. The reinvestment program was still underway when the Federal Reserve began implementing new QE programs in response to the COVID-19 pandemic in March 2020. |
| Maturity Extension Program | | September 21, 2011 | September 21, 2011 - December 2012 | It was established to replace short-term Treasury securities in the Federal Reserve's portfolio with long-term Treasury securities. Specifically, it involved the Federal Reserve selling or redeeming Treasury securities with maturities of 3 years or less and using the proceeds to buy an equivalent amount of Treasury securities with maturities of 6 to 30 years. When the program ended in December 2012, the Federal Reserve had sold or redeemed $667 billion of Treasury securities with maturities of 3 years or less and had bought an equivalent amount of Treasury securities with maturities of 6 to 30 years. |

## 3.2.2. Quantitative Easing (QE) Operations

A quantitative easing (QE) operation, program, or policy (or simply quantitative easing) is defined in **Section 2** as a central bank outright and large-scale purchase of domestic public or private assets that increases the quantity of bank reserves. With Lehman Brothers collapsing and the target federal funds rate reaching 1% in September and October 2008, respectively, the Federal Reserve announced on November 25, 2008, that it would engage in the purchase of long-term securities through a series of large-scale asset purchase programs. Based on the above definition of QE, these Federal Reserve large-scale asset purchase programs qualify as QE operations. The Federal Reserve's intentions with the QE operations or programs were to support economic activity by putting downward pressure on long-term interest rates. By the end of the financial crisis, the Federal Reserve had conducted a total of three QE programs. Together with their corresponding announcements or forward guidance, these three QE programs are described below.

### 3.2.2.1. First QE Program (QE1)
The Federal Reserve's first quantitative easing program (QE1) was announced on November 25, 2008 (Borio & Zabai, 2016; Federal Reserve System, 2008b). The Federal Reserve stated that over several quarters it would purchase up to $100 billion in direct obligations of the government-sponsored enterprises (GSE) Fannie Mae, Freddie Mac, and the Federal Home Loan Banks, and up to $500 billion in mortgage-backed securities guaranteed by Fannie Mae, Freddie Mac, and Ginnie Mae (Borio & Zabai, 2016; Federal Reserve System, 2008b).[39] Purchases of the GSE direct obligations (agency debt) were to be made with primary dealers through competitive auctions and were scheduled to begin a week after the announcement (Federal Reserve System, 2008b). On the other hand, purchases of the mortgage-backed securities (agency MBS) were to be made with asset managers selected through a competitive process and were scheduled to begin before year-end (Federal Reserve System, 2008b). The purchases of agency debt began on December 5, 2008, while those of agency MBS started on January 5, 2009 (Federal Reserve Bank of New York, 2022a).

---

[39] "Purchases of up to $100 billion in GSE direct obligations under the program will be conducted with the Federal Reserve's primary dealers through a series of competitive auctions and will begin next week. Purchases of up to $500 billion in MBS will be conducted by asset managers selected via a competitive process with a goal of beginning these purchases before year-end. Purchases of both direct obligations and MBS are expected to take place over several quarters" (Federal Reserve System, 2008b).

On March 18, 2009, due to persistent poor economic conditions, the Federal Reserve announced it would purchase $300 billion in long-term Treasury securities plus an additional $100 billion in agency debt and $750 billion in agency MBS (Borio & Zabai, 2016; Federal Reserve System, 2009a).[40] The long-term Treasury securities purchases began on March 25, 2009 (Federal Reserve Bank of New York, 2022a). QE1 was completed at the end of March 2010, with the Federal Reserve having purchased $175 billion in agency debt, $1.25 trillion in agency MBS, and $300 billion in long-term Treasury securities under the program (Federal Reserve Bank of New York, 2022a; Federal Reserve System, 2010b, 2016; Williamson, 2017).

*3.2.2.2. Second QE Program (QE2)*

Because the economy was slow to recover from the crisis, the Federal Reserve announced a second quantitative easing program (QE2) on November 03, 2010 (Federal Reserve System, 2010c). In the announcement, the Federal Reserve said it planned to increase its securities holdings by purchasing an additional $600 billion in long-term Treasury securities at a pace of approximately $75 billion per month before the end of 2011Q2 (Federal Reserve System, 2010c).⁴¹ The program started in the same month and ended in June 2011 (Federal Reserve Bank of New York, 2022a; Federal Reserve System, 2011b, 2016; Williamson, 2017). At the end of QE2, the Federal Reserve had purchased $600 billion in long-term Treasury securities under the program (Federal Reserve Bank of New York, 2022a; Federal Reserve System, 2016; Williamson, 2017).

---

[40] "To provide greater support to mortgage lending and housing markets, the Committee decided today to increase the size of the Federal Reserve's balance sheet further by purchasing up to an additional $750 billion of agency mortgage-backed securities, bringing its total purchases of these securities to up to $1.25 trillion this year, and to increase its purchases of agency debt this year by up to $100 billion to a total of up to $200 billion. Moreover, to help improve conditions in private credit markets, the Committee decided to purchase up to $300 billion of longer-term Treasury securities over the next six months" (Federal Reserve System, 2009a).

*3.2.2.3. Third QE Program (QE3)*

Despite the Federal Reserve's extensive efforts with its first two quantitative easing programs, the economy's recovery was still slow. Therefore, to reduce the persistent high unemployment rates recorded after QE2, the Federal Reserve announced a third quantitative easing program (QE3) on September 13, 2012 (Borio & Zabai, 2016; Federal Reserve System, 2012b, 2019). The Federal Reserve stated in the announcement that it would purchase additional agency MBS at a pace of $40 billion per month (Federal Reserve System, 2012b).⁴² QE3 differed from the first two QE programs in that it was open-ended—the Federal Reserve did not communicate the program's total size when it was announced (Borio & Zabai, 2016; Federal Reserve System, 2016). The purchases started in the same month that year. Four months later, judging that unemployment was still high, the Federal Reserve announced on January 30, 2013, that it would supplement the $40-billion-per-month purchases of agency MBS with purchases of long-term Treasury securities at a pace of $45 billion per month (Federal Reserve System, 2013b, 2016).⁴³

---

⁴¹ "(…) the Committee intends to purchase a further $600 billion of longer-term Treasury securities by the end of the second quarter of 2011, a pace of about $75 billion per month" (Federal Reserve System, 2010c).

⁴² "To support a stronger economic recovery and to help ensure that inflation, over time, is at the rate most consistent with its dual mandate, the Committee agreed today to increase policy accommodation by purchasing additional agency mortgage-backed securities at a pace of $40 billion per month" (Federal Reserve System, 2012b).

⁴³ "To support a stronger economic recovery and to help ensure that inflation, over time, is at the rate most consistent with its dual mandate, the Committee will continue purchasing additional agency mortgage-backed securities at a pace of $40 billion per month and longer-term Treasury securities at a pace of $45 billion per month" (Federal Reserve System, 2013b).

After each of the seven FOMC meetings between December 1, 2013, and September 30, 2014, the Federal Reserve reduced the pace of agency MBS and long-term Treasury securities purchases by $5 billion per month each (Federal Reserve Bank of New York, 2022a; Federal Reserve System, 2013a, 2014a, 2014b, 2014c, 2014d, 2014e, 2014g). Specifically, the pace of agency MBS and long-term Treasury securities purchases decreased from (i) $40 to $35 billion per month and $45 to $40 billion per month, respectively, after the December 18, 2013, FOMC meeting; (ii) $35 to $30 billion per month and $40 to $35 billion per month, respectively, after the January 29, 2014, FOMC meeting; (iii) $30 to $25 billion per month and $35 to $30 billion per month, respectively, after the March 19, 2014, FOMC meeting; (iv) $25 to $20 billion per month and $30 to $25 billion per month, respectively, after the April 30, 2014, FOMC meeting; (v) $20 to $15 billion per month and $25 to $20 billion per month, respectively, after the June 18, 2014, FOMC meeting; (vi) $15 to $10 billion per month and $20 to $15 billion per month, respectively, after the July 30, 2014, FOMC meeting; and (vii) $10 to $5 billion per month and $15 to $10 billion per month, respectively, after the September 17, 2014, FOMC meeting (Federal Reserve System, 2013a, 2014a, 2014b, 2014c, 2014d, 2014e, 2014g).

After gradually reducing the pace of agency MBS and long-term Treasury securities purchases between December 2013 and September 2014 (as discussed above), with the economy in good shape, the Federal Reserve finally terminated the purchases in October 2014, thereby concluding QE3 (Federal Reserve Bank of New York, 2022a; Federal Reserve System, 2014f, 2016, 2019; Williamson, 2017). At the end of QE3, the Federal Reserve had purchased $823 billion in agency MBS and $790 billion in long-term Treasury securities under the program (Federal Reserve Bank of New York, 2022a).

**Table 3** below summarizes the Federal Reserve's three QE operations (programs) conducted in response to the Global Financial Crisis. The table is followed by **Figure 8**, which shows the Federal Reserve's balance sheet during its QE and qQE programs conducted in response to the Global Financial Crisis.

**Table 3**

*The Federal Reserve's Quantitative Easing (QE) Programs Conducted in Response to the Global Financial Crisis*

| Quantitative Easing (QE) Program | Announcement Date | Period Active | Description |
|---|---|---|---|
| First QE Program (QE1) | November 25, 2008 | December 2008 - March 2010 | The Federal Reserve purchased $175 billion in agency debt, $1.25 trillion in agency mortgage-backed securities (agency MBS), and $300 billion in long-term Treasury securities. |
| Second QE Program (QE2) | November 03, 2010 | November 2010 - June 2011 | The Federal Reserve purchased $600 billion in long-term Treasury securities. |
| Third QE Program (QE3) | September 13, 2012 | September 2012 - October 2014 | The Federal Reserve purchased $823 billion in agency MBS and $790 billion in long-term Treasury securities. |

**Figure 8**

*The Federal Reserve's Balance Sheet During the QE and qQE Programs Conducted in Response to the Global Financial Crisis*

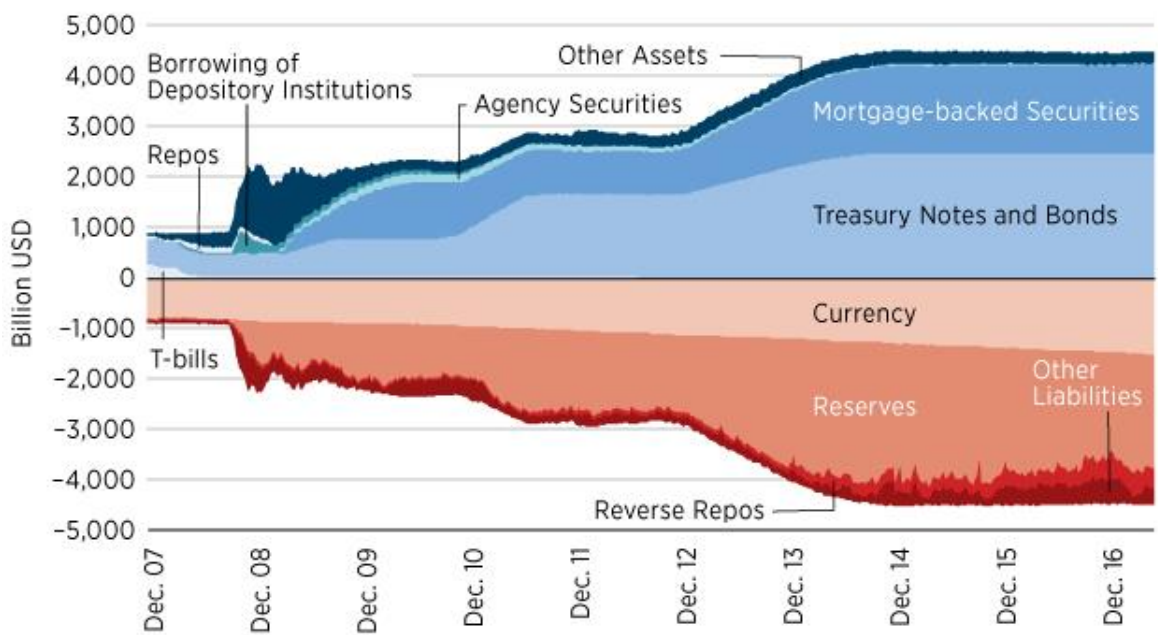

*Note*. Reprinted from Kamkoum (2022; see also 2023b), CC BY 4.0. Kamkoum (2022) reprinted with permission from Williamson (2017), copyright 2017 by the Federal Reserve Bank of St. Louis.

*3.2.3. Unusual But Not Unprecedented Monetary Policies*

Although many economists characterize the Federal Reserve's response to the Global Financial Crisis as unusual, according to Mishkin and White (2014), policies the Federal Reserve implemented during the crisis were not that unprecedented. As shown in **Table 4** below, they argue that almost every monetary policy the Federal Reserve implemented during the Global Financial Crisis had already been implemented at least once during the past crises.

**Table 4**

*Unprecedented Actions by Monetary Authorities in Financial Crises*

|  | Unusual Monetary Easing | Non-Bagehot Liquidity Facilities | International Central Bank Cooperation | Non-Conventional Monetary Policy | Rescue/ Orderly Liquidation of Financial Institution | Direct Treasury Collaboration/ Intervention/ Aid | Supervision |
|---|---|---|---|---|---|---|---|
|  | 1 | 2 | 3 | 4 | 5 | 6 | 7 |
| U.S. 2008 | X | X | X | X | X | X | X |
| U.K. 1866 |  |  |  |  |  | X |  |
| France 1889 |  |  |  |  | X | X |  |
| U.K. 1890 |  |  | X |  | X | X |  |
| U.S. 1907 |  | X | X |  |  | X |  |
| U.S. 1929 |  | X |  |  |  |  |  |
| U.S. 1930-1933 |  |  |  | X | X | X | X |
| Penn Central 1970 |  | X |  |  |  |  |  |
| Continental Illinois 1984 |  |  |  |  | X |  |  |
| Crash of 1987 |  | X |  |  |  |  |  |
| LTCM 1998 | X |  |  |  | X |  |  |

*Note*. LTCM stands for *Long-Term Capital Management*. From Mishkin and White (2014), copyright 2014 by the Federal Reserve Bank of Dallas (reprinting without permission is allowed).

## 3.3. Policy Normalization

### *3.3.1. Announcements and Their Consequences: Taper Tantrum*

In testimony on May 22, 2013, the then-chairman of the Federal Reserve, Ben Bernanke, announced to the U.S. Congress that economic conditions warranted a near-future start of monetary policy normalization whereby the Federal Reserve would gradually slow the pace of its asset purchases (Bernanke, 2013).[44] A central bank's gradual slowing of the pace of its asset purchases is commonly called *monetary tapering* (Milstein et al., 2021; Neely, 2014). Monetary tapering does not refer to a sale of assets; it only refers to a reduction in the pace at which assets are purchased (Milstein et al., 2021). In other words, monetary tapering is different from quantitative tightening: when a central bank does tapering, the size of its balance sheet still increases. A second announcement regarding the Federal Reserve's policy normalization was made on June 19, 2013, by Bernanke during a press conference where he again confirmed that economic conditions were good and that tapering was on its way (Neely, 2014).

Bernanke's announcements about tapering were each followed by high market volatility, increasing interest rates, U.S. dollar appreciation, and emerging-market capital outflows. This peculiar market reaction to Bernanke's announcements is commonly referred to as *taper tantrum* (Milstein et al., 2021; The Economist, 2021). As Panels A and B of **Figure 9** below show, following Bernanke's announcements about tapering, the long-term U.S. bond yields increased sharply and the U.S. dollar highly appreciated relative to other currencies. This shows that the announcements were unexpected by the public. The bond yields rose because the market concluded from the announcements that monetary policy would soon be tighter.

---

[44] "If we see continued improvement and we have confidence that that is going to be sustained, then we could in the next few meetings, take a step down in our pace of purchases" (Bernanke, 2013, p. 11).

**Figure 9**

A: *Asset Prices Around the June 19, 2013, FOMC Meeting and Press Conference*

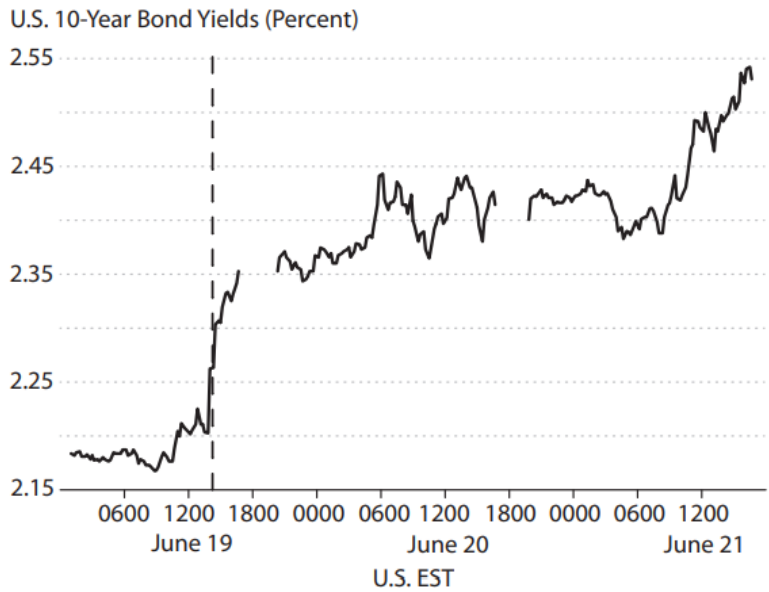

B: *Value of the Dollar in Foreign Exchange Markets*

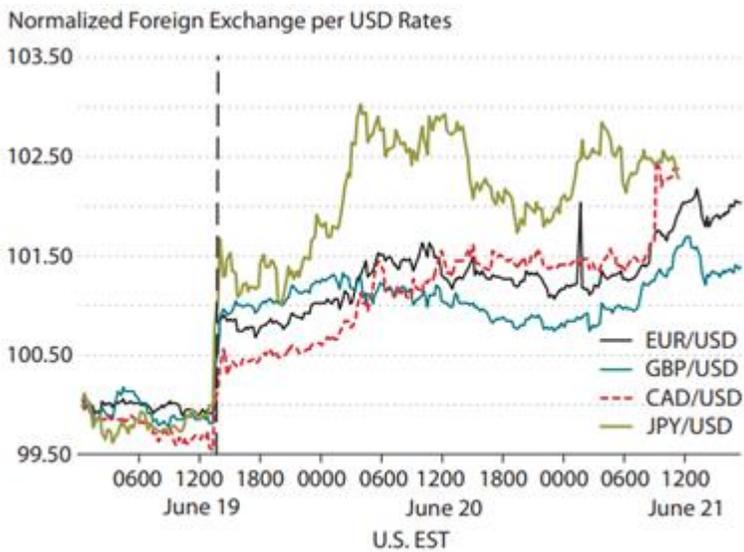

*Note*. Reprinted from Kamkoum (2022; see also 2023b), CC BY 4.0. Kamkoum (2022) reprinted with permission from Neely (2014), copyright 2014 by the Federal Reserve Bank of St. Louis.

The announcements also implied that the federal funds rate, which had been at the zero lower bound since December 2008 (for almost five years before the announcements on tapering), was soon going to start increasing.

Although the taper tantrum did not affect the U.S. economy severely, its effects on emerging markets were very harsh (Milstein et al., 2021). Indeed, as shown in **Figure 10** below, emerging market economies like Turkey, India, Indonesia, South Africa, and Brazil experienced high capital outflows, currency depreciation, equity price drops, and long-term bond yield increases.

**Figure 10**

*Variation of Asset Price Responses Across Countries Following the May–June 2013 Announcements on Tapering (z-scores)*

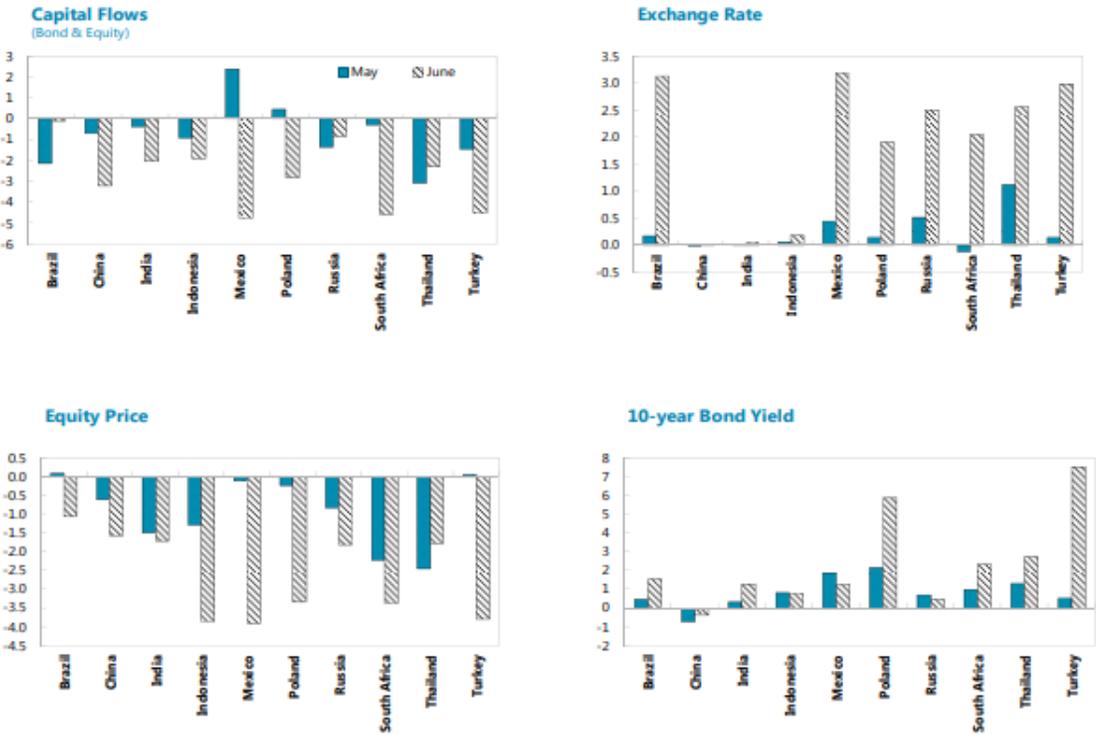

*Note*. Reprinted from Kamkoum (2022; see also 2023b), CC BY 4.0. Kamkoum (2022) reprinted with permission from Sahay et al. (2014), copyright 2014 by the International Monetary Fund (IMF).

### *3.3.2. Implementation: Monetary Tapering and Interest Rate Policy With Ample Reserves*

Despite the Taper Tantrum caused by Bernanke's announcements, the Federal Reserve communicated six months later, on December 18, 2013, that the planned monetary tapering would begin in January 2014.[45] The beginning of the tapering was intended to mark the start of the first part of the implementation of a twofold normalization process consisting of normalizing the balance sheet on the one hand and the federal funds rate on the other hand.

The *balance sheet normalization process* occurred as described in **Sections 3.2.2.3** and **3.2.1.2** above. That is, the Federal Reserve normalized its balance sheet by (i) reducing the pace of agency MBS and long-term Treasury securities purchases by $5 billion per month each after each of the seven FOMC meetings between December 1, 2013, and September 30, 2014; (ii) completely terminating the agency MBS and long-term Treasury securities purchases in October 2014; and (iii) decreasing reinvestments of principal payments from agency MBS and long-term Treasury securities in October 2017. In other words, the Federal Reserve implemented the balance sheet normalization process by first reducing the pace of its QE operations, then terminating the QE operations altogether, and finally decreasing the size of its reinvestment program.

The *federal funds rate normalization process*, on the other hand, began on December 16, 2015, when the Federal Reserve announced that it would start normalizing the level of the federal funds rate the same day by increasing its target range from the effective lower bound range of 0.00-0.25 to a range of 0.25-0.50.[46] This effectively meant that the federal funds rate target would change for the first time since it reached the zero lower bound exactly seven years earlier, on December 16, 2008. The Federal Reserve normalized the level of the federal funds rate mainly by modifying the interest rate it paid on *excess reserves* (commonly called the *IOER rate*). As **Figure 11** below shows, the Federal Reserve used the IOER rate as a monetary policy tool during the federal funds rate normalization period to exert downward and upward pressure on the federal funds rate.

---

[45] "In light of the cumulative progress toward maximum employment and the improvement in the outlook for labor market conditions, the Committee decided to modestly reduce the pace of its asset purchases. Beginning in January, the Committee will add to its holdings of agency mortgage-backed securities at a pace of $35 billion per month rather than $40 billion per month, and will add to its holdings of longer-term Treasury securities at a pace of $40 billion per month rather than $45 billion per month" (Federal Reserve System, 2013a).

**Figure 11**

*Behavior of the Effective Federal Funds Rate Relative to the Interest on Excess Reserves (IOER) and Interest on Required Reserves (IORR) Rates*

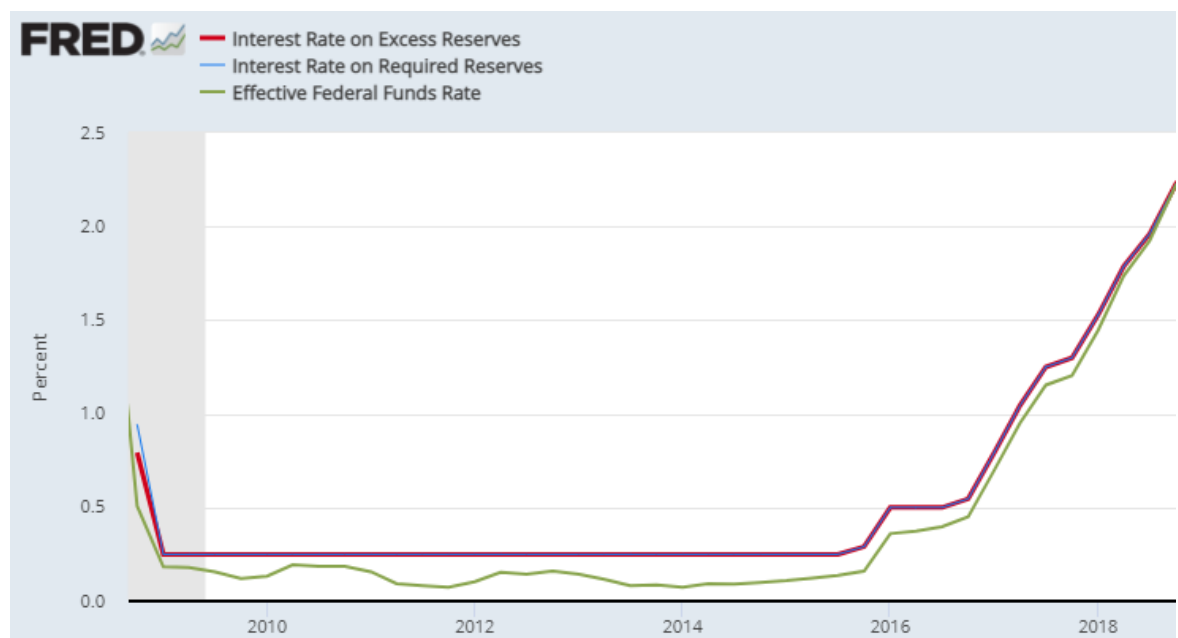

*Note*. From Federal Reserve System/FRED (2022b, 2022d, 2022e), in the public domain.

The graph shows that, during that period, the Federal Reserve progressively set the IOER rate at levels higher than those it set during the crisis. In other words, the Federal Reserve set the IOER rate low when it intended to maintain the effective federal funds rate low and increased it when it wanted the effective federal funds rate to be higher. Also, the rate on *required reserves* (commonly called the *IORR rate*) was appropriately changed when the cost to depository institutions of holding required balances increased or decreased.[47]

---

[46] "Given the economic outlook, and recognizing the time it takes for policy actions to affect future economic outcomes, the Committee decided to raise the target range for the federal funds rate to 1/4 to 1/2 percent" (Federal Reserve System, 2015a).

During the federal funds rate normalization process, in addition to paying interest on excess reserves, the Federal Reserve used another tool called *overnight reverse repurchase* (*ON RRP*) to control the federal funds rate. An ON RRP was conducted as follows: on a given day, an eligible financial counterparty supplied funds to the Federal Reserve in exchange for Treasury securities found on the Federal Reserve's balance sheet and was paid the ON RRP rate, then on the day that followed, the counterparty returned those securities to the Federal Reserve, and the Federal Reserve returned the funds to the counterparty (Federal Reserve System, 2016). The Federal Reserve could control the effective federal funds rate with an ON RRP because any financial institution eligible to participate in such an operation would not be willing to trade its funds with another institution at a rate lower than the ON RRP rate.

Therefore, the Federal Reserve conducted monetary policy during the federal funds rate normalization period mainly by setting the IOER and ON RRP rates equal to the upper and lower limits of its target range for the federal funds rate, respectively (Federal Reserve System, 2016; Ihrig & Wolla, 2020; Wolla, 2019). That is, starting from December 2015, because of ample reserves in the banking system, instead of influencing the federal funds rate through open market operations, the Federal Reserve did it by changing the IOER and ON RRP rates.

---

[47] Although the Federal Reserve was initially authorized by the U.S. Congress to pay interest on reserves only from 2011, it was later allowed to begin this operation in October 2008 due to the financial crisis (Federal Reserve System, 2016). By paying interest on *required reserves*, the Federal Reserve intended to remove the costs incurred by depository institutions when they hold these reserves. Not being able to lend required reserves to their customers was an example of such costs. On the other hand, the Federal Reserve used its policy of paying interest on *excess reserves* as an additional monetary tool. This is because, by modifying the interest rate it paid on excess reserves (commonly called the *IOER rate*), the Federal Reserve influenced depository institutions' opportunity costs of holding these reserves, and therefore affected the federal funds rate and other short-term interest rates (Federal Reserve System, 2016).

## 3.4. The Difference Between the Federal Reserve's and the Bank of Canada's Responses to the Global Financial Crisis

In response to the Global Financial Crisis, the Bank of Canada (the central bank of Canada) used a combination of its traditional liquidity tools and a set of newly created liquidity facilities. These included the following (Bank of Canada, 2009, 2022; Selody & Wilkins, 2010; Zorn et al., 2009): (i) the *Standing Liquidity Facility* (*SLF*) and *Emergency Lending Assistance* (*ELA*), through which the bank provided loans to direct clearers in the Large Value Transfer System (LVTS) and individual institutions that faced liquidity shortages (these are the bank's traditional liquidity facilities); (ii) the *Term Purchase and Resale Agreements* (*PRAs*), through which the bank supplied funds to counterparties in key financial markets; and (iii) the *Term Loan Facility* (*TLF*), through which it offered loans to LVTS counterparties.

Since these liquidity programs all involved temporary operations (term loans and term purchases), based on this paper's definition of quantitative easing (QE) provided in **Section 2**, they do not qualify as QE operations. They qualify as quasi-quantitative easing (qQE) policies. Therefore, in response to the Global Financial Crisis, while the Federal Reserve implemented both QE and qQE policies, the Bank of Canada conducted only the latter. In other words, in response to the crisis, *while the Federal Reserve conducted QE, the Bank of Canada did not*. This fundamental difference between the Federal Reserve's and the Bank of Canada's responses to the crisis forms the basis of my statistical analysis in **Section 4** below.

# 4. THE LONG-TERM EFFECTS OF THE FEDERAL RESERVE'S RESPONSE TO THE GLOBAL FINANCIAL CRISIS: AN INTERRUPTED TIME-SERIES ANALYSIS OF THE CAUSAL IMPACT OF ITS QUANTITATIVE EASING PROGRAMS

## 4.1. Introduction

The Federal Reserve's quantitative easing (QE) programs conducted in response to the Global Financial Crisis led to a large increase in the U.S. monetary base and money supply (see **Section 3**). Theory suggests that the increase in the monetary base and money supply should have helped in the fight against the crisis (see **Section 2**). Empirical studies on the subject using both new Keynesian structural and Granger-causality nonstructural models corroborate the theory (Bhattarai & Neely, 2016; Borio & Zabai, 2016, 2018; Cecioni et al., 2019). However, using a nonstructural approach based on an informal natural experiment, Williamson (2017) finds no evidence to support the hypothesis that the QE programs increased U.S. real GDP.[48]

Motivated by Williamson's (2017) peculiar cross-national empirical investigation and the stark contrast between results from his *informal* quasi-experiment and those from fully non-experimental methods, I decided to verify his findings using a *formal* design-based natural experimental technique.

This section proceeds as follows. **Section 4.2** briefly reviews Williamson's (2017) evidence on the macroeconomic effect of the Federal Reserve's QE programs on U.S. real GDP. To the best of my knowledge, Williamson's paper is the only existing study that examines, albeit informally, the macroeconomic impact of QE using a natural experimental approach. Therefore, the present study is the first paper to use a formal design-based natural experimental method to measure the macroeconomic effects of QE. Reviews of empirical work that use VARs and new Keynesian DSGE models to measure the macroeconomic effects of QE can be found in Borio and Zabai (2016, 2018), Bhattarai and Neely (2016), and Cecioni et al. (2019). **Section 4.3** presents a comprehensive overview of the natural experimental approach of interrupted time-series analysis. And **Section 4.4** analyzes the causal effects of the Federal Reserve's QE programs using the interrupted time-series analysis method.

---

[48] Although Williamson's (2017) study does not qualify as a *formal* design-based natural experiment (the definition of a design-based natural experiment is provided in **Section 2**), it may be considered an *informal* natural experiment since it uses a control and treatment group in its analysis.

## 4.2. A Brief Literature Review of Empirical Work on the Effects of the Federal Reserve's QE Programs on U.S. Real GDP

Because the U.S. and Canada are subject to similar economic shocks (since they are geographically close and have a similar economic status) and the U.S. conducted QE programs in response to the Global Financial Crisis while Canada did not, Williamson (2017) uses **Figure 12** below to informally investigate the effect of the Federal Reserve's QE programs on U.S. real GDP between 2007 and 2016 by comparing U.S. and Canada real GDP performance over this period. In other words, Williamson examines the impact of the Federal Reserve's QE programs on U.S. real GDP using an informal natural experimental approach in which Canada is the control group and the U.S. is the treatment group. He remarks from the figure that the difference between U.S. and Canada real GDP from 2007 to 2016 was not significantly large. Based on this, he concludes that there is no evidence on the positive effect of the Federal Reserve's QE programs on U.S. real GDP. Since conclusions from a simple graphical analysis may be inaccurate, it is essential to verify Williamson's findings using a more sophisticated natural experimental approach.

**Figure 12**

*U.S. Versus Canada Real GDP*

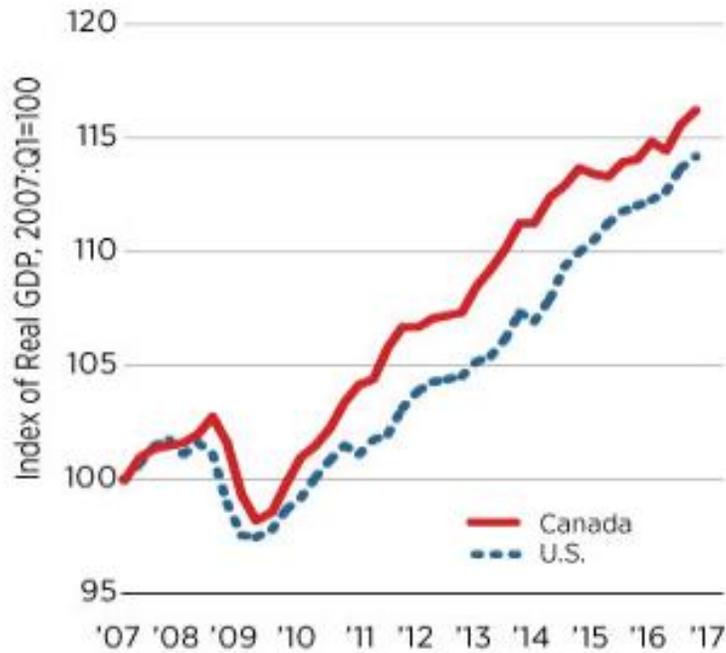

*Note*. Reprinted from Kamkoum (2022; see also 2023b), CC BY 4.0. Kamkoum (2022) reprinted with permission from Williamson (2017), copyright 2017 by the Federal Reserve Bank of St. Louis.

### 4.3. A Comprehensive Overview of Interrupted Time-Series Analysis

This paper examines the macroeconomic effects of the Federal Reserve's QE programs using a formal natural experimental approach based on the method of interrupted time-series analysis. The following subsections present a comprehensive overview of this methodology.

#### *4.3.1. An Introduction to the Interrupted Time-Series Analysis Approach*

A time series is a sequence of observations made on the same variable at successive and equally separated time points. An *interrupted time series* (*ITS*) is a special type of time series in which the series shows an interruption in its values and can be used to estimate the impact of an intervention (McDowall et al., 2019; Shadish et al., 2002). In other words, an interrupted time series is a segmented series whose segments represent pre- and post-intervention periods. The name *interrupted time-series analysis* (*ITS analysis* or *ITSA*) was coined about fifty years ago to describe the statistical methods and models of studies that use interrupted time series (McDowall et al., 2019).

ITS analysis is very popular in the evaluation of health-care policies (Bernal et al., 2017; Jandoc et al., 2015; Penfold & Zhang, 2013; Taljaard et al., 2014; S. L. Turner, Forbes, et al., 2021; S. L. Turner, Karahalios, et al., 2021; A. K. Wagner et al., 2002).[49] It is also commonly used in education policy research (Bloom, 1999, 2003; Bloom et al., 2001; Dee & Jacob, 2011, 2010; Wong et al., 2009, 2015; Zieger et al., 2022). Furthermore, the technique has been employed in various study areas of economics (Anderton & Carter, 2001; Barbieri & Levy, 1999; Bonham et al., 1992; J. A. Wagner et al., 1988). However, its use in estimating the causal effects of monetary policy on inflation and real output is extremely limited.[50] This is why by using the ITS analysis approach to measure the inflation and output effects of monetary policy, this paper significantly contributes to the literature.

The intervention in an ITS study may be a variation deliberately introduced by a researcher or a phenomenon over which the researcher has no control (McDowall et al., 2019; Shadish et al., 2002). While the first type of intervention is the subject of field and laboratory experiments, the second type is the subject of natural experiments. As discussed in **Section 2**, interventions in natural experiments include policy modifications, natural occurrences, and institutional changes.

---

[49] See also Biglan et al. (2000), Hategeka et al. (2020), Linden (2015), Santa-Ana-Tellez et al. (2013), Schaffer et al. (2021), and Valiyeva et al. (2008).

[50] Google Scholar (http://scholar.google.com/) and RePEc (https://ideas.repec.org/ and https://econpapers.repec.org/) searches for the keywords "interrupted time series + inflation," "interrupted time series + price," "interrupted time series + output," "interrupted time series + GDP," "interrupted time series + income," "interrupted time series + money," "interrupted time series + monetary policy," "interrupted time series + monetary" give only one relevant result: a paper by Hudu et al. (2016).

The effect of an intervention in an ITS study may be described in three ways: according to the effect's *form*, *immediacy*, and *permanence* (Shadish et al., 2002). According to its form, the effect may be a change in the level, trend, variance, or cyclicity of the time series. Changes in level (intercept) and trend (slope) are the most investigated types of effects, and are the only effects studied in this paper. According to its immediacy, the effect (change) may be either immediate or delayed. The effect is immediate if it occurs exactly at the first time point post-intervention; otherwise, it is delayed. A delayed effect should be justified by some theory (Shadish et al., 2002). Finally, according to its permanence, the effect may be either continuous or discontinuous. While a continuous effect persists over time, a discontinuous effect does not. In other words, if an intervention has a continuous level effect of $l$ units, this effect will be $l$ units at every time point that comes after the first point where the effect was first recorded; but if it has a discontinuous level effect, the effect will decrease or increase over time. A continuous level effect is only possible if there is no trend change post-intervention.

To investigate the impact of an intervention on a given outcome, an ITS study may use a single time series or both a time series and a suitable control for that time series. In an ITS study that uses a single time series (i.e., in a *simple or basic ITS study*), the data examined consist of several observations made on a single treatment group before and after the intervention. The causal effect of the intervention is, therefore, determined by comparing longitudinal trends for the same time series before and after the intervention. On the other hand, in an ITS study that uses both a time series and a control for that time series (i.e., in a *comparative, controlled, or multiple ITS study*), in addition to the treatment group data, there are data from a different but comparable group (called the comparison or control group) that was not subjected to the intervention. Thus, in this case, the causal effect of the intervention is measured by comparing longitudinal trends for two different time series before and after the intervention.

Despite there not being a specific figure for the minimum number of data points to use in an ITS study, it is recommended to have at least eight points pre- and eight points post-intervention in order to capture the trend of the outcome variable in each of these periods and have enough statistical power to estimate credible model parameters (Law, 2015; Penfold & Zhang, 2013). Also, it is not necessary to have the same number of data points pre- and post-intervention (Law, 2015). In other words, for a study with data points covering the period $t = 1, 2, 3, \ldots, k, k+1, k+2, k+3, \ldots, n$, where $t = 1, 2, 3, \ldots, k$ are the time points pre-intervention and $t = k+1, k+2, k+3, \ldots, n$ are the time points post-intervention, $k$ should not necessarily be equal to $n/2$.[51]

Although the effect of an intervention in an ITS study can be causally interpreted regardless of whether the data used are from a single time series or from a time series and a suitable control for that time series, it has been argued that if a researcher does not deliberately introduce the variation being investigated, they should make causal inferences solely from the second kind of data (McDowall et al., 2019). In other words, in an ITS natural experiment, a researcher is advised to draw causal inferences using data from both a treatment and a control time series, not from a single time series.

In a simple ITS study, the main assumption for causal inference is that absent the intervention, the existing level and trend of the outcome variable would have remained the same. Here, therefore, the *counterfactual*—what would have happened to the outcome variable absent the intervention—is constructed by projecting the pre-intervention level and trend of the outcome variable into the post-intervention period. *Threats to internal validity*—problems that make causal inference in the context of a specific study incorrect—regarding a simple ITS design consist of issues that render the above key assumption implausible. These include complications or biases related to *history*, *instrumentation*, *maturation*,

---

[51] It should be noted that the *post-* prefix in *post-intervention* means *after the beginning of*. It does not mean *after the end of*. In other words, the phrase *time points post-intervention* means *time points after the beginning of the intervention*. It does not mean *time points after the end of the intervention*.

*selection*, and *statistical regression* (Law, 2015; Shadish et al., 2002; Wong et al., 2015).[52] History is an issue when an event or policy that is unrelated to the intervention and can affect the outcome variable occurs simultaneously with the intervention. Instrumentation is a problem when there is a change in outcome measurement during the period under study. Maturation is a problem when there is a systematic or naturally occurring change over time that could be misinterpreted as a change resulting from the intervention. There may be a selection issue stemming from a change in the composition of units under study, such that the units are not comparable before and after the intervention.[53] Finally, there may be a statistical regression problem due to the average baseline outcome value for units under study being extremely low or high pre-intervention, such that apparent treatment effects on the outcome are attributable to the initial characteristics of the units and not to the intervention.

In a comparative ITS study, the key assumption for causal inference is that absent the intervention, the existing difference in level or trend between the treatment and control groups would have remained the same. That is, the key assumption is that absent the intervention, the level or trend change in the outcome for the treatment group would have been the same as the level or trend change in the outcome for the control group. Thus, the counterfactual is constructed by projecting the pre-intervention difference in level or trend between the treatment and control groups into the post-intervention period.

The existence of a control group in a comparative ITS study renders most of the threats to internal validity that are common in a simple ITS study implausible (Law, 2015; Shadish et al., 2002; Wong et al., 2015). Indeed, in a comparative ITS design, a threat to internal validity is plausible only if it occurs *both* simultaneously with the intervention *and* in just one of either the treatment or control groups (Law, 2015; Shadish et al., 2002; Wong et al., 2015). In other words, a threat to internal validity would be an issue only if it affects the treatment and control groups differentially around the same time as the intervention. Therefore, any history, instrumentation, maturation, selection, or statistical regression issues that occur simultaneously with the intervention but affect the treatment and control groups similarly are not considered threats to the internal validity of a comparative ITS design.[54] This feature of a comparative ITS design is what distinguishes it from a simple ITS design and makes it so powerful for establishing causal inference. The feature also helps a comparative ITS model overcome the endogeneity bias that is common in some non-experimental models since it renders the two leading causes of endogeneity—simultaneity and confounding variables—implausible. Specifically, the possibility that, in addition to being caused by the intervention, the outcome variable also affects the intervention (the problem of simultaneity or bidirectional causation) is ruled out because the intervention is absent in the control group. Moreover, without rejecting the possible existence of a variable that may affect both the outcome and the intervention but that is not included in the model, this feature of a comparative ITS design renders the problem of confounding (missing or omitted) variables unimportant by supposing that either (i) such variables would influence the treatment and control groups equally or (ii) any differential effects of such variables on the treatment and control groups would be the same pre- and post-intervention.

---

[52] Although there are other potential threats to internal validity (such as attrition and ambiguous temporal precedence (Shadish et al., 2002)), history, instrumentation, maturation, selection, and statistical regression are the most plausible ones (McCleary et al., 2017).

[53] This is not a widely used description of *selection*. Selection is generally viewed as an issue caused by a systematic (intrinsic or underlying) difference in the characteristics of units in the treatment and control groups pre-intervention that renders both groups incomparable. Interpreted this way, selection can be a problem only in a comparative ITS study (McCleary et al., 2017).

---

[54] Likewise, any history, instrumentation, maturation, selection, or statistical regression issues that affect the treatment and control groups differentially but do not occur simultaneously with the intervention are not considered threats to the internal validity of a comparative ITS design. However, this second consideration is more reasonable when measuring changes in level than when estimating changes in trend. This is because an issue that affects the treatment and control groups differentially long after the intervention period (i.e., an issue that does not occur simultaneously with the intervention) may render estimates of changes in trend inaccurate since such an issue would be incorrectly modeled by a comparative ITS design as having influenced both groups similarly. In other words, since the counterfactual trend in a comparative ITS design is constructed by assuming that *all* issues post-intervention (be it shortly or long after the intervention) influence the treatment and control groups similarly, an issue that affects the two groups differentially even long after the intervention would render the constructed counterfactual trend incorrect insofar as the differential effect occurs within the period under study.

Differential threats to internal validity between the treatment and control groups usually arise only from interactions among the various threats that occur in a simple ITS study (Law, 2015; Shadish et al., 2002). The most common include *selection-history*, *selection-instrumentation*, *selection-maturation*, and *selection-regression* (Law, 2015; Shadish et al., 2002). Selection-history or local history is a problem when an event that is unrelated to the intervention occurs concurrently with the intervention and affects the treatment and control groups differentially. Selection-instrumentation is an issue if there is a change in the outcome measurement procedure (technique) in just one of the treatment or control groups around the same time as the intervention. Selection-maturation is a threat if a systematic change over time that starts simultaneously with the intervention is present only in the treatment or control group, not in both. Lastly, selection-regression is a problem when the intervention occurs immediately after an extremely low or high observation in either the treatment or control group. The more comparable (similar) the treatment and control groups are, the less likely is the possibility of any of the above four threats.

### 4.3.2. Modeling an Interrupted Time Series Study

A simple ITS study can be modeled or designed as in **Equation 1** below (Huitema & McKean, 2000, 2007; Linden, 2015; S. L. Turner, Forbes, et al., 2021; S. L. Turner, Karahalios, et al., 2021; see also Law, 2015; Penfold & Zhang, 2013; A. K. Wagner et al., 2002).

$$Outcome_{it} = \beta_0 + \beta_1 time_t + \beta_2 intervention_i + \beta_3 intervention_i * (time_t - r) + \varepsilon_{it}$$

**(1)**

Where $Outcome_{it}$ is the outcome for intervention status $i$ at time $t$; $time_t = t$ is the time since the beginning of the study period, taking discrete values $t = 1, 2, 3, …, k, k+1, k+2, k+3, …, n$, with $t = 1, 2, 3, …, k$ representing the time points pre-intervention, $t = k+1, k+2, k+3, …, n$ representing the time points post-intervention, and $k$ is not necessarily equal to $n/2$ (thus, there are $k$ periods pre-intervention and $(n - k)$ periods post-intervention);

$intervention_i = i$ is a dummy (binary or indicator) variable that equals 1 post-intervention (at time points $t = k+1, k+2, k+3, \ldots, n$) and 0 pre-intervention (at time points $t = 1, 2, 3, \ldots, k$); $r = (k+1)$, such that $(time_t - r)$ captures the post-intervention time points at which the post-intervention trend change is *measurable*, therefore, in addition to the pre-intervention time points, $(time_t - r)$ is also equal to 0 at the first time point post-intervention ($time_t - r = t - (k+1) = (k+1) - (k+1) = 0$ at $t = k+1$) since $t = k+1$ represents the point at which there is a level change and *no trend change* (afterward, $time_t - r = t - (k+1) = (k+2) - (k+1) = 1$ at the second time point post-intervention, $time_t - r = t - (k+1) = (k+3) - (k+1) = 2$ at the third time point post-intervention, ..., and $time_t - r = t - (k+1) = n - (k+1) = (n - k) - 1$ at the $(n - k)$th time point post-intervention);[55] $intervention_i * (time_t - r)$ is an interaction term that equals 0 from $t = 1$ to $t = k+1$ and increases by one unit from 1 at $t = k+2$ to $(n - k - 1)$ at $t = n$ (i.e., although $intervention_i$ equals 1 at $t = k+1$, $intervention_i * (time_t - r)$ equals 0 at this point since $time_t - r = 0$);[56] $\beta_0$ is the intercept, representing the *level* of the outcome variable before the beginning of the study period (at $t = 0$); $\beta_1$ is the observed *trend* of the outcome variable pre-intervention, representing the change over time in the outcome variable between any two distinct time points pre-intervention; $\beta_2$ is the *level change* (*intercept change*) in the outcome variable at the first time point post-intervention, measured as the difference between the observed and counterfactual levels at $t = k+1$ and representing the immediate (short-term) causal effect of the intervention;[57] $\beta_3$ is the *trend change* of the outcome variable post-intervention, measured as the difference between the observed and counterfactual trends post-intervention and representing the change over time in the outcome variable between any two distinct time points post-intervention relative to the change over time that would have occurred absent the intervention;[58] and $\varepsilon_{it}$ is a normally distributed random error term, $\varepsilon_{it} \sim N(0, \sigma^2)$. The parameters of interest are $\beta_2$ and $\beta_3$. (A mathematical proof that $\beta_2$ is equal to the difference between the observed and counterfactual levels at $t = k+1$ is provided in **Appendix 1**. And a proof that $\beta_3$ is equal to the difference between the observed and counterfactual trends post-intervention is provided in **Appendix 2**.)

---

[55] The post-intervention *trend change* is *measurable* only from $t = k+2$ because (i) post-intervention *trend change* = post-intervention *trend* – pre-intervention trend and (ii) post-intervention *trend* can be calculated only when there are *at least two time points post-intervention*. In other words, the post-intervention *trend change* is measurable only from $t = k+2$ because the post-intervention *trend* is computable only from $t = k+2$.

[56] It should be noted that instead of $intervention_i * (time_t - r)$, some authors like A. K. Wagner et al. (2002), Penfold & Zhang (2013), and Law (2015) use an interaction term that is equivalent to $intervention_i * (time_t - k)$ (A. K. Wagner et al. and Law call this term "time after intervention" and Penfold & Zhang call it "time after program"). According to these authors, there are both a level change and a trend change at the first time point post-intervention (at $t = k+1$). Based on my understanding and interpretation of the subject, the explanation that $t = k+1$ represents a point at which there is a level change and *no trend change*, and the mathematical proofs provided in the appendices, these authors are incorrect. My view is shared by Huitema & McKean (2000, 2007), Linden (2015), S. L. Turner, Forbes, et al. (2021), and S. L. Turner, Karahalios, et al. (2021).

Therefore, according to the primary assumption for causal inference in a simple ITS design, $\beta_2$ and $\beta_3$ would be zero in the absence of the intervention. In other words, hypothesis testing in a simple ITS study is based on the following null and alternative hypotheses.

$$H_0: \beta_2 = \beta_3 = 0$$
$$H_1: \beta_2 \neq 0, \beta_3 \neq 0, or\ both$$

Whereby the existence of enough evidence to support the claim that the intervention had an impact on the level or trend of the outcome leads to our rejection of the null hypothesis.

---

[57] $\beta_2$ may be equivalently measured as the *difference-in-differences* between the observed and counterfactual levels at $t = k+1$ and $t = k$. This reasoning is based on the following proof.

> Let the difference-in-differences between the observed and counterfactual levels at $t = k+1$ and $t = k$ be equal to $\lambda$.
> 
> We know that the difference between the observed and counterfactual levels at $t = k+1$ is equal to $\beta_2$.
> 
> Since the counterfactual level is equal to the observed level at any time point pre-intervention, the difference between the observed and counterfactual levels at $t = k$ is equal to 0.
> 
> Therefore,
> 
> $\lambda = \beta_2 - 0 = \beta_2$
> 
> Quod erat demonstrandum (Q.E.D.).

[58] Similarly, $\beta_3$ may be equivalently measured as the *difference-in-differences* between the observed and counterfactual trends post-intervention and pre-intervention (where the difference between the observed and counterfactual trends pre-intervention is equal to 0). The reasoning is the same as that in **Footnote 57** above.

The variables and parameters of a simple ITS design—the variables and coefficients in **Equation 1** above—can be represented graphically as in **Figure 13** below.

**Figure 13**

*Graphical Representation of a Simple ITS Design*

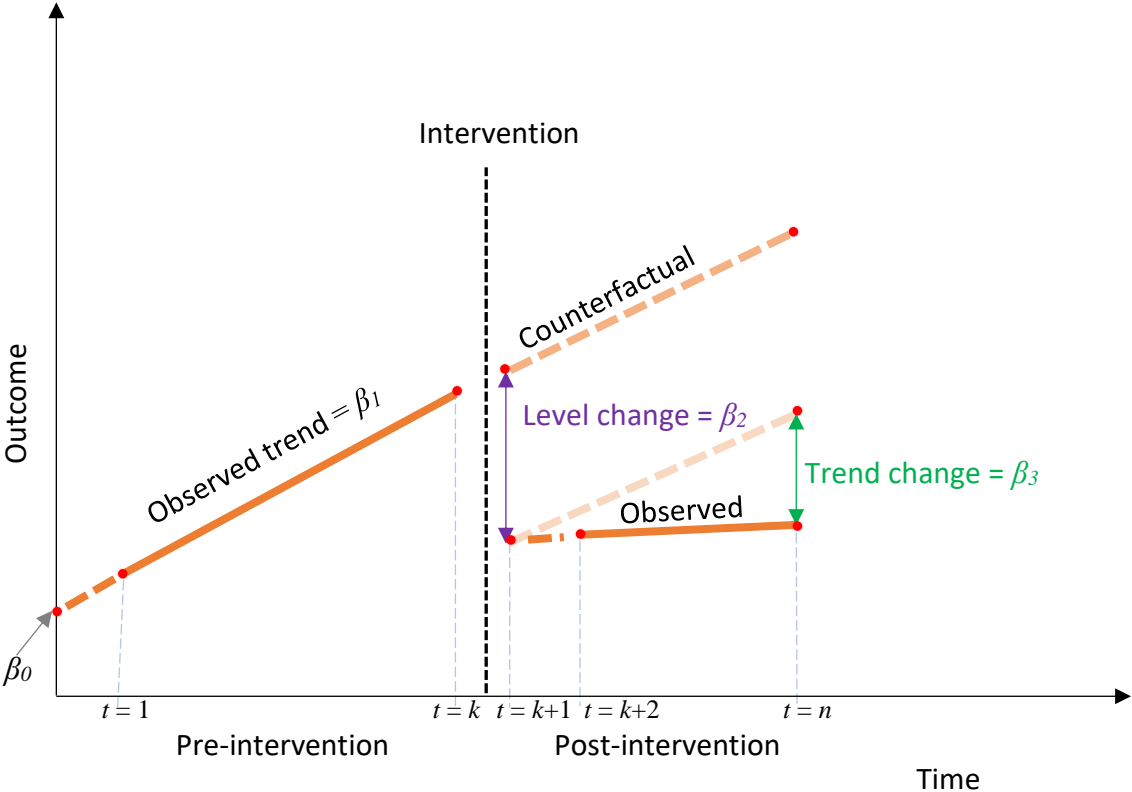

*Note.* $\beta_0$ is the *intercept* or *level* of the outcome variable before the beginning of the study period (at $t = 0$); $\beta_1$ is the observed *trend* of the outcome variable pre-intervention; $\beta_2$ is the *level change* (*intercept change*) in the outcome variable at the first time point post-intervention, measured as the difference between the observed and counterfactual levels at $t = k+1$ and representing the short-term causal effect of the intervention; $\beta_3$ is the *trend change* of the outcome variable post-intervention, measured as the difference between the observed and counterfactual trends post-intervention and representing the change over time in the outcome variable between any two distinct time points post-intervention relative to the

change over time that would have occurred absent the intervention. Trends from $t = 0$ to $t = 1$ and $t = k+1$ to $t = k+2$ are backward extrapolations of the pre-intervention and post-intervention trends, respectively (since the pre-intervention and post-intervention trends are *measurable* only from $t = 1$ and $t = k+2$, respectively). The figure is based on the author's understanding and interpretation of the subject.

**Figure 13** shows that, as described earlier, the counterfactual level and trend of the outcome are formed by projecting (extrapolating) their pre-intervention values into the post-intervention period. The level impact of the intervention is calculated as the difference between the observed and counterfactual outcome values at the first time point post-intervention. And the trend impact over a specific period is estimated as the difference between the observed and counterfactual trends over that period.

On the other hand, a comparative ITS study can be modeled as in **Equation 2** below (Huitema & McKean, 2000, 2007; Linden, 2015; S. L. Turner, Forbes, et al., 2021; S. L. Turner, Karahalios, et al., 2021; see also Law, 2015; Wong et al., 2015).

$$Outcome_{igt} = \beta_0 + \beta_1 time_t + \beta_2 intervention_i + \beta_3 intervention_i * (time_t - r)$$
$$+ \beta_4 group_g + \beta_5 (group_g * time_t) + \beta_6 (group_g * intervention_i)$$
$$+ \beta_7 group_g * intervention_i * (time_t - r) + \varepsilon_{igt}$$

(2)

Where $Outcome_{igt}$ is the outcome for intervention status *i*, in group *g*, at time *t*; $time_t$, $intervention_i$, $time_t - r$, and $intervention_i * (time_t - r)$ are as earlier defined; $group_g$ = *g* is a dummy variable that equals 1 for the treatment group time series and 0 for the control group time series; $group_g * time_t$, $group_g * intervention_i$, and $group_g * intervention_i * (time_t - r)$ are interaction terms between the defined variables; *β₀* is the level of the outcome for the control group time series before the beginning of the study period (at $t = 0$); *β₁* is the observed trend of the outcome for the control group time series pre-intervention; $\beta_2$ is the level change in the outcome for the control group time series at the first time point post-intervention, measured as the difference between the observed and counterfactual levels of the outcome for the control time series at $t = k+1$ and representing an extraneous level change; $\beta_3$ is the trend change in the outcome for the control group time series post-intervention, measured as the difference between the observed and counterfactual trends of the outcome for the control time series post-intervention and representing an extraneous trend change; $\beta_4$ is the level difference between the outcome for the treatment and control group time series before the beginning of the study period (at $t = 0$); $\beta_5$ is the trend difference between the outcome for the treatment and control group time series pre-intervention; $\beta_6$ is the level change difference between the outcome for the treatment and control group time series at the first time point post-intervention, measured as the difference-in-differences between the observed and counterfactual levels of the outcome for the treatment and control time series at $t = k+1$;[59] $\beta_7$ is the trend change difference between the outcome for the treatment and control group time series post-intervention, measured as the difference-in-differences between the observed and counterfactual trends of the outcome for the treatment and control time series post-intervention;[60] and $\varepsilon_{igt}$ is a normally distributed random error term, $\varepsilon_{igt} \sim N(0, \sigma^2)$. The parameters of interest are $\beta_6$ and $\beta_7$. (A mathematical proof that $\beta_6$ is equal to the difference-in-differences between the observed and counterfactual levels of the outcome for the treatment and control time series at $t = k+1$ is provided in **Appendix 3**. And a proof that $\beta_7$ is equal to the difference-in-differences between the observed and counterfactual trends of the outcome for the treatment and control time series post-intervention is provided in **Appendix 4**.) (Henceforth, unless otherwise stated, regardless of whether it is a simple or comparative ITS analysis, the terms *level change* (level effect or level impact) and *trend change* (trend effect or trend impact) will be used to describe the *overall* (as opposed to group-specific) level and trend causal effects of an intervention, respectively. In other words, in a comparative ITS analysis, instead of using the terms *level change difference* and *trend change difference* to refer to the overall level and trend causal effects of an intervention, the terms *level change* and *trend change* will be used.)

---

[59] That is, $\beta_6$ is the level change in the outcome for the treatment group time series at the first time point post-intervention *relative* to the level change in the outcome for the control group time series at that point.

[60] That is, $\beta_7$ is the trend change in the outcome for the treatment group time series post-intervention *relative* to the trend change in the outcome for the control group time series post-intervention.

Thus, according to the key assumption for causal inference in a comparative ITS design, $\beta_6$ and $\beta_7$ would be zero in the absence of the intervention. That is, hypothesis testing in a comparative ITS study is based on the following null and alternative hypotheses.

$$H_0: \beta_6 = \beta_7 = 0$$
$$H_1: \beta_6 \neq 0, \beta_7 \neq 0, or\ both$$

Whereby the existence of enough evidence to support the claim that the intervention had an impact on the level or trend of the outcome leads to our rejection of the null hypothesis.

The variables and parameters of a comparative ITS design (in **Equation 2** above) can be represented graphically as in **Figure 14** below.

As previously explained, **Figure 14** shows that the counterfactual level and trend of the outcome are formed by projecting the pre-intervention difference in level or trend between the treatment and control groups into the post-intervention period. The level impact of the intervention is calculated as the difference-in-differences between the observed and counterfactual levels of the outcome for the treatment and control time series at the first time point post-intervention (at $t = k+1$). And the trend impact over a specific period is estimated as the difference-in-differences between the observed and counterfactual trends of the outcome for the treatment and control time series over that period.

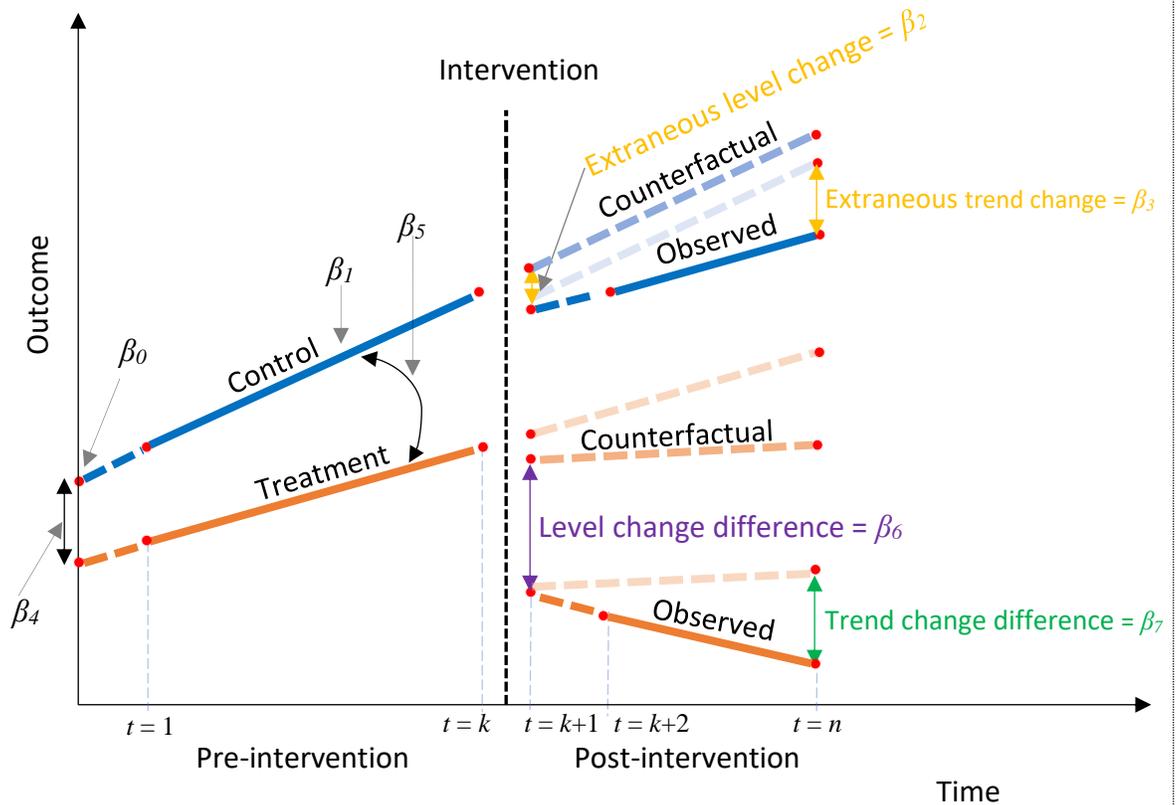

**Figure 14**

*Graphical Representation of a Comparative ITS Design*

*Note.* $\beta_0$ is the *level* of the outcome for the control group time series before the beginning of the study period (at $t = 0$); $\beta_1$ is the observed *trend* of the outcome for the control group time series pre-intervention; $\beta_2$ is the *level change* in the outcome for the control group time series at the first time point post-intervention (at $t = k+1$), representing an extraneous level change unrelated to the intervention; $\beta_3$ is the *trend change* in the outcome for the control group time series post-intervention, representing an extraneous trend change unrelated to the intervention; $\beta_4$ is the *level difference* between the outcome for the treatment and control group time series before the beginning of the study period (at $t = 0$); $\beta_5$ is the *trend difference* between the outcome for the treatment and control group time series pre-intervention; $\beta_6$ is the *level change difference* between the outcome for the treatment and control group time series at the first time point post-intervention; and $\beta_7$ is the *trend change difference* between the outcome for the treatment and control group time series post-intervention. Trends from

$t = 0$ to $t = 1$ and $t = k+1$ to $t = k+2$ are backward extrapolations of the pre-intervention and post-intervention trends, respectively (since the pre-intervention and post-intervention trends are *measurable* only from $t = 1$ and $t = k+2$, respectively). The figure is based on the author's understanding and interpretation of the subject.

Although most ITS designs involve a single intervention like in **Equations 1** and **2** and their corresponding graphical representations, it is possible to have two or more interventions. A design with multiple interventions is modeled and represented graphically following the same logic and principles as one with a single intervention. More specifically, a design with multiple interventions must have the recommended minimum number of data points (observations) pre- and post-intervention for each intervention. A simple ITS design with two interventions and a comparative ITS design with two interventions can be modeled as in **Equations 3** and **4** below, respectively.

$$Outcome_{ijt} = \beta_0 + \beta_1 time_t + \beta_2 intervention_i + \beta_3 intervention_i * (time_t - r)$$
$$+ \beta_4 intervention_j + \beta_5 intervention_j * (time_t - s) + \varepsilon_{ijt}$$

(3)

$$Outcome_{ijgt} = \beta_0 + \beta_1 time_t + \beta_2 intervention_i + \beta_3 intervention_i * (time_t - r)$$
$$+ \beta_4 intervention_j + \beta_5 intervention_j * (time_t - s) + \beta_6 group_g$$
$$+ \beta_7 (group_g * time_t) + \beta_8 (group_g * intervention_i)$$
$$+ \beta_9 group_g * intervention_i * (time_t - r)$$
$$+ \beta_{10} (group_g * intervention_j)$$
$$+ \beta_{11} group_g * intervention_j * (time_t - s) + \varepsilon_{ijgt}$$

(4)

Where $Outcome_{ijt}$ is the outcome for intervention statuses $i$ and $j$ at time $t$; $Outcome_{ijgt}$ is the outcome for intervention statuses $i$ and $j$, in group $g$, at time $t$; $intervention_i$ and $intervention_j$ are the first and second interventions, respectively; $time_t = t$ is the time since the beginning of the study period, taking discrete values $t$ = 1, 2, 3, ..., $k$, $k$+1, $k$+2, $k$+3, ...,$m$, $m$+1, $m$+2, $m$+3, ..., $n$, with $t$ = 1, 2, 3, ..., $k$ representing the time points before the start of the first intervention, $t = k$+1, $k$+2, $k$+3, ..., $m$ representing the time points after the start of the first intervention but before the beginning of the second intervention, and $t$ = $m$+1, $m$+2, $m$+3, ..., $n$ representing the time points after the beginning of the second intervention (therefore, there are $k$ periods before the first intervention, $m$ periods before the second intervention, ($n - k$) periods after the first intervention, and ($n - m$) periods after the second intervention); $r$ = ($k$+1), such that ($time_t - r$) captures the time points after the beginning of the first intervention at which the trend change is measurable, therefore, in addition to the time points before the first intervention, ($time_t - r$) is also equal to 0 at $t$ = $k$+1 (afterward, it increases by one unit from 1 at $t = k$+2 to ($n - k - 1$) at $t = n$); $s$ = ($m$+1), such that ($time_t - s$) captures the time points after the start of the second intervention where the trend change is measurable, thus, in addition to the time points before the second intervention, ($time_t - s$) is also equal to 0 at $t = m$+1 (following that, it increases by one unit from 1 at $t = m$+2 to ($n - m - 1$) at $t = n$); $intervention_i * (time_t - r)$ is an interaction term that equals 0 from $t$ = 1 to $t = k$+1 and increases by one unit from 1 at $t = k$+2 to ($n - k - 1$) at $t = n$; $intervention_j * (time_t - s)$ is an interaction term that equals 0 from $t$ = 1 to $t = m$+1 and increases by one unit from 1 at $t = m$+2 to ($n - m - 1$) at $t = n$; $group_g = g$ equals 1 for the treatment group time series and 0 for the control group time series; $group_g * time_t$, $group_g * intervention_i$, $group_g * intervention_i * (time_t - r)$, $group_g * intervention_j$, and $group_g * intervention_j * (time_t - s)$ are interaction terms between the defined variables. In **Equation 3**, the coefficients of interest are $β_2$, $β_3$, $β_4$, and $β_5$. Here, $β_2$ and $β_4$ denote the level change in the outcome variable at the first time point after the start of the first and second interventions, respectively (at $t = k$+1 and $t = m$+1). $β_3$ and $β_5$ correspond to the trend change in the outcome variable after the beginning of the first and second interventions, respectively. Therefore, $β_4$ represents the level impact of the second intervention relative to that of the first intervention. And $β_5$ represents the trend impact of the second intervention relative to that of the first intervention. In **Equation 4**, the coefficients of interest are $\beta_8$, $\beta_9$, $\beta_{10}$, and $\beta_{11}$. Whereby $\beta_8$ and $\beta_{10}$ stand for the level change difference between the outcome for the treatment and control group time series at the first time point after the start of the first and second interventions, respectively. $\beta_9$ and $\beta_{11}$ show the trend change difference between the outcome for the treatment and control group time series after the beginning of the first and second interventions, respectively. Therefore, $\beta_{10}$ represents the level impact of the second intervention relative to that of the first intervention. And $\beta_{11}$ represents the trend impact of the second intervention relative to that of the first intervention. (A mathematical proof that the $\beta_4$ coefficient of **Equation 3** is equal to the difference between the observed and counterfactual levels at $t = m+1$ is provided in **Appendix 5**, a proof that the $\beta_5$ coefficient of **Equation 3** is equal to the difference between the observed and counterfactual trends post-second-intervention (after the beginning of the second intervention) is provided in **Appendix 6**, a proof that the $\beta_{10}$ coefficient of **Equation 4** is equal to the difference-in-differences between the observed and counterfactual levels of the outcome for the treatment and control time series at $t = m+1$ is provided in **Appendix 7**, and a proof that the $\beta_{11}$ coefficient of **Equation 4** is equal to the difference-in-differences between the observed and counterfactual trends of the outcome for the treatment and control time series post-second-intervention is provided in **Appendix 8**. Mathematical proofs for the $\beta_2$ and $\beta_3$ coefficients of **Equation 3** and the $\beta_8$ and $\beta_9$ coefficients of **Equation 4** follow the same logic as the proofs provided in **Appendices 1**, **2**, **3**, and **4**.)

The interrupted time-series designs in **Equations 1**, **2**, **3**, and **4** represent segmented linear regressions (Taljaard et al., 2014; S. L. Turner, Forbes, et al., 2021; S. L. Turner, Karahalios, et al., 2021; A. K. Wagner et al., 2002). Each of these segmented regression models can be estimated by various statistical methods. The *Ordinary least squares* (*OLS*) method is among the most used techniques when data points or observations are uncorrelated. Since ITS models involve time-series data, the data for the equations above are most likely correlated, and using OLS may yield inaccurate results. If the data are tested and confirmed to be correlated, alternative statistical methods that can be used to estimate the equations include *generalized least squares* (*GLS*), *restricted maximum likelihood* (*REML*), *autoregressive moving averages* (*ARMA*), *autoregressive integrated moving average* (*ARIMA*), *maximum* likelihood autoregressive moving averages (*ARMAX*), and *maximum likelihood autoregressive integrated moving averages* (*ARIMAX*), among others (McCleary et al., 2017; McDowall et al., 2019; S. L. Turner, Forbes, et al., 2021; S. L. Turner, Karahalios, et al., 2021).

Contrary to the short-term causal effect of the intervention (the level change at the first time point post-intervention), the medium- and long-term impacts (level changes at time points that are several periods away from the point of intervention) are not represented directly by parameters of an ITS model. The post-intervention trend change coefficient only helps to estimate them indirectly through additional calculations.

More specifically, in a simple and comparative ITS model with one intervention (in **Equations 1** and **2** above), the level change at any time point post-intervention (relative to what it would have been absent the intervention) is given by **Equations 5** and **6** below, respectively (Huitema & McKean, 2000; see also Law, 2015; A. K. Wagner et al., 2002). (Mathematical proofs for these formulas are provided in **Appendices 9** and **10**.)

$$\Delta \widehat{Level}_t = \hat{\beta}_2 + \hat{\beta}_3(t - r) \qquad (5)$$

$$\Delta \widehat{Level}_t = \hat{\beta}_6 + \hat{\beta}_7(t - r) \qquad (6)$$

As seen in **Equation 5**, at the first time point post-intervention (at $t = k+1 = r$), where it is assumed that there is a level change and *no trend change*, the calculated level change is simply equal to the one obtained directly by estimating **Equation 1** (i.e., an estimate for *β₂*). And in **Equation 6**, the level change at $t = k+1 = r$ is simply equal to that obtained directly by estimating **Equation 2** (an estimate for *β₆*).

In addition, in a simple ITS model with two interventions (in **Equation 3**), while the level change at any time point from $t = k+1$ to $t = m$ (i.e., after the beginning of the first but before the start of the second intervention) is given by **Equation 5** above, the level change at any time point from $t = m+1$ to $t = n$ (i.e., after the start of the second intervention) is given by **Equation 7** below. (A mathematical proof for **Equation 7** is provided in **Appendix 11**.)

$$\Delta \widehat{Level}_t = \hat{\beta}_4 + \hat{\beta}_5(t - s) \qquad (7)$$

Finally, in a comparative ITS model with two interventions (in **Equation 4**), while the level change at any time point from $t = k+1$ to $t = m$ is given by **Equation 8 below**, the level change at any time point from $t = m+1$ to $t = n$ is given by **Equation 9**. (A mathematical proof for **Equation 9** is provided in **Appendix 12**. A mathematical proof for **Equation 8** follows the same logic as that for **Equation 6** provided in **Appendix 10**.)

$$\Delta \widehat{Level}_t = \hat{\beta}_8 + \hat{\beta}_9(t - r) \qquad (8)$$

$$\Delta \widehat{Level}_t = \hat{\beta}_{10} + \hat{\beta}_{11}(t - s) \qquad (9)$$

### *4.3.3. Stages in the Conduction of an Interrupted Time-Series Study*

There are thirteen main stages in the conduction of an ITS study. These include (Bernal et al., 2017; Jandoc et al., 2015; Law, 2015; S. L. Turner, Karahalios, et al., 2021; A. K. Wagner et al., 2002):

1. *Determining if an ITS analysis is suitable for the research topic*. This involves knowing the theory of ITS analysis as presented in the previous subsections. More specifically, a researcher who chooses the ITS analysis approach should know the specific time points at which the intervention began and ended, the nature of the outcome variable, the theoretical relationship between the intervention and the outcome, and the type of data available. Moreover, they should justify their preference for the ITS design over alternative research methods.
2. *Designing a theoretical model for the study*. Based on the availability or nonavailability of a suitable comparison group, the historical context, past empirical findings, and knowledge about the theoretical relationship between the intervention and the outcome variable, the researcher should design the theoretical model they intend to use for their study. The availability or non-availability of a reasonable control group guides the decision on whether to use a simple or comparative ITS model. The historical context helps the researcher decide whether to consider a single interruption or multiple interruptions. Past empirical findings and knowledge about the theoretical relationship between the intervention and the outcome variable inform the researcher about whether the investigated impact should be immediate or delayed.

3. *Coding the data based on the variables of the theoretical model*. Next, the researcher must code (transform or set up) the data according to the variables of the theoretical ITS model they have chosen since the ITS approach has a unique way in which raw data must be coded and organized depending on the type of model design selected.

4. *Graphically representing and visually inspecting the coded data*. At this stage, the researcher should graphically represent the coded organized data in order to detect and correct any anomalies they may have (such as containing outliers).

5. *Specifying and estimating a preliminary model*. After coding, graphically representing, and visually inspecting the data, the researcher should specify a preliminary model based on the theoretical model they designed in *Step 2*. Then, they should estimate this preliminary model using the ordinary least squares (OLS) method.

6. *Checking the preliminary model for autocorrelation*. Here, the standard OLS analysis results should be used to determine whether or not there is autocorrelation in the error terms of the preliminary regression model. More specifically, the preliminary model should be checked for autoregression (AR) and moving averages (MA).

7. *Specifying and estimating the final model*. If it is determined in *Step 6* that the data being used is generated from an autoregressive (AR) or moving average (MA) process (i.e., if the data are autocorrelated), the researcher should address this by including the AR or MA process in the preliminary model to obtain the final model.

Then, they should estimate the specified final model using the generalized least squares (GLS) method or any other technique that is suitable for correlated data. On the other hand, if it is determined in *Step 6* that the data are not autocorrelated, then the final model is simply the preliminary model.

8. *Conducting a robustness check on the final model*. Next, the researcher should check the robustness of the final model by testing whether or not adding more AR and MA parameters to it increases its goodness of fit. If it is confirmed that alternative model specifications with more parameters are not statistically better than the final model in estimating the outcome variable, then the estimated results of the final model in *Step 7* are indeed the final results. Otherwise, the final model should be updated and re-estimated accordingly.

9. *Plotting the results of the final model*. Like most natural experimental methods, an ITS analysis is incomplete without an accompanying graph visually depicting the results. Plotting the results makes them more captivating and convincing, especially to less technical readers or audiences. This stage involves presenting plots for the fitted regression lines of the various segments of the final model. The graph here is different from that in *Step 4* in the sense that it shows the observed and fitted (predicted) values of the final model while the plot in *Step 4* represents the coded data values based on the theoretical model.

10. *Estimating Level Changes at time points other than the first time point post-intervention*. When there is a trend change post-intervention, level changes at different time points are different. In other words, when the trend changes post-intervention, the intervention's medium- and long-term causal effects are different from its short-term impact. The only level change measured directly by a parameter of an ITS model is the one exactly at the first time point post-intervention (representing the short-term impact of the intervention). If the researcher wants to know level changes at other time points, they must estimate them through additional calculations using either **Equation 5**, **6**, **7**, **8**, or **9** provided above (depending on whether the study design is a simple or comparative ITS model with one or several interventions).

11. *Describing the policy implications of the results*. The next step in the researcher's ITS analysis is drawing policy conclusions from the study's results.
12. *Discussing the strengths and limitations of the study*.
13. *Offering suggestions for future research*.

In the general organization of an ITS research paper, *Steps 1* to *2* would be grouped under the *study design* section, *3* to *4* under the *data organization*, *5* to *10* under the *statistical analysis and results*, and *11* to *13* under the *discussion* sections. Alternatively, Steps *1* to *6* could be grouped under a section titled *research methodology*, *7* to *10* under *results*, and *11* to *13* under *discussion*.

### *4.3.4. A Comparison of Interrupted Time-Series Analysis With Regression Discontinuity Design and Regression Discontinuity in Time*

ITS analysis is conceptually similar to the *regression discontinuity* (*RD*) design (Shadish et al., 2002). Indeed, ITS analysis may be regarded as a special type of the RD design whereby the running variable is time and the threshold is a specific time point at which the intervention begins or is expected to have an impact (Law, 2015).[61] However, while the traditional RD design focuses on examining a change in level (change at the threshold or cutoff), the ITS analysis approach typically studies changes in both level and trend (Law, 2015).[62] ITS analysis is also methodologically different from the traditional RD design since it relies on time-series models and techniques, not cross-sectional data analysis procedures.

---

[61] ITS analysis may also be considered as a generalized form of either the difference-in-differences (DiD) design (in the case of a comparative ITS analysis) (Wong et al., 2015; Zieger et al., 2022) or the pre-post design (in the case of a simple ITS analysis), whereby instead of only one time point pre-intervention and one time point post-intervention (only two time points in total), there are several time points both pre- and post-intervention.

[62] A traditional (cross-sectional) RD design compares units just below and just above the threshold, and provides little information about those far from it (Law, 2015). The reason being that, in a traditional RD design, since at a given point in time all units above the threshold are subject to the same condition (rule or policy), and all units below the threshold are subject to the same condition, comparing outcomes for two or more units on the same side of the threshold is unimportant. In an ITS analysis, on the other hand, although two or more units on the same side of the threshold are subject to the same condition, they are so at different points in time: making a comparison between or among them meaningful.

Some authors use the name *regression discontinuity in time* (*RDiT*) to describe an approach in which traditional RD design procedures (bandwidth specification, etc.) are applied to an RD design where the running variable is time and the threshold is a specific time point (Hausman & Rapson, 2018; see also Godard et al., 2019; Gray et al., 2021; Meng, 2022). Yes, the RDiT design is similar to the ITS analysis approach by its use of time as the running variable and a specific time point as the threshold. However, like the traditional RD design, the RDiT design differs from the ITS technique through its focus on investigating a change near the threshold. Consequently, while the RDiT design is typically used to measure only the short-term impact of an intervention (Hausman & Rapson, 2018), the ITS method can be employed to estimate both short-term and long-term effects.[63] Because of the previous point, while time series in the RDiT design are generally short and high-frequency (hourly or daily) (Hausman & Rapson, 2018), they are typically long and low-frequency (quarterly or yearly) in ITS studies (this point is what also distinguishes an ITS analysis from an *event study*). Furthermore, the appellation *regression discontinuity in time* is very new: it originated in 2018 from Hausman and Rapson (2018). As opposed to the name *interrupted time-series analysis*, which dates back to the 1970s.[64]

---

[63] It should be noted that, even though the traditional RD design is used to investigate a change near the threshold, it differs from the RDiT design since, just like the ITS method, it can also be used to measure the long-term impact of an intervention. This is because, in the traditional RD design, if the dependent variable represents outcomes several years after the intervention, then the effect of the intervention shows its long-term impact regardless of whether we are close to or far from the threshold (since the threshold is not a specific time point). On the other hand, in the RDiT design, since the threshold is a specific time point, the right-hand-side neighborhood of the threshold represents short periods after the intervention: and, therefore, all values of the dependent variable at time points within this neighborhood show only the short-term effects of the intervention.

[64] Although the *regression discontinuity in time* appellation was coined in 2018, the idea of using the RD framework in a design where the running variable is time and the threshold is a specific time point started around 2008 (Hausman & Rapson, 2018). Moreover, traditional (cross-sectional) RD designs originated as early as in the 1960s (Imbens & Lemieux, 2008; Lee & Lemieux, 2010; see also Cook, 2008).

## 4.4. An Interrupted Time-Series Analysis of the Long-Term Inflation and Real GDP Effects of the Federal Reserve's Quantitative Easing Programs Conducted in Response to the Global Financial Crisis

As described in **Section 3**, in response to the Global Financial Crisis, while the Federal Reserve conducted quantitative easing, the Bank of Canada (the central bank of Canada) did not. More specifically, while the Federal Reserve implemented both quantitative easing (QE) and quasi-quantitative easing (qQE) programs, the Bank of Canada conducted only the latter. Because the U.S. and Canada are geographical neighbors, are both high-income economies, and implemented identical fiscal and macroprudential policies in response to the crisis, it is justifiable to measure the causal effects of the Federal Reserve's QE programs on U.S. inflation and real GDP using an ITS approach in which the treatment group is the U.S. and the control group is Canada.

**Figure 15** below shows U.S. and Canada CPI inflation and real GDP per capita from 1995 to 2021. The data are from the Organization for Economic Co-operation and Development (2022a, 2022b) and are seasonally adjusted. The data are also accessible from Kamkoum (2023a). Taking into consideration the 2001 and COVID-19 recessions, the data used for the statistical analysis of this study consist of the 2002Q1-2018Q4 subset of the data depicted in **Figure 15**.

I organize the ITS study such that the 13 steps outlined in **Section 4.3.3** are grouped under the headings *study design* (*Steps 1* to *2*), *data organization* (*Steps 3* to *4*), *statistical analysis and results* (*Steps 5* to *10*), and *discussion* (*Steps 11* to *13*).

**Figure 15**

A: *U.S. and Canada Real GDP per Capita*

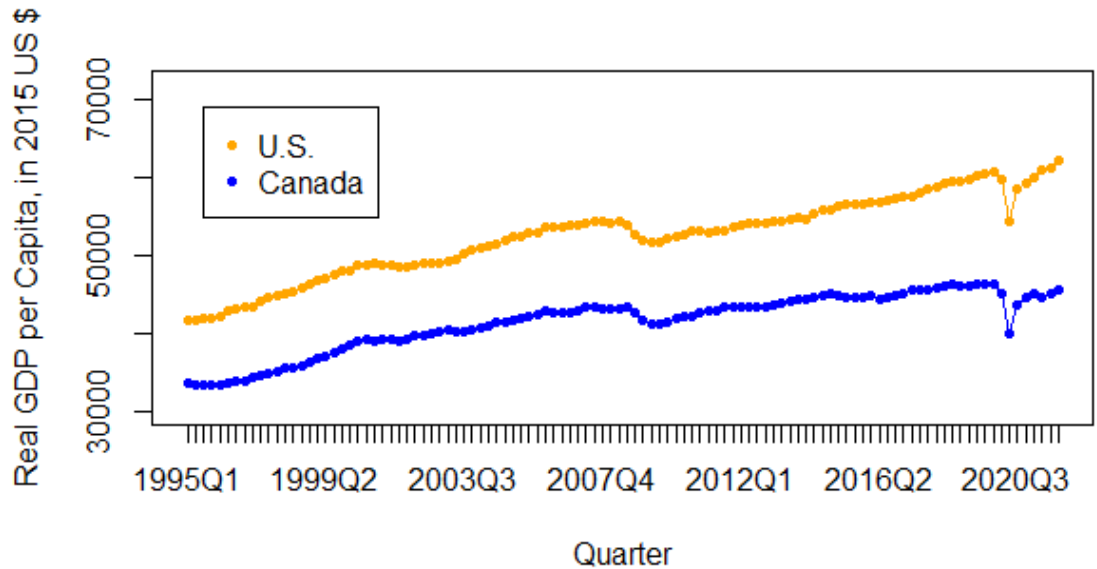

B: *U.S. and Canada CPI Inflation*

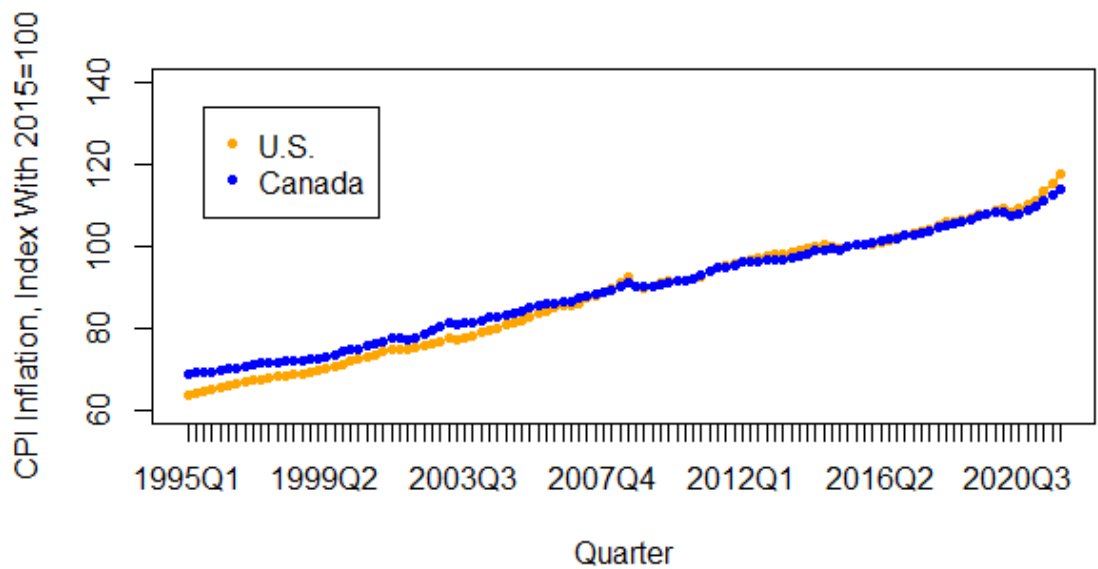

*Note*. All four series are seasonally adjusted. Drawn by the author using data from the Organization for Economic Co-operation and Development (2022a, 2022b). The data are also accessible from Kamkoum (2023a).

*4.4.1. Study Design*

*4.4.1.1. Suitability of the ITS Analysis Approach*

The ITS analysis approach is suitable for estimating the inflation and real GDP causal effects of the Federal Reserve's QE operations conducted in response to the Global Financial Crisis because (i) the specific time points at which the QE programs began and ended are known (December 2008 and October 2014, respectively), (ii) the implementation of the QE programs was the major policy change in the U.S. during that period, (iii) there is a reasonable comparison group (Canada), and (iv) there is enough inflation and real GDP data for statistical analysis.

*4.4.1.2. The Theoretical Model*

Because there is a suitable control group, the ideal theoretical model for this study should be the comparative ITS design.

Moreover, since the financial crisis itself affected the outcome variables (inflation and real GDP), the historical context suggests that, in addition to the intervention of interest (QE implementation), it may be necessary to specify an additional interruption (intervention) representing the crisis. Because there is evidence on the differential effects of the crisis on the U.S. and Canada (Ball, 2014; Ollivaud & D. Turner, 2015), the crisis is indeed specified as a different interruption.[65] In other words, there are two interventions in this study: (i) the crisis, which represents an interruption resulting from a natural occurrence, and (ii) the QE implementation, which represents an interruption resulting from a policy change. (Although the Federal Reserve implemented three QE programs, the three programs are considered a single intervention because of the long lags in the effects of monetary policy and the short intervals between the programs.)

---

[65] The financial crisis would not have been specified as a different interruption if it had affected the U.S. and Canada similarly (see **Section 4.3.1**).

Furthermore, because monetary policy affects inflation and real GDP with long and variable lags (Friedman, 1961, 1972; Mishkin, 2019; see also Barwell, 2016; Batini & Nelson, 2001; Rudebusch, 1995), the post-intervention period for the QE implementation should be modeled far apart from the time point at which QE started. Existing evidence suggests that the lags in the effects of monetary policy in the U.S. are over a year for real GDP and over two years for inflation (Mishkin, 2019). The lags are even longer when the policy rate is close to the zero lower bound. Another consequence of the long and variable lags in the effects of monetary policy is that the estimated *level* changes in inflation and real GDP at the *first time point post-QE* should be of little interest since these changes are too dependent on the lag length selected. What should interest us most should be the estimated *trend* changes in inflation and real GDP post-QE because they are calculated using data from various time points. Estimates for level changes in inflation and real GDP at time points that are *several quarters away from the specified QE start date* (i.e., estimates for the medium- and long-term level changes) should also be relevant since they are measured based on the trend changes and are less susceptible to the lag-length issue.

Based on the above facts and with the objective to estimate the causal effects of the QE programs independently of the impact of the crisis, instead of 2008Q4, this study specifies the QE intervention as starting between 2010Q4 and 2011Q1 for both the model in which real GDP is the outcome variable and that in which inflation is the outcome variable. This way, in both models, the first time point after the *specified* QE start date is in reality several time points after the *actual* QE start date.

In conclusion, the ideal theoretical model for this study is the comparative ITS model in **Equation 4** of **Section 4.3.2**. Therefore, with real GDP per capita and CPI inflation as separate outcome variables, the U.S. and Canada as treatment and control groups (respectively), and the Global Financial Crisis and QE implementation as first and second interventions (respectively), the study's two models are as follows.

$$GDP_{ijgt} = \beta_0 + \beta_1 time_t + \beta_2 crisis_i + \beta_3 crisis_i * (time_t - r) + \beta_4 QE_j$$
$$+ \beta_5 QE_j * (time_t - s) + \beta_6 US_g + \beta_7 (US_g * time_t)$$
$$+ \beta_8 (US_g * crisis_i) + \beta_9 US_g * crisis_i * (time_t - r)$$
$$+ \beta_{10}(US_g * QE_j) + \beta_{11} US_g * QE_j * (time_t - s) + \varepsilon_{ijgt}$$

**(Model 1)**

$$CPI_{ijgt} = \beta_0 + \beta_1 time_t + \beta_2 crisis_i + \beta_3 crisis_i * (time_t - r) + \beta_4 QE_j$$
$$+ \beta_5 QE_j * (time_t - s) + \beta_6 US_g + \beta_7 (US_g * time_t)$$
$$+ \beta_8 (US_g * crisis_i) + \beta_9 US_g * crisis_i * (time_t - r)$$
$$+ \beta_{10}(US_g * QE_j) + \beta_{11} US_g * QE_j * (time_t - s) + \varepsilon_{ijgt}$$

**(Model 2)**

Where the variables and coefficients are interpreted following the description provided in **Section 4.3.2**. More specifically, in both models, $crisis_i = i$ is a dummy variable that equals 1 after and 0 before the beginning of the crisis; $QE_j = j$ is a dummy variable that equals 1 after and 0 before the start of the QE implementation; $time_t = t$ is the time since the beginning of the study period, taking discrete values $t = 1, 2, 3, …, k, k+1, k+2, k+3, …, m, m+1, m+2, m+3, …, n$, with $t = 1, 2, 3, …, k$ representing the time points before the start of the crisis, $t = k+1, k+2, k+3, …, m$ representing the time points after the start of the crisis but before the beginning of the QE implementation, and $t = m+1, m+2, m+3, …, n$ representing the time points after the beginning of the QE implementation; $r = (k+1)$, such that $(time_t - r)$ captures the time points after the beginning of the crisis at which the trend change is measurable (i.e., in addition to the time points before the crisis, $(time_t - r)$ is also equal to 0 at $t = k+1$, it then increases by one unit from 1 at $t = k+2$ to $(n - k - 1)$ at $t = n$); $s = (m+1)$, such that $(time_t - s)$ captures the time points after the start of the QE implementation where the trend change is measurable (i.e., in addition to the time points before the QE implementation, $(time_t - s)$ is also equal to 0 at $t = m+1$, it then increases by one unit from 1 at $t = m+2$ to $(n - m - 1)$ at $t = n$); $US_g = g$ is a dummy variable that equals 1 for the U.S. and 0 for Canada; $crisis_i * (time_t - r)$, $QE_j * (time_t - s)$, $US_g * time_t$, $US_g * crisis_i$, $US_g * crisis_i * (time_t - r)$, $US_g * QE_j$, and $US_g * QE_j * (time_t - s)$ are interaction terms between the defined variables. In **Model 1**, $β_0$ is the intercept or level of real GDP per capita in Canada before the beginning of the study period (at $t = 0$); $β_1$ is the observed trend of real GDP per capita in Canada before the crisis; $β_2$ and $β_4$ denote the level change in real GDP per capita in Canada at the modeled first time point after the start of the crisis and QE implementation, respectively (at $t = k+1$ and $t = m+1$). $β_3$ and $β_5$ correspond to the trend change in real GDP per capita in Canada after the start of the crisis and QE implementation, respectively. $B_6$ is the level difference in real GDP per capita between the U.S. and Canada before the beginning of the study period (at $t = 0$). $B_7$ is the trend difference in real GDP per capita between the U.S. and Canada before the start of the crisis. $β_8$ and $β_{10}$ stand for the level change difference in real GDP per capita between the U.S. and Canada at the modeled first time point after the start of the crisis and QE implementation, respectively. $β_9$ and $β_{11}$ show the trend change difference in real GDP per capita between the U.S. and Canada after the start of the crisis and QE implementation, respectively. Thus, $β_{10}$ represents the level impact of the QE implementation relative to that of the crisis. And $β_{11}$ represents the trend impact of the QE implementation relative to that of the crisis. The coefficients of **Model 2** are interpreted similarly by simply considering that the outcome variable is CPI inflation instead of real GDP per capita. The parameters of interest in both models are $β_9$ and $β_{11}$ (as mentioned earlier, estimates for $β_8$ and $β_{10}$ are not very relevant considering the long and variable lags in the effects of monetary policy on inflation and real GDP—their importance is limited to their use in estimating longer-term level changes). In both models, it is important to note that the U.S. post-QE counterfactual represents what would have happened in the U.S. absent the QE implementation since QE was not conducted in Canada. However, the U.S. post-crisis counterfactual does not represent what would have happened in the U.S. absent the crisis since the crisis also occurred in Canada. In other words, while the U.S. post-QE counterfactual shows what would have happened in the U.S. absent the QE implementation, the U.S. post-crisis counterfactual shows what would have happened in the U.S. if the crisis' effect on the U.S. were the same as that on Canada.

*4.4.2. Data Organization*

*4.4.2.1. Data Coding*

The raw data for this ITS study consist of the 2002Q1-2018Q4 subset of the data depicted in **Figure 15** above. Based on the variables of **Models 1** and **2**, and considering the facts presented earlier, for the 2002Q1-2018Q4 study interval, (i) the pre- and post-crisis periods are 2002Q1-2007Q3 and 2007Q4-2018Q4, respectively (because the crisis began in August 2007, i.e., between 2007Q3 and 2007Q4); (ii) the pre- and post-QE periods are 2002Q1-2010Q4 and 2011Q1-2018Q4, respectively (because the *specified* QE start date is between 2010Q4 and 2011Q1); (iii) the trend change impact of the crisis is valid from 2008Q1, even though the post-crisis period begins in 2007Q4; (iv) the trend change impact of the QE implementation is valid from 2011Q2, even though the post-QE period starts in 2011Q1; (v) $k = 23$, such that $r = k+1 = 24$; and (vi) $m = 36$, such that $s = m+1 = 37$. Thus, following the variables' definitions provided in the previous subsection, the raw data are coded as in **Table 5** below. The raw and coded data can be freely downloaded from Kamkoum (2023a).

**Table 5**

*Coded Data*

| yearqtr | us | gdp | cpi | time | crisis | crisis*(time-24) | us*time | us*crisis | us*crisis*(time-24) | qe | qe*(time-37) | us*qe | us*qe*(time-37) |
|---|---|---|---|---|---|---|---|---|---|---|---|---|---|
| 2002Q1 | 1 | 48873 | 75.1 | 1 | 0 | 0 | 1 | 0 | 0 | 0 | 0 | 0 | 0 |
| 2002Q2 | 1 | 49064 | 75.7 | 2 | 0 | 0 | 2 | 0 | 0 | 0 | 0 | 0 | 0 |
| 2002Q3 | 1 | 49138 | 76.1 | 3 | 0 | 0 | 3 | 0 | 0 | 0 | 0 | 0 | 0 |
| 2002Q4 | 1 | 49082 | 76.6 | 4 | 0 | 0 | 4 | 0 | 0 | 0 | 0 | 0 | 0 |
| 2003Q1 | 1 | 49233 | 77.4 | 5 | 0 | 0 | 5 | 0 | 0 | 0 | 0 | 0 | 0 |
| 2003Q2 | 1 | 49563 | 77.2 | 6 | 0 | 0 | 6 | 0 | 0 | 0 | 0 | 0 | 0 |
| 2003Q3 | 1 | 50264 | 77.8 | 7 | 0 | 0 | 7 | 0 | 0 | 0 | 0 | 0 | 0 |
| 2003Q4 | 1 | 50723 | 78.1 | 8 | 0 | 0 | 8 | 0 | 0 | 0 | 0 | 0 | 0 |
| 2004Q1 | 1 | 50914 | 78.8 | 9 | 0 | 0 | 9 | 0 | 0 | 0 | 0 | 0 | 0 |
| 2004Q2 | 1 | 51199 | 79.4 | 10 | 0 | 0 | 10 | 0 | 0 | 0 | 0 | 0 | 0 |
| 2004Q3 | 1 | 51553 | 79.9 | 11 | 0 | 0 | 11 | 0 | 0 | 0 | 0 | 0 | 0 |
| 2004Q4 | 1 | 51950 | 80.8 | 12 | 0 | 0 | 12 | 0 | 0 | 0 | 0 | 0 | 0 |
| 2005Q1 | 1 | 52416 | 81.2 | 13 | 0 | 0 | 13 | 0 | 0 | 0 | 0 | 0 | 0 |
| 2005Q2 | 1 | 52559 | 81.7 | 14 | 0 | 0 | 14 | 0 | 0 | 0 | 0 | 0 | 0 |
| 2005Q3 | 1 | 52837 | 82.9 | 15 | 0 | 0 | 15 | 0 | 0 | 0 | 0 | 0 | 0 |
| 2005Q4 | 1 | 53004 | 83.7 | 16 | 0 | 0 | 16 | 0 | 0 | 0 | 0 | 0 | 0 |
| 2006Q1 | 1 | 53602 | 84.2 | 17 | 0 | 0 | 17 | 0 | 0 | 0 | 0 | 0 | 0 |
| 2006Q2 | 1 | 53612 | 84.9 | 18 | 0 | 0 | 18 | 0 | 0 | 0 | 0 | 0 | 0 |
| 2006Q3 | 1 | 53554 | 85.7 | 19 | 0 | 0 | 19 | 0 | 0 | 0 | 0 | 0 | 0 |
| 2006Q4 | 1 | 53868 | 85.4 | 20 | 0 | 0 | 20 | 0 | 0 | 0 | 0 | 0 | 0 |
| 2007Q1 | 1 | 53907 | 86.2 | 21 | 0 | 0 | 21 | 0 | 0 | 0 | 0 | 0 | 0 |
| 2007Q2 | 1 | 54130 | 87.2 | 22 | 0 | 0 | 22 | 0 | 0 | 0 | 0 | 0 | 0 |
| 2007Q3 | 1 | 54316 | 87.7 | 23 | 0 | 0 | 23 | 0 | 0 | 0 | 0 | 0 | 0 |
| 2007Q4 | 1 | 54508 | 88.8 | 24 | 1 | 0 | 24 | 1 | 0 | 0 | 0 | 0 | 0 |
| 2008Q1 | 1 | 54167 | 89.8 | 25 | 1 | 1 | 25 | 1 | 1 | 0 | 0 | 0 | 0 |
| 2008Q2 | 1 | 54358 | 90.9 | 26 | 1 | 2 | 26 | 1 | 2 | 0 | 0 | 0 | 0 |
| 2008Q3 | 1 | 53940 | 92.3 | 27 | 1 | 3 | 27 | 1 | 3 | 0 | 0 | 0 | 0 |
| 2008Q4 | 1 | 52639 | 90.2 | 28 | 1 | 4 | 28 | 1 | 4 | 0 | 0 | 0 | 0 |
| 2009Q1 | 1 | 51920 | 89.6 | 29 | 1 | 5 | 29 | 1 | 5 | 0 | 0 | 0 | 0 |
| 2009Q2 | 1 | 51726 | 90.1 | 30 | 1 | 6 | 30 | 1 | 6 | 0 | 0 | 0 | 0 |
| 2009Q3 | 1 | 51794 | 90.9 | 31 | 1 | 7 | 31 | 1 | 7 | 0 | 0 | 0 | 0 |
| 2009Q4 | 1 | 52226 | 91.6 | 32 | 1 | 8 | 32 | 1 | 8 | 0 | 0 | 0 | 0 |
| 2010Q1 | 1 | 52383 | 91.7 | 33 | 1 | 9 | 33 | 1 | 9 | 0 | 0 | 0 | 0 |
| 2010Q2 | 1 | 52789 | 91.7 | 34 | 1 | 10 | 34 | 1 | 10 | 0 | 0 | 0 | 0 |
| 2010Q3 | 1 | 53084 | 91.9 | 35 | 1 | 11 | 35 | 1 | 11 | 0 | 0 | 0 | 0 |
| 2010Q4 | 1 | 53245 | 92.7 | 36 | 1 | 12 | 36 | 1 | 12 | 0 | 0 | 0 | 0 |
| 2011Q1 | 1 | 53022 | 93.7 | 37 | 1 | 13 | 37 | 1 | 13 | 1 | 0 | 1 | 0 |
| 2011Q2 | 1 | 53285 | 94.7 | 38 | 1 | 14 | 38 | 1 | 14 | 1 | 1 | 1 | 1 |
| 2011Q3 | 1 | 53150 | 95.4 | 39 | 1 | 15 | 39 | 1 | 15 | 1 | 2 | 1 | 2 |
| 2011Q4 | 1 | 53635 | 95.8 | 40 | 1 | 16 | 40 | 1 | 16 | 1 | 3 | 1 | 3 |
| 2012Q1 | 1 | 53978 | 96.3 | 41 | 1 | 17 | 41 | 1 | 17 | 1 | 4 | 1 | 4 |
| 2012Q2 | 1 | 54127 | 96.5 | 42 | 1 | 18 | 42 | 1 | 18 | 1 | 5 | 1 | 5 |
| 2012Q3 | 1 | 54109 | 97 | 43 | 1 | 19 | 43 | 1 | 19 | 1 | 6 | 1 | 6 |
| 2012Q4 | 1 | 54056 | 97.6 | 44 | 1 | 20 | 44 | 1 | 20 | 1 | 7 | 1 | 7 |
| 2013Q1 | 1 | 54437 | 98 | 45 | 1 | 21 | 45 | 1 | 21 | 1 | 8 | 1 | 8 |
| 2013Q2 | 1 | 54420 | 97.9 | 46 | 1 | 22 | 46 | 1 | 22 | 1 | 9 | 1 | 9 |
| 2013Q3 | 1 | 54737 | 98.4 | 47 | 1 | 23 | 47 | 1 | 23 | 1 | 10 | 1 | 10 |
| 2013Q4 | 1 | 55009 | 98.8 | 48 | 1 | 24 | 48 | 1 | 24 | 1 | 11 | 1 | 11 |
| 2014Q1 | 1 | 54718 | 99.4 | 49 | 1 | 25 | 49 | 1 | 25 | 1 | 12 | 1 | 12 |
| 2014Q2 | 1 | 55319 | 99.9 | 50 | 1 | 26 | 50 | 1 | 26 | 1 | 13 | 1 | 13 |
| 2014Q3 | 1 | 55843 | 100.2 | 51 | 1 | 27 | 51 | 1 | 27 | 1 | 14 | 1 | 14 |
| 2014Q4 | 1 | 55972 | 99.9 | 52 | 1 | 28 | 52 | 1 | 28 | 1 | 15 | 1 | 15 |
| 2015Q1 | 1 | 56330 | 99.3 | 53 | 1 | 29 | 53 | 1 | 29 | 1 | 16 | 1 | 16 |
| 2015Q2 | 1 | 56555 | 100 | 54 | 1 | 30 | 54 | 1 | 30 | 1 | 17 | 1 | 17 |
| 2015Q3 | 1 | 56617 | 100.4 | 55 | 1 | 31 | 55 | 1 | 31 | 1 | 18 | 1 | 18 |
| 2015Q4 | 1 | 56580 | 100.3 | 56 | 1 | 32 | 56 | 1 | 32 | 1 | 19 | 1 | 19 |
| 2016Q1 | 1 | 56812 | 100.3 | 57 | 1 | 33 | 57 | 1 | 33 | 1 | 20 | 1 | 20 |
| 2016Q2 | 1 | 56883 | 101.1 | 58 | 1 | 34 | 58 | 1 | 34 | 1 | 21 | 1 | 21 |
| 2016Q3 | 1 | 57108 | 101.5 | 59 | 1 | 35 | 59 | 1 | 35 | 1 | 22 | 1 | 22 |
| 2016Q4 | 1 | 57281 | 102.2 | 60 | 1 | 36 | 60 | 1 | 36 | 1 | 23 | 1 | 23 |
| 2017Q1 | 1 | 57463 | 102.9 | 61 | 1 | 37 | 61 | 1 | 37 | 1 | 24 | 1 | 24 |
| 2017Q2 | 1 | 57696 | 103 | 62 | 1 | 38 | 62 | 1 | 38 | 1 | 25 | 1 | 25 |
| 2017Q3 | 1 | 58007 | 103.5 | 63 | 1 | 39 | 63 | 1 | 39 | 1 | 26 | 1 | 26 |
| 2017Q4 | 1 | 58454 | 104.3 | 64 | 1 | 40 | 64 | 1 | 40 | 1 | 27 | 1 | 27 |
| 2018Q1 | 1 | 58828 | 105.2 | 65 | 1 | 41 | 65 | 1 | 41 | 1 | 28 | 1 | 28 |
| 2018Q2 | 1 | 59240 | 105.8 | 66 | 1 | 42 | 66 | 1 | 42 | 1 | 29 | 1 | 29 |
| 2018Q3 | 1 | 59433 | 106.2 | 67 | 1 | 43 | 67 | 1 | 43 | 1 | 30 | 1 | 30 |
| 2018Q4 | 1 | 59478 | 106.6 | 68 | 1 | 44 | 68 | 1 | 44 | 1 | 31 | 1 | 31 |
| 2002Q1 | 0 | 39711 | 77.8 | 1 | 0 | 0 | 0 | 0 | 0 | 0 | 0 | 0 | 0 |
| 2002Q2 | 0 | 39838 | 78.6 | 2 | 0 | 0 | 0 | 0 | 0 | 0 | 0 | 0 | 0 |
| 2002Q3 | 0 | 40046 | 79.5 | 3 | 0 | 0 | 0 | 0 | 0 | 0 | 0 | 0 | 0 |
| 2002Q4 | 0 | 40148 | 80.2 | 4 | 0 | 0 | 0 | 0 | 0 | 0 | 0 | 0 | 0 |
| 2003Q1 | 0 | 40333 | 81.2 | 5 | 0 | 0 | 0 | 0 | 0 | 0 | 0 | 0 | 0 |
| 2003Q2 | 0 | 40184 | 80.8 | 6 | 0 | 0 | 0 | 0 | 0 | 0 | 0 | 0 | 0 |
| 2003Q3 | 0 | 40216 | 81.2 | 7 | 0 | 0 | 0 | 0 | 0 | 0 | 0 | 0 | 0 |
| 2003Q4 | 0 | 40373 | 81.5 | 8 | 0 | 0 | 0 | 0 | 0 | 0 | 0 | 0 | 0 |
| 2004Q1 | 0 | 40613 | 81.9 | 9 | 0 | 0 | 0 | 0 | 0 | 0 | 0 | 0 | 0 |
| 2004Q2 | 0 | 41007 | 82.6 | 10 | 0 | 0 | 0 | 0 | 0 | 0 | 0 | 0 | 0 |
| 2004Q3 | 0 | 41369 | 82.8 | 11 | 0 | 0 | 0 | 0 | 0 | 0 | 0 | 0 | 0 |
| 2004Q4 | 0 | 41537 | 83.4 | 12 | 0 | 0 | 0 | 0 | 0 | 0 | 0 | 0 | 0 |
| 2005Q1 | 0 | 41633 | 83.7 | 13 | 0 | 0 | 0 | 0 | 0 | 0 | 0 | 0 | 0 |
| 2005Q2 | 0 | 41850 | 84.1 | 14 | 0 | 0 | 0 | 0 | 0 | 0 | 0 | 0 | 0 |
| 2005Q3 | 0 | 42223 | 84.9 | 15 | 0 | 0 | 0 | 0 | 0 | 0 | 0 | 0 | 0 |
| 2005Q4 | 0 | 42495 | 85.3 | 16 | 0 | 0 | 0 | 0 | 0 | 0 | 0 | 0 | 0 |
| 2006Q1 | 0 | 42785 | 85.9 | 17 | 0 | 0 | 0 | 0 | 0 | 0 | 0 | 0 | 0 |
| 2006Q2 | 0 | 42708 | 86.2 | 18 | 0 | 0 | 0 | 0 | 0 | 0 | 0 | 0 | 0 |
| 2006Q3 | 0 | 42695 | 86.3 | 19 | 0 | 0 | 0 | 0 | 0 | 0 | 0 | 0 | 0 |
| 2006Q4 | 0 | 42721 | 86.5 | 20 | 0 | 0 | 0 | 0 | 0 | 0 | 0 | 0 | 0 |
| 2007Q1 | 0 | 42946 | 87.4 | 21 | 0 | 0 | 0 | 0 | 0 | 0 | 0 | 0 | 0 |
| 2007Q2 | 0 | 43272 | 88 | 22 | 0 | 0 | 0 | 0 | 0 | 0 | 0 | 0 | 0 |
| 2007Q3 | 0 | 43317 | 88.1 | 23 | 0 | 0 | 0 | 0 | 0 | 0 | 0 | 0 | 0 |
| 2007Q4 | 0 | 43218 | 88.7 | 24 | 1 | 0 | 0 | 0 | 0 | 0 | 0 | 0 | 0 |
| 2008Q1 | 0 | 43188 | 89 | 25 | 1 | 1 | 0 | 0 | 0 | 0 | 0 | 0 | 0 |
| 2008Q2 | 0 | 43243 | 90.2 | 26 | 1 | 2 | 0 | 0 | 0 | 0 | 0 | 0 | 0 |
| 2008Q3 | 0 | 43441 | 91.2 | 27 | 1 | 3 | 0 | 0 | 0 | 0 | 0 | 0 | 0 |
| 2008Q4 | 0 | 42776 | 90.3 | 28 | 1 | 4 | 0 | 0 | 0 | 0 | 0 | 0 | 0 |
| 2009Q1 | 0 | 41741 | 90 | 29 | 1 | 5 | 0 | 0 | 0 | 0 | 0 | 0 | 0 |
| 2009Q2 | 0 | 41182 | 90.2 | 30 | 1 | 6 | 0 | 0 | 0 | 0 | 0 | 0 | 0 |
| 2009Q3 | 0 | 41222 | 90.4 | 31 | 1 | 7 | 0 | 0 | 0 | 0 | 0 | 0 | 0 |
| 2009Q4 | 0 | 41545 | 91 | 32 | 1 | 8 | 0 | 0 | 0 | 0 | 0 | 0 | 0 |
| 2010Q1 | 0 | 41983 | 91.5 | 33 | 1 | 9 | 0 | 0 | 0 | 0 | 0 | 0 | 0 |
| 2010Q2 | 0 | 42101 | 91.5 | 34 | 1 | 10 | 0 | 0 | 0 | 0 | 0 | 0 | 0 |
| 2010Q3 | 0 | 42255 | 92.1 | 35 | 1 | 11 | 0 | 0 | 0 | 0 | 0 | 0 | 0 |
| 2010Q4 | 0 | 42569 | 93 | 36 | 1 | 12 | 0 | 0 | 0 | 0 | 0 | 0 | 0 |
| 2011Q1 | 0 | 42847 | 93.7 | 37 | 1 | 13 | 0 | 0 | 0 | 1 | 0 | 0 | 0 |
| 2011Q2 | 0 | 42850 | 94.6 | 38 | 1 | 14 | 0 | 0 | 0 | 1 | 1 | 0 | 0 |
| 2011Q3 | 0 | 43303 | 94.9 | 39 | 1 | 15 | 0 | 0 | 0 | 1 | 2 | 0 | 0 |
| 2011Q4 | 0 | 43494 | 95.5 | 40 | 1 | 16 | 0 | 0 | 0 | 1 | 3 | 0 | 0 |
| 2012Q1 | 0 | 43448 | 96 | 41 | 1 | 17 | 0 | 0 | 0 | 1 | 4 | 0 | 0 |
| 2012Q2 | 0 | 43494 | 96 | 42 | 1 | 18 | 0 | 0 | 0 | 1 | 5 | 0 | 0 |
| 2012Q3 | 0 | 43401 | 96.1 | 43 | 1 | 19 | 0 | 0 | 0 | 1 | 6 | 0 | 0 |
| 2012Q4 | 0 | 43338 | 96.5 | 44 | 1 | 20 | 0 | 0 | 0 | 1 | 7 | 0 | 0 |
| 2013Q1 | 0 | 43668 | 96.9 | 45 | 1 | 21 | 0 | 0 | 0 | 1 | 8 | 0 | 0 |
| 2013Q2 | 0 | 43827 | 96.7 | 46 | 1 | 22 | 0 | 0 | 0 | 1 | 9 | 0 | 0 |
| 2013Q3 | 0 | 44027 | 97.1 | 47 | 1 | 23 | 0 | 0 | 0 | 1 | 10 | 0 | 0 |
| 2013Q4 | 0 | 44327 | 97.4 | 48 | 1 | 24 | 0 | 0 | 0 | 1 | 11 | 0 | 0 |
| 2014Q1 | 0 | 44352 | 98.2 | 49 | 1 | 25 | 0 | 0 | 0 | 1 | 12 | 0 | 0 |
| 2014Q2 | 0 | 44663 | 98.9 | 50 | 1 | 26 | 0 | 0 | 0 | 1 | 13 | 0 | 0 |
| 2014Q3 | 0 | 44946 | 99.2 | 51 | 1 | 27 | 0 | 0 | 0 | 1 | 14 | 0 | 0 |
| 2014Q4 | 0 | 45103 | 99.3 | 52 | 1 | 28 | 0 | 0 | 0 | 1 | 15 | 0 | 0 |
| 2015Q1 | 0 | 44835 | 99.2 | 53 | 1 | 29 | 0 | 0 | 0 | 1 | 16 | 0 | 0 |
| 2015Q2 | 0 | 44667 | 99.8 | 54 | 1 | 30 | 0 | 0 | 0 | 1 | 17 | 0 | 0 |
| 2015Q3 | 0 | 44710 | 100.4 | 55 | 1 | 31 | 0 | 0 | 0 | 1 | 18 | 0 | 0 |
| 2015Q4 | 0 | 44594 | 100.6 | 56 | 1 | 32 | 0 | 0 | 0 | 1 | 19 | 0 | 0 |
| 2016Q1 | 0 | 44778 | 100.7 | 57 | 1 | 33 | 0 | 0 | 0 | 1 | 20 | 0 | 0 |
| 2016Q2 | 0 | 44435 | 101.3 | 58 | 1 | 34 | 0 | 0 | 0 | 1 | 21 | 0 | 0 |
| 2016Q3 | 0 | 44720 | 101.6 | 59 | 1 | 35 | 0 | 0 | 0 | 1 | 22 | 0 | 0 |
| 2016Q4 | 0 | 44786 | 102 | 60 | 1 | 36 | 0 | 0 | 0 | 1 | 23 | 0 | 0 |
| 2017Q1 | 0 | 45204 | 102.6 | 61 | 1 | 37 | 0 | 0 | 0 | 1 | 24 | 0 | 0 |
| 2017Q2 | 0 | 45561 | 102.7 | 62 | 1 | 38 | 0 | 0 | 0 | 1 | 25 | 0 | 0 |
| 2017Q3 | 0 | 45569 | 103 | 63 | 1 | 39 | 0 | 0 | 0 | 1 | 26 | 0 | 0 |
| 2017Q4 | 0 | 45586 | 103.9 | 64 | 1 | 40 | 0 | 0 | 0 | 1 | 27 | 0 | 0 |
| 2018Q1 | 0 | 45888 | 104.7 | 65 | 1 | 41 | 0 | 0 | 0 | 1 | 28 | 0 | 0 |
| 2018Q2 | 0 | 46123 | 105 | 66 | 1 | 42 | 0 | 0 | 0 | 1 | 29 | 0 | 0 |
| 2018Q3 | 0 | 46232 | 105.7 | 67 | 1 | 43 | 0 | 0 | 0 | 1 | 30 | 0 | 0 |
| 2018Q4 | 0 | 46159 | 106 | 68 | 1 | 44 | 0 | 0 | 0 | 1 | 31 | 0 | 0 |

*4.4.2.2. Graphical Representation of the Coded Data*

**Figure 16** below visually depicts the coded data.

**Figure 16**

A: *U.S. and Canada Real GDP per Capita*

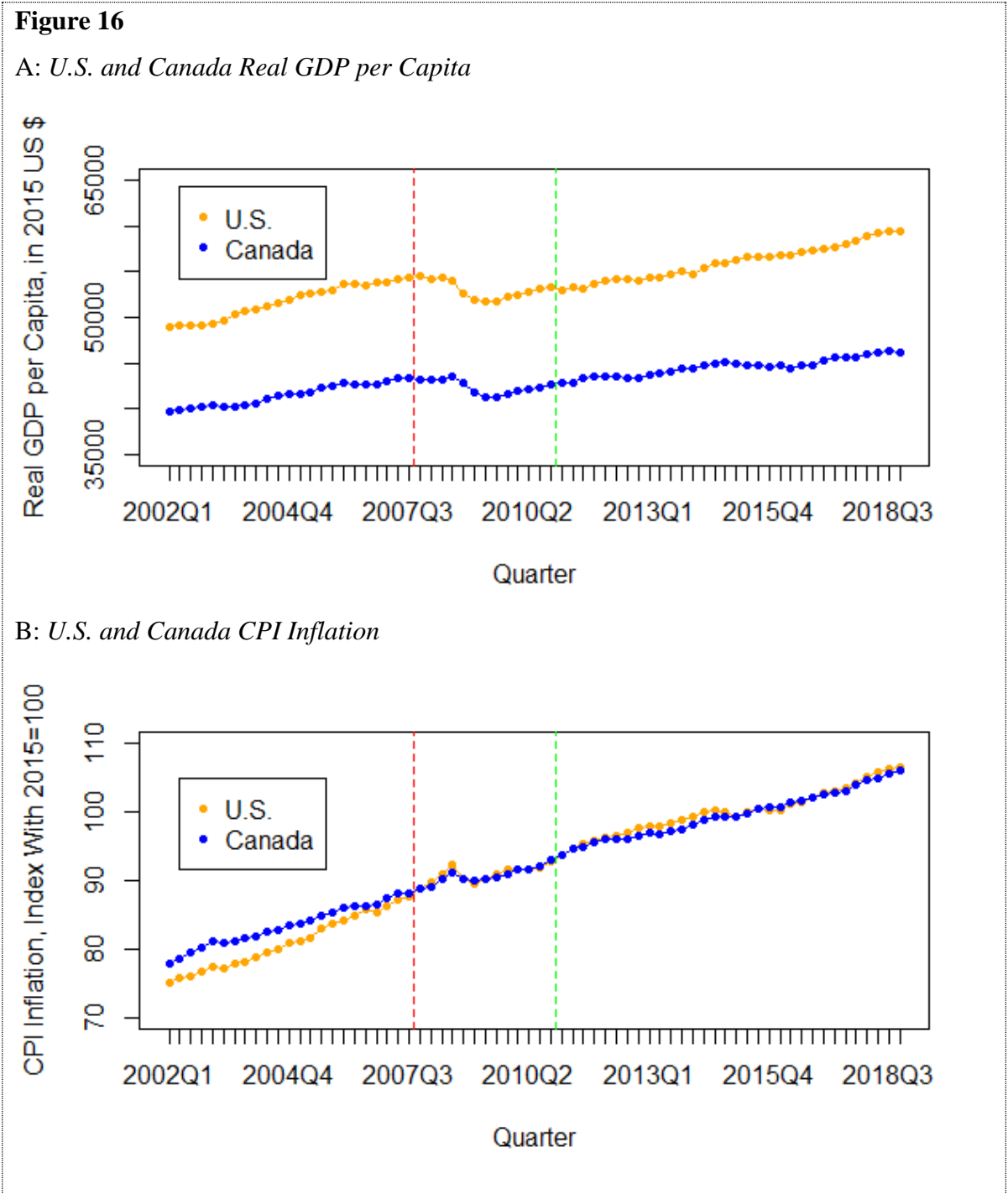

B: *U.S. and Canada CPI Inflation*

*Note.* All series are seasonally adjusted. The dashed red and green lines represent this study's specified dates for the Global Financial Crisis and QE implementation, respectively. Instead of 2008Q4, this study specifies the QE implementation as starting between 2010Q4 and 2011Q1 for both the model in which real GDP is the outcome variable and that in which inflation is the outcome variable since monetary policy affects inflation and real GDP with long lags (i.e., 2002Q1-2010Q4 is the pre-QE period and 2011Q1-2018Q4 is the post-QE period in both models). Drawn by the author using data from the Organization for Economic Co-operation and Development (2022a, 2022b). The data are also accessible from Kamkoum (2023a). R codes for the figure are found in **Appendix 14** and can be freely downloaded from Kamkoum (2023a).

### *4.4.3. Statistical Analysis and Results*

All the statistical analyses in this study are done in R Version 4.2.0. The R codes for the statistical analyses (model estimation results or tables) and all the figures in the study are found in **Appendix 14**. The R codes can also be freely downloaded from Kamkoum (2023a).

#### *4.4.3.1. The Preliminary Model*

Using the data in **Table 5** to estimate **Models 1** and **2** (where $r = 24$ and $s = 37$) by the OLS method gives the results in **Table 6** below.

In Panel A of **Table 6**, estimates for the two parameters of interest (estimates for $β_9$ and $β_{11}$, the coefficients of the $us.crisis.time\_minus\_24$ and $us.qe.time\_minus\_37$ interaction terms) are both statistically significant. The coefficient of the $us.crisis.time\_minus\_24$ interaction term ($\hat{β}_9$) shows that the post-crisis trend change difference in real GDP per capita between the U.S. and Canada was -124 US dollars. In other words, for the 2008Q1-2018Q4 period, real GDP per capita decreased by an average of 124 US dollars per quarter in the U.S. relative to Canada due to the crisis.

**Table 6**

A: *OLS Results for Model 1* (gdp = time + crisis + crisis.(time-24) + qe + qe.(time-37) + us + us.time + us.crisis + us.crisis.(time-24) + us.qe + us.qe.(time-37))

```
Coefficients:
                         Estimate Std. Error t value Pr(>|t|)
(Intercept)              39339.04    177.54  221.574  < 2e-16 ***
time                       178.34     12.95   13.773  < 2e-16 ***
crisis                    -600.61    279.53   -2.149  0.03361 *
crisis.time_minus_24      -290.88     33.17   -8.770 1.15e-14 ***
qe                        1439.51    281.04    5.122 1.12e-06 ***
qe.time_minus_37           211.51     31.54    6.707 6.29e-10 ***
us                        9034.97    251.08   35.984  < 2e-16 ***
us.time                     99.77     18.31    5.448 2.63e-07 ***
us.crisis                 -644.47    395.32   -1.630  0.10559
us.crisis.time_minus_24   -124.00     46.90   -2.644  0.00926 **
us.qe                     -636.25    397.45   -1.601  0.11196
us.qe.time_minus_37        131.02     44.60    2.938  0.00394 **
---
Signif. codes:  0 '***' 0.001 '**' 0.01 '*' 0.05 '.' 0.1 ' ' 1

Residual standard error: 411.9 on 124 degrees of freedom
Multiple R-squared:  0.9955,    Adjusted R-squared:  0.9951
F-statistic:  2516 on 11 and 124 DF,  p-value: < 2.2e-16
```

B: *OLS Results for Model 2* (cpi = time + crisis + crisis.(time-24) + qe + qe.(time-37) + us + us.time + us.crisis + us.crisis.(time-24) + us.qe + us.qe.(time-37))

```
Coefficients:
                         Estimate Std. Error t value Pr(>|t|)
(Intercept)              78.06877    0.23077 338.292  < 2e-16 ***
time                      0.44318    0.01683  26.331  < 2e-16 ***
crisis                    0.37288    0.36334   1.026 0.306761
crisis.time_minus_24     -0.17285    0.04311  -4.010 0.000104 ***
qe                        1.27815    0.36530   3.499 0.000649 ***
qe.time_minus_37          0.09730    0.04099   2.374 0.019143 *
us                       -4.05573    0.32636 -12.427  < 2e-16 ***
us.time                   0.14160    0.02380   5.949 2.57e-08 ***
us.crisis                 1.21885    0.51384   2.372 0.019228 *
us.crisis.time_minus_24  -0.19545    0.06097  -3.206 0.001713 **
us.qe                     0.80835    0.51661   1.565 0.120195
us.qe.time_minus_37       0.04268    0.05797   0.736 0.462928
---
Signif. codes:  0 '***' 0.001 '**' 0.01 '*' 0.05 '.' 0.1 ' ' 1

Residual standard error: 0.5354 on 124 degrees of freedom
Multiple R-squared:  0.9964,    Adjusted R-squared:  0.996
F-statistic:  3077 on 11 and 124 DF,  p-value: < 2.2e-16
```

The coefficient of the $us.qe.time\_minus\_37$ interaction term ($\hat{\beta}_{11}$) shows that the post-QE trend change difference in real GDP per capita between the U.S. and Canada was 131 US dollars. That is, from 2011Q2 to 2018Q4, real GDP per capita increased by an average of 131 US dollars per quarter in the U.S. relative to Canada due to the QE implementation. Since the change in Canada represents the U.S. counterfactual, this result means that *real GDP per capita in the U.S. increased by an average of 131 dollars per quarter post-QE implementation relative to how it would have changed had the QE not been conducted*. In other words, had the U.S. not implemented its QE programs, real GDP per capita in the U.S. would have been lower by an average of 131 dollars per quarter from 2011Q2 to 2018Q4.

In Panel B of **Table 6**, only one of the two coefficients of interest is statistically significant. We notice that while the financial crisis differentially affected CPI inflation in the U.S. and Canada, QE implementation did not. That is, the result indicates that *the Federal Reserve's QE programs did not affect U.S. inflation*. (To interpret the other coefficients in the table, see the description provided in **Section 4.4.1.2**.)

For the results of **Table 6** to be valid, residuals of **Models 1** and **2** should not be autocorrelated. That is why the next step in this ITS study is checking the models for autocorrelation. This is done in the following subsection.

*4.4.3.2. Autocorrelation Checks on Models 1 and 2*
**Figures 17** and **18** show the results of autocorrelation checks on **Models 1** and **2**. The figures show that, taken together, the Durbin-Watson and Breusch-Godfrey tests for residual autocorrelation and ACF and Partial ACF plots suggest that the data used to estimate **Models 1** and **2** are generated from autoregressive processes. The Figures suggest that **Models 1** and **2** should be re-estimated with AR(13) and AR(10) terms, respectively. This is done in the following paragraphs using the GLS method.

**Figure 17**

A: *Durbin-Watson and Breusch-Godfrey Residual Autocorrelation Tests on Model 1*

```
> dwtest(model_1)

        Durbin-Watson test

data:  model_1
DW = 0.58583, p-value < 2.2e-16
alternative hypothesis: true autocorrelation is greater than 0

> bgtest(model_1, order = 16)

        Breusch-Godfrey test for serial correlation of order up to 16

data:  model_1
LM test = 95.541, df = 16, p-value = 2.357e-13
```

B: *ACF and Partial-ACF for Residuals of Model 1*

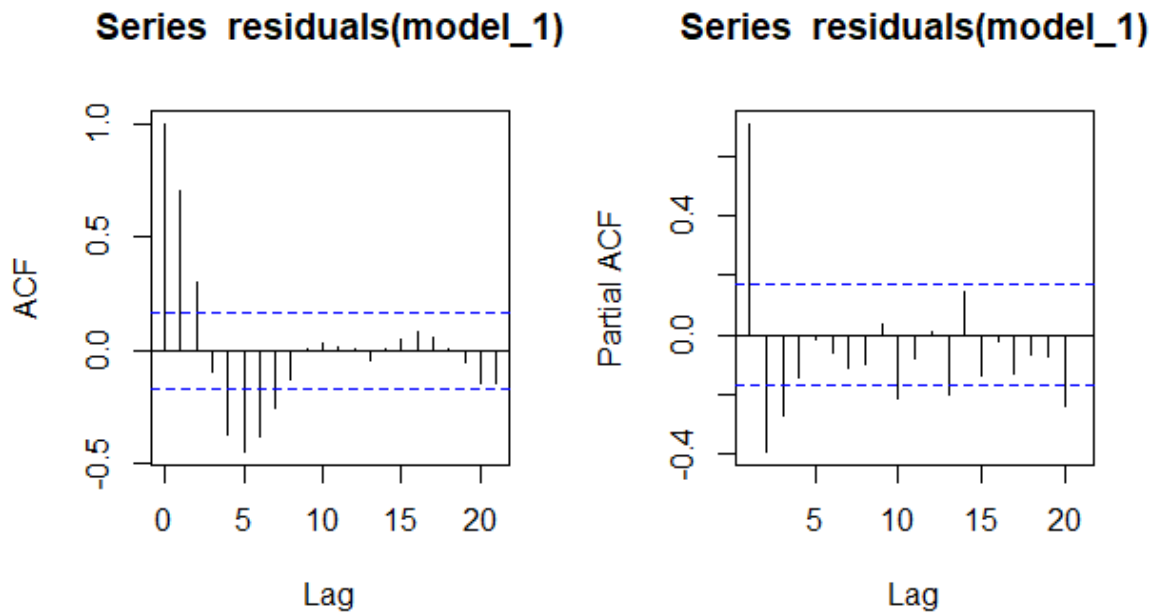

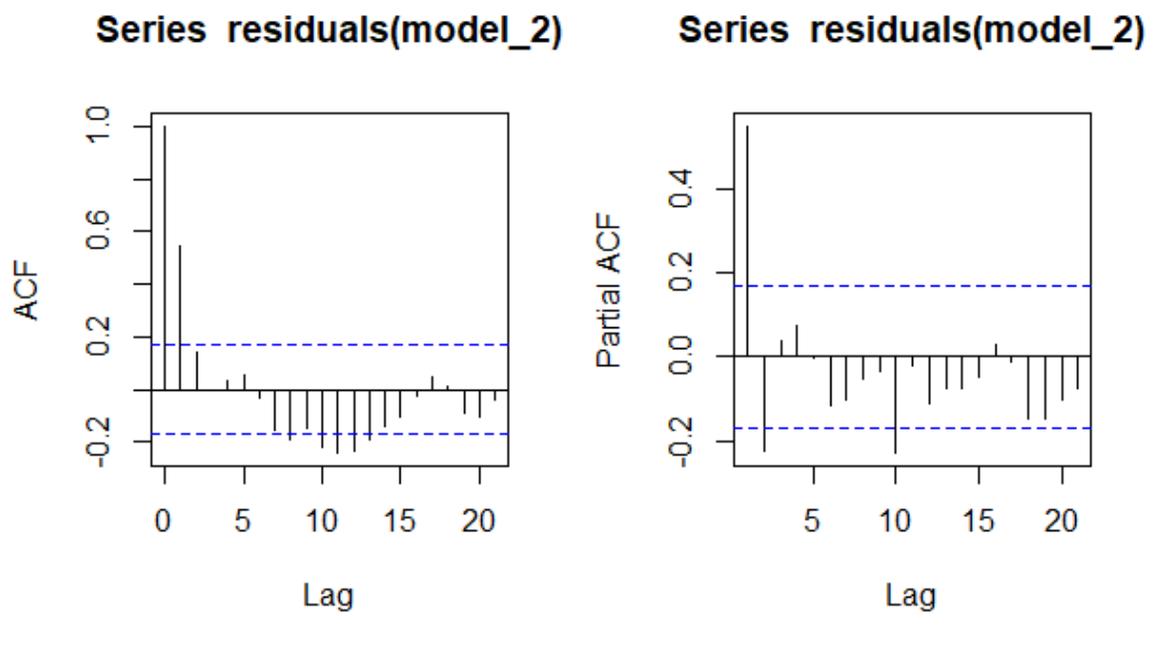

**Figure 18**

A: *Durbin-Watson and Breusch-Godfrey Residual Autocorrelation Tests on Model 2*

```
> dwtest(model_2)

        Durbin-Watson test

data:  model_2
DW = 0.88353, p-value = 1.782e-14
alternative hypothesis: true autocorrelation is greater than 0

> bgtest(model_2, order = 16)

        Breusch-Godfrey test for serial correlation of order up to 16

data:  model_2
LM test = 58.605, df = 16, p-value = 8.977e-07
```

B: *ACF and Partial-ACF for Residuals of Model 2*

*4.4.3.3. The Final Model*

Estimating **Models 1** and **2** by the GLS method with the AR and MA terms taken into consideration gives the results in **Table 7** below.

**Table 7**

A: *GLS Results for Model 1 With AR(13) Terms* (gdp = time + crisis + crisis.(time-24) + qe + qe.(time-37) + us + us.time + us.crisis + us.crisis.(time-24) + us.qe + us.qe.(time-37))

```
Correlation Structure: ARMA(13,0)
 Formula: ~time | us
 Parameter estimate(s):
      Phi1        Phi2        Phi3        Phi4        Phi5        Phi6        Phi7
 1.18064554 -0.35127517 -0.12189850 -0.04379184  0.17405623 -0.09050343 -0.16432124
      Phi8        Phi9       Phi10       Phi11       Phi12       Phi13
-0.05000054  0.30563944 -0.14943534 -0.22651040  0.24242074 -0.15252212

Coefficients:
                            Value Std.Error   t-value p-value
(Intercept)              39701.01  287.5699 138.05688  0.0000
time                       143.95   20.0237   7.18885  0.0000
crisis                    -230.47  201.4462  -1.14409  0.2548
crisis.time_minus_24      -182.16   45.1816  -4.03174  0.0001
qe                         237.10  206.9896   1.14545  0.2542
qe.time_minus_37           153.67   37.6754   4.07883  0.0001
us                        9182.44  406.6853  22.57875  0.0000
us.time                     85.85   28.3177   3.03150  0.0030
us.crisis                  204.84  284.8880   0.71902  0.4735
us.crisis.time_minus_24   -210.31   63.8964  -3.29149  0.0013
us.qe                        3.76  292.7275   0.01285  0.9898
us.qe.time_minus_37        231.15   53.2810   4.33833  0.0000
```

B: *GLS Results for Model 2 With AR(10) Terms* (cpi = time + crisis + crisis.(time-24) + qe + qe.(time-37) + us + us.time + us.crisis + us.crisis.(time-24) + us.qe + us.qe.(time-37))

```
Correlation Structure: ARMA(10,0)
 Formula: ~time | us
 Parameter estimate(s):
      Phi1        Phi2        Phi3        Phi4        Phi5        Phi6        Phi7
 0.83677832 -0.30576851  0.05597644 -0.12927626  0.17763357 -0.13364026 -0.09440921
      Phi8        Phi9       Phi10
-0.08600032  0.11274732 -0.21125886

Coefficients:
                           Value Std.Error   t-value p-value
(Intercept)             78.00186 0.2932711 265.97191  0.0000
time                     0.44628 0.0212370  21.01427  0.0000
crisis                   0.13205 0.3836870   0.34415  0.7313
crisis.time_minus_24    -0.11269 0.0520599  -2.16459  0.0323
qe                       0.51257 0.3885620   1.31915  0.1895
qe.time_minus_37         0.04110 0.0466333   0.88127  0.3799
us                      -4.16446 0.4147479 -10.04093  0.0000
us.time                  0.15734 0.0300336   5.23869  0.0000
us.crisis                0.17494 0.5426154   0.32239  0.7477
us.crisis.time_minus_24 -0.10542 0.0736238  -1.43193  0.1547
us.qe                    0.32383 0.5495096   0.58930  0.5567
us.qe.time_minus_37     -0.07078 0.0659495  -1.07331  0.2852
```

**Table 7** shows that, when autocorrelation is taken into account in the estimation of **Models 1** and **2**, the coefficients of interest are statistically significant only in **Model 1**. In other words, *while the crisis and QE implementation significantly affected U.S. real GDP over time, they did not impact U.S. inflation*. The results show that, for the 2008Q1-2018Q4 period, real GDP per capita decreased by an average of 210 US dollars per quarter in the U.S. relative to Canada due to the crisis (as seen in **Figure 19** ahead, the actual decrease was between 2008Q1 and 2010Q4, the decline between 2011Q1 and 2018Q4 was offset by an increase resulting from the QE implementation). And for the 2011Q2-2018Q4 period, real GDP per capita increased by an average of 231 US dollars per quarter in the U.S. relative to Canada due to the QE implementation. In other words, *real GDP per capita in the U.S. increased by an average of 231 dollars per quarter post-QE implementation relative to how it would have changed had the QE programs not been conducted*.

Therefore, the results contradict Williamson's (2017) informal natural experimental evidence and confirm the conclusions of VARs and new Keynesian DSGE models that the Federal Reserve's QE policies positively affected U.S. real GDP.

To see whether adding more AR and MA terms to **AR(13) Model 1** and **AR(10) Model 2** increases their goodness of fit, robustness checks are done on these models and presented in the next subsection.

*4.4.3.4. Robustness Checks on the Final Model*
Results of robustness checks on the GDP model with AR(13) terms and the CPI model with AR(10) terms are presented **Table 8** below.

**Table 8**

A: *Robustness Checks on AR(13) Model 1*

```
> anova(model_1_p13, model_1_p13q1)
             Model df      AIC      BIC   logLik   Test  L.Ratio p-value
model_1_p13      1 26 1885.957 1961.686 -916.9786
model_1_p13q1    2 27 1888.436 1967.078 -917.2182 1 vs 2 0.4792351  0.4888
> anova(model_1_p13, model_1_p14)
            Model df      AIC      BIC   logLik   Test  L.Ratio p-value
model_1_p13     1 26 1885.957 1961.686 -916.9786
model_1_p14     2 27 1887.236 1965.877 -916.6179 1 vs 2 0.7213872  0.3957
```

B: *Robustness Checks on AR(10) Model 2*

```
> anova(model_2_p10, model_2_p10q1)
              Model df      AIC      BIC   logLik   Test   L.Ratio p-value
model_2_p10       1 23 155.4927 222.4837 -54.74634
model_2_p10q1     2 24 157.4167 227.3204 -54.70835 1 vs 2 0.07597645  0.7828
> anova(model_2_p10, model_2_p11)
            Model df      AIC      BIC   logLik   Test   L.Ratio p-value
model_2_p10     1 23 155.4927 222.4837 -54.74634
model_2_p11     2 24 157.4701 227.3739 -54.73507 1 vs 2 0.02254645  0.8806
```

*Note*. model_1_p13, model_1_p13q1, and model_1_p14 are **AR(13) Model 1**, **ARMA(13,1) Model 1**, and **AR(14) Model 1**, respectively. model_2_p10, model_2_p10q1, and model_2_p11 are **AR(10) Model 2**, **ARMA(10,1) Model 2**, and **AR(11) Model 2**, respectively.

Panel A of **Table 8** shows that when **AR(13) Model 1** is compared with **ARMA(13,1) Model 1** and **AR(14) Model 1**, these alternative model specifications with more parameters are not statistically better than **AR(13) Model 1**. Thus, **AR(13) Model 1** is rightfully our final model for real GDP per capita. Also, Panel B shows that when **AR(10) Model 2** is compared with **ARMA(10,1) Model 2** and **AR(11) Model 2**, the alternative model specifications are not statistically better. Therefore, **AR(10) Model 2** is an accurate final model for CPI inflation.

*4.4.3.5. Graphical Representation of the Final Model*

The segmented regression plots of **AR(13) Model 1** and **AR(10) Model 2** are represented in **Figure 19** below.

**Figure 19**

A: *Segmented Regression Plot of AR(13) Model 1*

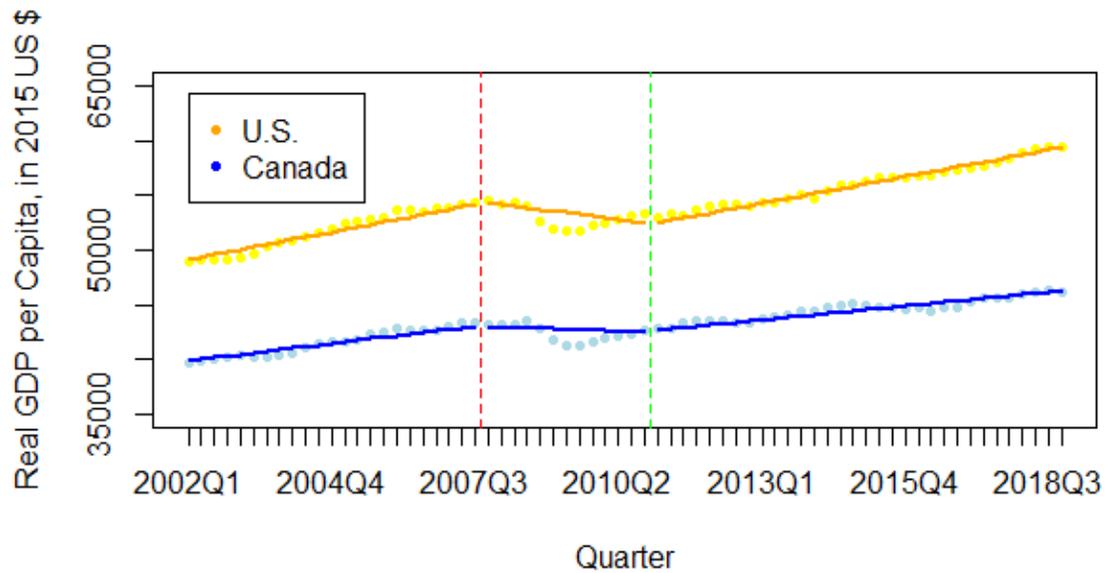

B: Segmented Regression Plot of AR(10) Model 2

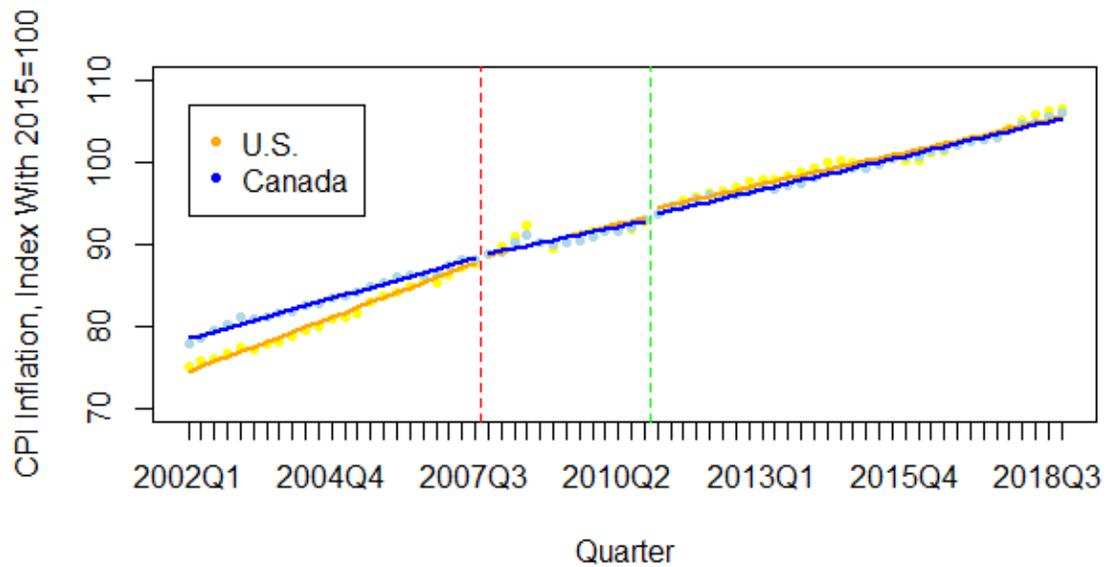

*Note*. All series are seasonally adjusted. The dashed red and green lines represent this study's specified dates for the Global Financial Crisis and QE implementation, respectively. Instead of 2008Q4, this study specifies the QE implementation as starting between 2010Q4 and 2011Q1 for both the model in which real GDP is the outcome variable and that in which inflation is the outcome variable since monetary policy affects inflation and real GDP with long lags (i.e., 2002Q1-2010Q4 is the pre-QE period and 2011Q1-2018Q4 is the post-QE period in both models). The figure is based on the author's calculations using data from the Organization for Economic Co-operation and Development (2022a, 2022b). The data are also accessible from Kamkoum (2023a). R codes for the figure are found in **Appendix 14** and can be freely downloaded from Kamkoum (2023a).

To visually compare the trend changes in real GDP per capita in the U.S. post-crisis and post-QE implementation to what the changes in real GDP per capita would have been had the crisis and QE not taken place, it is worth considering the segmented regression plot in **Figure 20** below (An alternative representation of the figure is presented in **Appendix 13**).

**Figure 20**

*Segmented Regression Plot of Observed and Counterfactual Trends in Real GDP per Capita in the U.S. and Canada*

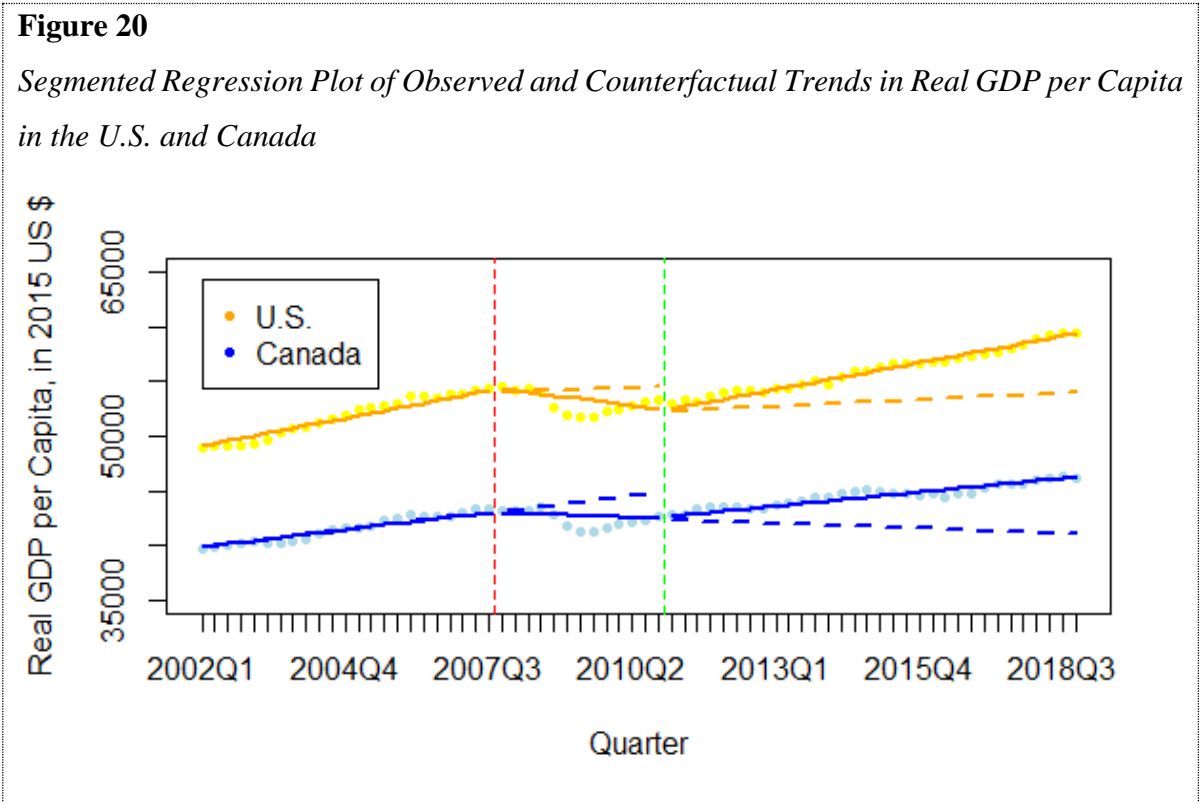

*Note*. While the solid yellow and blue lines show observed trends, the dashed yellow and blue lines represent counterfactual trends. Moreover, the dashed red and green lines represent this study's specified dates for the Global Financial Crisis and QE implementation, respectively. (An alternative representation of the figure is presented in **Appendix 13**). The figure is based on the author's calculations using data from the Organization for Economic Co-operation and Development (2022a, 2022b). The data are also accessible from Kamkoum (2023a). R codes for the figure are found in **Appendix 14** and can be freely downloaded from Kamkoum (2023a).

*4.4.3.6. Level Change Estimates at Time Points Other Than the First Time Point Post-Intervention*

As noted earlier, contrary to the level change at the first time point post-intervention (the short-term causal effect of the intervention), the level changes at time points that are several periods away from the point of intervention (the medium- and long-term causal effects of the intervention) are not represented directly by coefficients of an ITS model. They are estimated indirectly using the coefficients of the model.

Based on the results of **Table 7** and using **Equation 9** in **Section 4.3.2** in which $t = 68$ (i.e., in 2018Q4, which is 32 and 40 quarters after the *specified* and *actual* QE start dates, respectively) and $s = m+1 = 37$, it is estimated that *real GDP per capita in the U.S. was approximately 7169 dollars higher in 2018Q4 relative to what it would have been at that time had the QE programs not been implemented* ($\Delta \widehat{Level}_t = \hat{\beta}_{10} + \hat{\beta}_{11}(t - s) = 3.76 + 231.15\,(68 - 37) = 3.76 + 231.15\,(31) = 7169.41$).

Because the observed real GDP per capita in the U.S. was 59478 dollars in 2018Q4 (see **Table 5** in **Section 4.4.2**), the change of 7169 dollars represents an increase of approximately 14% in real GDP per capita in the U.S. in 2018Q4 relative to what it would have been during that quarter in the absence of the QE implementation ($59478 - Counterfactual = 7169 \Rightarrow Counterfactual = 52309 \Rightarrow \frac{Observed - Counterfactual}{Counterfactual} =$ $\frac{59478-Counterfactual}{Counterfactual} = \frac{7169}{52309} = 0.13705 \simeq 14\%$).[66] That is, had the U.S. not implemented its QE programs, its real GDP per capita would have been approximately 14% lower in 2018Q4. In conclusion, *in 2018Q4, ten years after the beginning of the Federal Reserve's QE programs, real GDP per capita in the U.S. was 14% higher relative to what it would have been during that quarter had there not been the QE programs.*

*4.4.4. Discussion*

The results obtained above contradict Williamson's (2017) informal natural experimental evidence and confirm the conclusions of VARs and new Keynesian DSGE models that the Federal Reserve's QE policies positively affected U.S. real GDP.

Williamson's findings are not accurate since graphical analysis alone is insufficient for causal inferences. Furthermore, his results are incorrect since he did not take into account the differential effects of the crisis on the U.S. and Canada in his analysis.

This paper's results indicate that QE is a monetary policy tool central banks should consider when the policy rate is too low. Also, the results suggest that the current U.S. and worldwide high inflation rates are likely not because of the QE programs implemented in response to the financial crisis that accompanied the COVID-19 pandemic. They are likely due to the unprecedentedly large fiscal stimulus packages used, the peculiar nature of the financial downturn itself, the negative supply shocks from the war in Ukraine, or a combination of these factors. To the best of my knowledge, this paper is the first study to measure the macroeconomic effects of QE using a design-based natural experimental approach.

---

[66] The increase of 7169 dollars or 14% in real GDP per capita is a cumulative change for the entire 2011Q1-2018Q4 post-QE period. In other words, it is a cumulative change for the 32 quarters in the 2011Q1-2018Q4 period. It is essential to remember that while the *trend change of 231 dollars* obtained in **Section 4.4.3.3** is for the *2011Q2-2018Q4 post-QE period*, the *level change of 7169 dollars or 14%* calculated here in **Section 4.4.3.6** is for the *2011Q1-2018Q4 post-QE period*. That is, while the trend change is for *31 quarters*, the level change is for *32 quarters*. The trend change is for only 31 quarters because, as explained in **Sections 4.3.2** and **4.4.1.2** and shown in **Table 5**, there is no trend change in the first quarter post-QE (in 2011Q1).

# 5. CONCLUSION

In response to the Global Financial Crisis, the Federal Reserve implemented various quantitative easing (QE) programs, created various liquidity facilities, and engaged in many forward guidance operations. These unconventional monetary policies, and specifically the QE programs, led to a significant increase in the U.S. monetary base and money supply. Theory suggests that the increase in the monetary base and money supply should have helped in the fight against the crisis. Empirical studies on the subject that use VARs and new Keynesian DSGE models corroborate the theory. However, using an informal natural experimental approach, Williamson (2017) finds no evidence to support the hypothesis that the QE programs increased U.S. real GDP.

This paper uses the formal design-based natural experimental method of interrupted time-series (ITS) analysis to examine the long-term causal effects of the QE programs on U.S. inflation and real GDP. The results show that the QE operations positively affected U.S. real GDP but did not significantly impact U.S. inflation. Specifically, it is found that, for the 2011Q2-2018Q4 post-QE period, real GDP per capita in the U.S. increased by an average of 231 dollars per quarter relative to how it would have changed had the QE programs not been conducted. Moreover, the results show that, in 2018Q4, ten years after the beginning of the QE programs, real GDP per capita in the U.S. was 14% higher relative to what it would have been during that quarter had there not been the QE programs.

Therefore, this paper's findings contradict Williamson's (2017) informal natural experimental evidence and confirm the conclusions of VARs and new Keynesian DSGE models that the Federal Reserve's QE policies positively affected U.S. real GDP.

These results indicate that QE is a monetary policy tool central banks should consider when the policy rate is too low. Also, the results suggest that the current U.S. and worldwide high inflation rates are likely not because of the QE programs implemented in response to the financial crisis that accompanied the COVID-19 pandemic. They are likely due to the unprecedentedly large fiscal stimulus packages used, the peculiar nature of the financial downturn itself, the negative supply shocks from the war in Ukraine, or a combination of these factors. To the best of my knowledge, this paper is the first study to measure the macroeconomic effects of QE using a design-based natural experimental approach.

# APPENDICES

**Appendix 1: A Mathematical Proof That the *β₂* Coefficient of *Equation 1* Is Equal to the *Difference* Between the Observed and Counterfactual Levels at *t* = *k*+1**

For $Outcome_{it}$ of **Equation 1**, the observed and counterfactual levels at $t = k+1$ are given by $Outcome_{i=1, t=k+1}$ and $Outcome_{i=0, t=k+1}$, respectively. We calculate that

$$\begin{aligned} Outcome_{i=1,t=k+1} &= \beta_0 + \beta_1 * (k+1) + \beta_2 * (1) + \beta_3 * (1) * [(k+1) - r] \\ &= \beta_0 + \beta_1 * (k+1) + \beta_2 + \beta_3 * [(k+1) - (k+1)] \\ &= \beta_0 + \beta_1 * (k+1) + \beta_2, \end{aligned}$$

and

$$\begin{aligned} Outcome_{i=0,t=k+1} &= \beta_0 + \beta_1 * (k+1) + \beta_2 * (0) + \beta_3 * (0) * [(k+1) - r] \\ &= \beta_0 + \beta_1 * (k+1). \end{aligned}$$

Therefore,

$$\begin{aligned} Level\ Difference &= Outcome_{i=1,t=k+1} - Outcome_{i=0,t=k+1} \\ &= [\beta_0 + \beta_1 * (k+1) + \beta_2] - [\beta_0 + \beta_1 * (k+1)] \\ &= \beta_2 \end{aligned}$$

Q.E.D.

**Appendix 2: A Mathematical Proof That the *β₃* Coefficient of *Equation 1* Is Equal to the *Difference* Between the Observed and Counterfactual Trends Post-Intervention**

Given **Equation 1**, for any two distinct time points, $t_1$ and $t_2$, between $t = k+2$ and $t = n$ (with $k+2$ and $n$ included; as mentioned earlier, the post-intervention trend is *measurable* from $t = k+2$, not $t = k+1$), the observed and counterfactual trends post-intervention are represented by $\frac{Outcome_{i=1,t=t_2} - Outcome_{i=1,t=t_1}}{t_2 - t_1}$ and $\frac{Outcome_{i=0,t=t_2} - Outcome_{i=0,t=t_1}}{t_2 - t_1}$, respectively. We calculate that

$$Outcome_{i=1,t=t_2} - Outcome_{i=1,t=t_1}$$
$$= [\beta_0 + \beta_1 * (t_2) + \beta_2 + \beta_3 * (t_2 - r)] - [\beta_0 + \beta_1 * (t_1) + \beta_2 + \beta_3 * (t_1 - r)]$$
$$= \beta_1 * (t_2 - t_1) + \beta_3 * (t_2 - t_1),$$

such that

$$\frac{Outcome_{i=1,t=t_2} - Outcome_{i=1,t=t_1}}{t_2 - t_1} = \beta_1 + \beta_3.$$

And

$$Outcome_{i=0,t=t_2} - Outcome_{i=0,t=t_1} = [\beta_0 + \beta_1 * (t_2)] - [\beta_0 + \beta_1 * (t_1)]$$
$$= \beta_1 * (t_2 - t_1),$$

such that

$$\frac{Outcome_{i=0,t=t_2} - Outcome_{i=0,t=t_1}}{t_2 - t_1} = \beta_1.$$

Therefore,

$$\text{Trend Difference} = \left[\frac{Outcome_{i=1,t=t_2} - Outcome_{i=1,t=t_1}}{t_2-t_1}\right] - \left[\frac{Outcome_{i=0,t=t_2} - Outcome_{i=0,t=t_1}}{t_2-t_1}\right]$$

$$= (\beta_1 + \beta_3) - \beta_1$$

$$= \beta_3$$

Q.E.D.

**Appendix 3: A Mathematical Proof That the $\beta_6$ Coefficient of *Equation 2* Is Equal to the *Difference-in-Differences* Between the Observed and Counterfactual Levels of the Outcome for the Treatment and Control Group Time Series at $t = k+1$**

For $Outcome_{igt}$ of **Equation 2**, the <u>treatment group</u> observed and counterfactual levels at $t = k+1$ are given by $Outcome_{i=1,g=1,t=k+1}$ and $Outcome_{i=0,g=1,t=k+1}$, respectively. And the <u>control group</u> observed and counterfactual levels at $t = k+1$ are given by $Outcome_{i=1,g=0,t=k+1}$ and $Outcome_{i=0,g=0,t=k+1}$, respectively. We calculate that

$$\begin{aligned}Outcome_{i=1,g=1,t=k+1} &= \beta_0 + \beta_1 * (k+1) + \beta_2 + \beta_3 * [(k+1) - r] + \beta_4 \\&\quad + \beta_5 * (k+1) + \beta_6 + \beta_7 * [(k+1) - r] \\&= \beta_0 + \beta_1 * (k+1) + \beta_2 + \beta_3 * [(k+1) - (k+1)] + \beta_4 \\&\quad + \beta_5 * (k+1) + \beta_6 + \beta_7 * [(k+1) - (k+1)] \\&= \beta_0 + \beta_1 * (k+1) + \beta_2 + \beta_4 + \beta_5 * (k+1) + \beta_6,\end{aligned}$$

and

$$Outcome_{i=0,g=1,t=k+1} = \beta_0 + \beta_1 * (k+1) + \beta_4 + \beta_5 * (k+1).$$

Therefore,

$$\begin{aligned}
&Treatment\ Group\ Level\ Difference\\
&\quad = Outcome_{i=1,g=1,t=k+1} - Outcome_{i=0,g=1,t=k+1}\\
&\quad = [\beta_0 + \beta_1 * (k+1) + \beta_2 + \beta_4 + \beta_5 * (k+1) + \beta_6]\\
&\qquad - [\beta_0 + \beta_1 * (k+1) + \beta_4 + \beta_5 * (k+1)]\\
&\quad = \beta_2 + \beta_6.
\end{aligned}$$

Also,

$$Outcome_{i=1,g=0,t=k+1} = \beta_0 + \beta_1 * (k+1) + \beta_2,$$

and

$$Outcome_{i=0,g=0,t=k+1} = \beta_0 + \beta_1 * (k+1).$$

Therefore,

$$\begin{aligned}
&Control\ Group\ Level\ Difference\\
&\quad = Outcome_{i=1,g=0,t=k+1} - Outcome_{i=0,g=0,t=k+1}\\
&\quad = [\beta_0 + \beta_1 * (k+1) + \beta_2] - [\beta_0 + \beta_1 * (k+1)]\\
&\quad = \beta_2.
\end{aligned}$$

And thus,

$$\begin{aligned}
&Level\ Difference-in-Differences\\
&\quad = Treatment\ Group\ Level\ Difference\\
&\qquad - Control\ Group\ Level\ Difference\\
&\quad = (\beta_2 + \beta_6) - \beta_2\\
&\quad = \beta_6
\end{aligned}$$

Q.E.D.

**Appendix 4: A Mathematical Proof That the *β₇* Coefficient of *Equation 2* Is Equal to the *Difference-in-Differences* Between the Observed and Counterfactual Trends of the Outcome for the Treatment and Control Group Time Series Post-Intervention**

Given **Equation 2**, for any two distinct time points, $t_1$ and $t_2$, between $t = k+2$ and $t = n$ (with $k+2$ and $n$ included), the <u>treatment group</u> observed and counterfactual trends post-intervention are represented by $\frac{Outcome_{i=1,g=1,t=t_2} - Outcome_{i=1,g=1,t=t_1}}{t_2-t_1}$ and $\frac{Outcome_{i=0,g=1,t=t_2} - Outcome_{i=0,g=1,t=t_1}}{t_2-t_1}$, respectively. And the <u>control group</u> observed and counterfactual trends post-intervention are represented by $\frac{Outcome_{i=1,g=0,t=t_2} - Outcome_{i=1,g=0,t=t_1}}{t_2-t_1}$ and $\frac{Outcome_{i=0,g=0,t=t_2} - Outcome_{i=0,g=0,t=t_1}}{t_2-t_1}$, respectively. We calculate that

$$\frac{Outcome_{i=1,g=1,t=t_2} - Outcome_{i=1,g=1,t=t_1}}{t_2 - t_1} =$$

$$\frac{[\beta_0+\beta_1*(t_2)+\beta_2+\beta_3*(t_2-r)+\beta_4+\beta_5*(t_2)+\beta_6+\beta_7*(t_2-r)] - [\beta_0+\beta_1*(t_1)+\beta_2+\beta_3*(t_1-r)+\beta_4+\beta_5*(t_1)+\beta_6+\beta_7*(t_1-r)]}{t_2-t_1}$$

$$= \frac{\beta_1*(t_2-t_1) + \beta_3*(t_2-t_1) + \beta_5*(t_2-t_1) + \beta_7*(t_2-t_1)}{t_2-t_1}$$

$$= \boldsymbol{\beta_1 + \beta_3 + \beta_5 + \beta_7} \qquad (*),$$

$$\frac{Outcome_{i=0,g=1,t=t_2} - Outcome_{i=0,g=1,t=t_1}}{t_2 - t_1}$$

$$= \frac{[\beta_0 + \beta_1*(t_2) + \beta_4 + \beta_5*(t_2)] - [\beta_0 + \beta_1*(t_1) + \beta_4 + \beta_5*(t_1)]}{t_2-t_1}$$

$$= \frac{\beta_1 * (t_2 - t_1) + \beta_5 * (t_2 - t_1)}{t_2 - t_1}$$

$$= \boldsymbol{\beta_1 + \beta_5} \qquad (**),$$

$$\frac{Outcome_{i=1,g=0,t=t_2} - Outcome_{i=1,g=0,t=t_1}}{t_2 - t_1}$$

$$= \frac{[\beta_0 + \beta_1 * (t_2) + \beta_2 + \beta_3 * (t_2 - r)] - [\beta_0 + \beta_1 * (t_1) + \beta_2 + \beta_3 * (t_1 - r)]}{t_2 - t_1}$$

$$= \frac{\beta_1 * (t_2 - t_1) + \beta_3 * (t_2 - t_1)}{t_2 - t_1}$$

$$= \boldsymbol{\beta_1 + \beta_3} \qquad (***),$$

and

$$\frac{Outcome_{i=0,g=0,t=t_2} - Outcome_{i=0,g=0,t=t_1}}{t_2 - t_1}$$

$$= \frac{[\beta_0 + \beta_1 * (t_2)] - [\beta_0 + \beta_1 * (t_1)]}{t_2 - t_1}$$

$$= \frac{\beta_1 * (t_2 - t_1)}{t_2 - t_1}$$

$$= \boldsymbol{\beta_1} \qquad (****).$$

Therefore,

$$Treatment\ Group\ Trend\ Difference = (*) - (**) = \beta_3 + \beta_7,$$

and

$$Control\ Group\ Trend\ Difference = (***) - (****) = \beta_3.$$

Thus,

$$\begin{aligned}Trend\ &Difference-in-Differences\\ &= Treatment\ Group\ Trend\ Difference\\ &\quad - Control\ Group\ Trend\ Difference\\ &= [(*) - (**)] - [(***) - (****)]\\ &= (\beta_3 + \beta_7) - \beta_3\\ &= \beta_7\end{aligned}$$

Q.E.D.

## Appendix 5: A Mathematical Proof That the $\beta_4$ Coefficient of *Equation 3* Is Equal to the *Difference* Between the Observed and Counterfactual Levels at $t = m+1$

For $Outcome_{ijt}$ of **Equation 3**, the observed and counterfactual levels at $t = m+1$ are given by $Outcome_{i=1,j=1,t=m+1}$ and $Outcome_{i=1,j=0,t=m+1}$, respectively. We calculate that

$$\begin{aligned}Outcome_{i=1,j=1,t=m+1} &= \beta_0 + \beta_1 * (m+1) + \beta_2 + \beta_3 * [(m+1) - r] + \beta_4\\ &\quad + \beta_5 * [(m+1) - s]\\ &= \beta_0 + \beta_1 * (m+1) + \beta_2 + \beta_3 * [(m+1) - (k+1)]\end{aligned}$$

$$+ \beta_4 + \beta_5 * [(m+1) - (m+1)]$$
$$= \beta_0 + \beta_1 * (m+1) + \beta_2 + \beta_3 * (m-k) + \beta_4,$$

and

$$Outcome_{i=1,j=0,t=m+1} = \beta_0 + \beta_1 * (m+1) + \beta_2 + \beta_3 * [(m+1) - r]$$
$$= \beta_0 + \beta_1 * (m+1) + \beta_2 + \beta_3 * [(m+1) - (k+1)]$$
$$= \beta_0 + \beta_1 * (m+1) + \beta_2 + \beta_3 * (m-k).$$

Therefore,

$$Level\ Difference = Outcome_{i=1,j=1,t=m+1} - Outcome_{i=1,j=0,t=m+1}$$
$$= [\beta_0 + \beta_1 * (m+1) + \beta_2 + \beta_3 * (m-k) + \beta_4]$$
$$- [\beta_0 + \beta_1 * (m+1) + \beta_2 + \beta_3 * (m-k)]$$
$$= \beta_4$$

Q.E.D.

**Appendix 6: A Mathematical Proof That the *β₅* Coefficient of *Equation 3* Is Equal to the *Difference* Between the Observed and Counterfactual Trends Post-Second-Intervention**

Given **Equation 3**, for any two distinct time points, $t_1$ and $t_2$, between $t = m+2$ and $t = n$ (with $m+2$ and $n$ included), the observed and counterfactual trends post-second-intervention (after the beginning of the second intervention) are represented by $\frac{Outcome_{i=1,j=1,t=t_2} - Outcome_{i=1,j=1,t=t_1}}{t_2 - t_1}$ and $\frac{Outcome_{i=1,j=0,t=t_2} - Outcome_{i=1,j=0,t=t_1}}{t_2 - t_1}$, respectively. We calculate that

$$Outcome_{i=1,j=1,t=t_2} - Outcome_{i=1,j=1,t=t_1}$$
$$= [\beta_0 + \beta_1 * (t_2) + \beta_2 + \beta_3 * (t_2 - r) + \beta_4 + \beta_5 * (t_2 - s)]$$
$$- [\beta_0 + \beta_1 * (t_1) + \beta_2 + \beta_3 * (t_1 - r) + \beta_4 + \beta_5 * (t_1 - s)]$$
$$= \beta_1 * (t_2 - t_1) + \beta_3 * (t_2 - t_1) + \beta_5 * (t_2 - t_1),$$

such that

$$\frac{Outcome_{i=1,j=1,t=t_2} - Outcome_{i=1,j=1,t=t_1}}{t_2 - t_1} = \beta_1 + \beta_3 + \beta_5.$$

And

$$Outcome_{i=1,j=0,t=t_2} - Outcome_{i=1,j=0,t=t_1}$$
$$= [\beta_0 + \beta_1 * (t_2) + \beta_2 + \beta_3 * (t_2 - r)]$$
$$- [\beta_0 + \beta_1 * (t_1) + \beta_2 + \beta_3 * (t_1 - r)]$$
$$= \beta_1 * (t_2 - t_1) + \beta_3 * (t_2 - t_1),$$

such that

$$\frac{Outcome_{i=1,j=0,t=t_2} - Outcome_{i=1,j=0,t=t_1}}{t_2 - t_1} = \beta_1 + \beta_3.$$

Therefore,

$$Trend\ Difference\ Post-Second-Intervention =$$
$$\left[\frac{Outcome_{i=1,j=1,t=t_2} - Outcome_{i=1,j=1,t=t_1}}{t_2 - t_1}\right] - \left[\frac{Outcome_{i=1,j=0,t=t_2} - Outcome_{i=1,j=0,t=t_1}}{t_2 - t_1}\right]$$
$$= (\beta_1 + \beta_3 + \beta_5) - (\beta_1 + \beta_3)$$
$$= \beta_5$$

**Appendix 7: A Mathematical Proof That the *β₁₀* Coefficient of *Equation 4* Is Equal to the *Difference-in-Differences* Between the Observed and Counterfactual Levels of the Outcome for the Treatment and Control Group Time Series at *t = m*+1**

For $Outcome_{ijgt}$ of **Equation 4**, the <u>treatment group</u> observed and counterfactual levels at $t = m+1$ are given by $Outcome_{i=1,j=1,g=1,t=m+1}$ and $Outcome_{i=1,j=0,g=1,t=m+1}$, respectively. And the <u>control group</u> observed and counterfactual levels at $t = m+1$ are given by $Outcome_{i=1,j=1,g=0,t=m+1}$ and $Outcome_{i=1,j=0,g=0,t=m+1}$, respectively. We calculate that

$Outcome_{i=1,j=1,g=1,t=m+1}$
$= \beta_0 + \beta_1 * (m+1) + \beta_2 + \beta_3 * [(m+1) - r] + \beta_4 + \beta_5 * [(m+1) - s] + \beta_6$
$\quad + \beta_7 * (m+1) + \beta_8 + \beta_9 * [(m+1) - r] + \beta_{10} + \beta_{11} * [(m+1) - s]$
$= \beta_0 + \beta_1 * (m+1) + \beta_2 + \beta_3 * [(m+1) - (k+1)] + \beta_4$
$\quad + \beta_5 * [(m+1) - (m+1)] + \beta_6 + \beta_7 * (m+1) + \beta_8$
$\quad + \beta_9 * [(m+1) - (k+1)] + \beta_{10} + \beta_{11} * [(m+1) - (m+1)]$
$= \boldsymbol{\beta_0 + \beta_1 * (m+1) + \beta_2 + \beta_3 * (m-k) + \beta_4 + \beta_6 + \beta_7 * (m+1) + \beta_8}$
$\quad \boldsymbol{+ \beta_9 * (m-k) + \beta_{10}}$ \hfill (*),

$Outcome_{i=1,j=0,g=1,t=m+1}$
$= \boldsymbol{\beta_0 + \beta_1 * (m+1) + \beta_2 + \beta_3 * (m-k) + \beta_6 + \beta_7 * (m+1) + \beta_8}$
$\quad \boldsymbol{+ \beta_9 * (m-k)}$ \hfill (**),

$Outcome_{i=1,j=1,g=0,t=m+1}$
$= \boldsymbol{\beta_0 + \beta_1 * (m+1) + \beta_2 + \beta_3 * (m-k) + \beta_4}$ \hfill (***),

and

$$Outcome_{i=1,j=0,g=0,t=m+1}$$
$$= \boldsymbol{\beta_0} + \boldsymbol{\beta_1} * (\boldsymbol{m}+\boldsymbol{1}) + \boldsymbol{\beta_2} + \boldsymbol{\beta_3} * (\boldsymbol{m}-\boldsymbol{k}) \qquad (****).$$

Therefore,

$$Treatment\ Group\ Level\ Difference$$
$$= Outcome_{i=1,j=1,g=1,t=m+1} - Outcome_{i=1,j=0,g=1,t=m+1}$$
$$= (*) - (**)$$
$$= \beta_4 + \beta_{10},$$

and

$$Control\ Group\ Level\ Difference$$
$$= Outcome_{i=1,j=1,g=0,t=m+1} - Outcome_{i=1,j=0,g=0,t=m+1}$$
$$= (***) - (****)$$
$$= \beta_4.$$

Thus,

$$Level\ Difference-in-Differences$$
$$= Treatment\ Group\ Level\ Difference$$
$$\quad - Control\ Group\ Level\ Difference$$
$$= [(*) - (**)] - [(***) - (****)]$$
$$= \beta_{10}$$

Q.E.D.

**Appendix 8: A Mathematical Proof That the *β₁₁* Coefficient of *Equation 4* Is Equal to the *Difference-in-Differences* Between the Observed and Counterfactual Trends of the Outcome for the Treatment and Control Group Time Series Post-Second-Intervention**

Given **Equation 4**, for any two distinct time points, $t_1$ and $t_2$, between $t = m+2$ and $t = n$ (with $m+2$ and $n$ included), the <u>treatment group</u> observed and counterfactual trends post-second-intervention (after the beginning of the second intervention) are represented by $\frac{Outcome_{i=1,j=1,g=1,t=t_2} - Outcome_{i=1,j=1,g=1,t=t_1}}{t_2-t_1}$ and $\frac{Outcome_{i=1,j=0,g=1,t=t_2} - Outcome_{i=1,j=0,g=1,t=t_1}}{t_2-t_1}$, respectively. And the <u>control group</u> observed and counterfactual trends post-second-intervention are represented by $\frac{Outcome_{i=1,j=1,g=0,t=t_2} - Outcome_{i=1,j=1,g=0,t=t_1}}{t_2-t_1}$ and $\frac{Outcome_{i=1,j=0,g=0,t=t_2} - Outcome_{i=1,j=0,g=0,t=t_1}}{t_2-t_1}$, respectively. We calculate that

$$\begin{aligned}
Outcome_{i=1,j=1,g=1,t=t_2} &- Outcome_{i=1,j=1,g=1,t=t_1} \\
= [\beta_0 + \beta_1 &* (t_2) + \beta_2 + \beta_3 * (t_2 - r) + \beta_4 + \beta_5 * (t_2 - s) + \beta_6 + \beta_7 * (t_2) \\
&+ \beta_8 + \beta_9 * (t_2 - r) + \beta_{10} + \beta_{11} * (t_2 - s)] \\
- [\beta_0 + \beta_1 &* (t_1) + \beta_2 + \beta_3 * (t_1 - r) + \beta_4 + \beta_5 * (t_1 - s) + \beta_6 + \beta_7 * (t_1) \\
&+ \beta_8 + \beta_9 * (t_1 - r) + \beta_{10} + \beta_{11} * (t_1 - s)] \\
= \beta_1 * (t_2 - t_1) &+ \beta_3 * (t_2 - t_1) + \beta_5 * (t_2 - t_1) + \beta_7 * (t_2 - t_1) \\
&+ \beta_9 * (t_2 - t_1) + \beta_{11} * (t_2 - t_1),
\end{aligned}$$

such that

$$\frac{Outcome_{i=1,j=1,g=1,t=t_2} - Outcome_{i=1,j=1,g=1,t=t_1}}{t_2 - t_1} = \boldsymbol{\beta_1 + \beta_3 + \beta_5 + \beta_7 + \beta_9 + \beta_{11}}$$

(*);

$$Outcome_{i=1,j=0,g=1,t=t_2} - Outcome_{i=1,j=0,g=1,t=t_1}$$
$$= \beta_1 * (t_2 - t_1) + \beta_3 * (t_2 - t_1) + \beta_7 * (t_2 - t_1) + \beta_9 * (t_2 - t_1),$$

such that

$$\frac{Outcome_{i=1,j=0,g=1,t=t_2} - Outcome_{i=1,j=0,g=1,t=t_1}}{t_2 - t_1} = \boldsymbol{\beta_1 + \beta_3 + \beta_7 + \beta_9}$$

(**);

$$Outcome_{i=1,j=1,g=0,t=t_2} - Outcome_{i=1,j=1,g=0,t=t_1}$$
$$= \beta_1 * (t_2 - t_1) + \beta_3 * (t_2 - t_1) + \beta_5 * (t_2 - t_1),$$

such that

$$\frac{Outcome_{i=1,j=1,g=0,t=t_2} - Outcome_{i=1,j=1,g=0,t=t_1}}{t_2 - t_1} = \boldsymbol{\beta_1 + \beta_3 + \beta_5}$$

(***);

and

$$Outcome_{i=1,j=0,g=0,t=t_2} - Outcome_{i=1,j=0,g=0,t=t_1}$$
$$= \beta_1 * (t_2 - t_1) + \beta_3 * (t_2 - t_1),$$

such that

$$\frac{Outcome_{i=1,j=0,g=0,t=t_2} - Outcome_{i=1,j=0,g=0,t=t_1}}{t_2 - t_1} = \boldsymbol{\beta_1 + \beta_3}$$

(****).

Therefore,

$$Treatment\ Group\ Trend\ Difference\ Post-Second-Intervention$$
$$= (*) - (**)$$
$$= \beta_5 + \beta_{11},$$

and

$$Control\ Group\ Trend\ Difference\ Post-Second-Intervention$$
$$= (***) - (****)$$
$$= \beta_5.$$

Thus,

$$Trend\ Difference-in-Differences\ Post-Second-Intervention$$
$$= (Treatment\ Group\ Trend\ Difference\ Post-Second-Intervention)$$
$$\quad - (Control\ Group\ Trend\ Difference\ Post-Second-Intervention)$$
$$= [(*) - (**)] - [(***) - (****)]$$
$$= \beta_{11}$$

Q.E.D.

**Appendix 9: A Mathematical Proof for *Equation 5***

The proof in **Appendix 1** shows that the $\beta_2$ coefficient of **Equation 1** is the level change at $t = k+1$ (the *first time point* post-intervention). The following proof generalizes the proof of **Appendix 1** to include level changes at all $t$ where $k+1 \leq t \leq n$ (*all time points* post-intervention).

For $Outcome_{it}$ of **Equation 1**, the observed and counterfactual levels at any time point, $t = t$, post-intervention are given by $Outcome_{i=1,t=t}$ and $Outcome_{i=0,t=t}$, respectively. We calculate that

$$Outcome_{i=1,t=t} = \beta_0 + \beta_1 * (t) + \beta_2 * (1) + \beta_3 * (1) * (t - r)$$
$$= \beta_0 + \beta_1 * t + \beta_2 + \beta_3 * (t - r),$$

and

$$Outcome_{i=0,t=t} = \beta_0 + \beta_1 * (t) + \beta_2 * (0) + \beta_3 * (0) * (t - r)$$
$$= \beta_0 + \beta_1 * t.$$

Therefore,

$$\Delta\, Level_t = (Level\ Difference)_t$$
$$= Outcome_{i=1,t=t} - Outcome_{i=0,t=t}$$
$$= [\beta_0 + \beta_1 * t + \beta_2 + \beta_3 * (t - r)] - [\beta_0 + \beta_1 * t]$$
$$= \beta_2 + \beta_3 * (t - r)$$

Q.E.D.

And thus, when $t = k+1 = r$, we get the proof in **Appendix 1**.

**Appendix 10: A Mathematical Proof for *Equation 6***

The proof in **Appendix 3** shows that the $β_6$ coefficient of **Equation 2** is the level change at $t = k+1$ (the *first time point* post-intervention). The following proof generalizes the proof of **Appendix 3** to include level changes at all $t$ where $k+1 \leq t \leq n$ (*all time points* post-intervention).

For $Outcome_{igt}$ of **Equation 2**, the treatment group observed and counterfactual levels at any time point, $t = t$, post-intervention are given by $Outcome_{i=1,g=1,t=t}$ and $Outcome_{i=0,g=1,t=t}$, respectively. And the control group observed and counterfactual levels at any time point, $t = t$, post-intervention are given by $Outcome_{i=1,g=0,t=t}$ and $Outcome_{i=0,g=0,t=t}$, respectively. We calculate that

$$Outcome_{i=1,g=1,t=t} = β_0 + β_1 * (t) + β_2 + β_3 * (t - r) + β_4$$
$$+ β_5 * (t) + β_6 + β_7 * (t - r)$$
$$= β_0 + β_1 * t + β_2 + β_3 * (t - r) + β_4$$
$$+ β_5 * t + β_6 + β_7 * (t - r)$$

and

$$Outcome_{i=0,g=1,t=t} = β_0 + β_1 * t + β_4 + β_5 * t.$$

Therefore,

$$Δ\,(Treatment\ Group\ Level)_t$$
$$= (Treatment\ Group\ Level\ Difference)_t$$
$$= Outcome_{i=1,g=1,t=t} - Outcome_{i=0,g=1,t=t}$$

$$= [\beta_0 + \beta_1 * t + \beta_2 + \beta_3 * (t - r) + \beta_4 + \beta_5 * t + \beta_6 + \beta_7 * (t - r)]$$
$$- [\beta_0 + \beta_1 * t + \beta_4 + \beta_5 * t]$$
$$= \beta_2 + \beta_3 * (t - r) + \beta_6 + \beta_7 * (t - r).$$

Also,

$$Outcome_{i=1,g=0,t=t} = \beta_0 + \beta_1 * t + \beta_2 + \beta_3 * (t - r),$$

and

$$Outcome_{i=0,g=0,t=t} = \beta_0 + \beta_1 * t.$$

Therefore,

$$\Delta\,(Control\ Group\ Level)_t$$
$$= (Control\ Group\ Level\ Difference)_t$$
$$= Outcome_{i=1,g=0,t=t} - Outcome_{i=0,g=0,t=t}$$
$$= [\beta_0 + \beta_1 * t + \beta_2 + \beta_3 * (t - r)] - (\beta_0 + \beta_1 * t)$$
$$= \beta_2 + \beta_3 * (t - r).$$

And thus,

$$\Delta\,Level_t = (Level\ Difference - in - Differences)_t$$
$$= (Treatment\ Group\ Level\ Difference)_t$$
$$\quad - (Control\ Group\ Level\ Difference)_t$$
$$= [\beta_2 + \beta_3 * (t - r) + \beta_6 + \beta_7 * (t - r)] - [\beta_2 + \beta_3 * (t - r)]$$
$$= \beta_6 + \beta_7 * (t - r)$$

Q.E.D.

And thus, when $t = k+1 = r$, we get the proof in **Appendix 3**.

**Appendix 11: A Mathematical Proof for *Equation 7***

The proof in **Appendix 5** shows that the *β₄* coefficient of **Equation 3** is the level change at $t = m+1$ (the *first time point* after the beginning of the second intervention). The following proof generalizes the proof of **Appendix 5** to include level changes at all *t* where $m+1 \leq t \leq n$ (*all time points* after the beginning of the second intervention).

For $Outcome_{ijt}$ of **Equation 3**, the observed and counterfactual levels at any time point, $t = t$, after the beginning of the second intervention are given by $Outcome_{i=1,j=1,t=t}$ and $Outcome_{i=1,j=0,t=t}$, respectively. We calculate that

$$Outcome_{i=1,j=1,t=t} = \beta_0 + \beta_1 * t + \beta_2 + \beta_3 * (t - r) + \beta_4 + \beta_5 * (t - s)$$

and

$$Outcome_{i=1,j=0,t=t} = \beta_0 + \beta_1 * t + \beta_2 + \beta_3 * (t - r).$$

Therefore,

$$\begin{aligned}\Delta\, Level_t &= (Level\ Difference)_t \\ &= Outcome_{i=1,j=1,t=t} - Outcome_{i=1,j=0,t=t} \\ &= \beta_4 + \beta_5 * (t - s)\end{aligned}$$

Q.E.D.

And thus, when $t = m+1 = s$, we get the proof in **Appendix 5**.

**Appendix 12: A Mathematical Proof for *Equation 9***

The proof in **Appendix 7** shows that the $\beta_{10}$ coefficient of **Equation 4** is the level change at $t = m+1$ (the *first time point* after the beginning of the second intervention). The following proof generalizes the proof of **Appendix 7** to include level changes at all $t$ where $m+1 \leq t \leq n$ (*all time points* after the beginning of the second intervention).

For $Outcome_{ijgt}$ of **Equation 4**, the <u>treatment group</u> observed and counterfactual levels at any time point, $t = t$, after the beginning of the second intervention are given by $Outcome_{i=1,j=1,g=1,t=t}$ and $Outcome_{i=1,j=0,g=1,t=t}$, respectively. And the <u>control group</u> observed and counterfactual levels at any time point, $t = t$, after the beginning of the second intervention are given by $Outcome_{i=1,j=1,g=0,t=t}$ and $Outcome_{i=1,j=0,g=0,t=t}$, respectively. We calculate that

$$Outcome_{i=1,j=1,g=1,t=t}$$
$$= \beta_0 + \beta_1 * (t) + \beta_2 + \beta_3 * (t-r) + \beta_4 + \beta_5 * (t-s) + \beta_6 + \beta_7 * (t) + \beta_8$$
$$+ \beta_9 * (t-r) + \beta_{10} + \beta_{11} * (t-s) \qquad (*),$$

$$Outcome_{i=1,j=0,g=1,t=t}$$
$$= \beta_0 + \beta_1 * (t) + \beta_2 + \beta_3 * (t-r) + \beta_6 + \beta_7 * (t) + \beta_8 + \beta_9 * (t-r) \qquad (**),$$

$$Outcome_{i=1,j=1,g=0,t=t}$$
$$= \beta_0 + \beta_1 * (t) + \beta_2 + \beta_3 * (t-r) + \beta_4 + \beta_5 * (t-s) \qquad (***),$$

and

$$Outcome_{i=1,j=0,g=0,t=t}$$
$$= \beta_0 + \beta_1 * (t) + \beta_2 + \beta_3 * (t-r) \qquad (****).$$

Therefore,

$$\Delta \, (Treatment\ Group\ Level)_t$$
$$= (Treatment\ Group\ Level\ Difference)_t$$
$$= Outcome_{i=1,j=1,g=1,t=t} - Outcome_{i=1,j=0,g=1,t=t}$$
$$= (*) - (**)$$
$$= \beta_4 + \beta_5 * (t-s) + \beta_{10} + \beta_{11} * (t-s),$$

$$\Delta \, (Control\ Group\ Level)_t$$
$$= (Control\ Group\ Level\ Difference)_t$$
$$= Outcome_{i=1,j=1,g=0,t=t} - Outcome_{i=1,j=0,g=0,t=t}$$
$$= (***) - (****)$$
$$= \beta_4 + \beta_5 * (t-s).$$

Thus,

$$\Delta \, Level_t = (Level\ Difference-in-Differences)_t$$
$$= (Treatment\ Group\ Level\ Difference)_t$$
$$\quad - (Control\ Group\ Level\ Difference)_t$$
$$= [(*) - (**)] - [(***) - (****)]$$
$$= \beta_{10} + \beta_{11} * (t-s)$$

Q.E.D.

And thus, when $t = m+1 = s$, we get the proof in **Appendix 7**.

**Appendix 13: An Alternative Representation of Figure 20**

In **Figure 20** of **Section 4.4.3.5**, the U.S. post-crisis and post-QE counterfactuals are constructed *based on* the observed changes in Canada post-crisis and post-QE, respectively. This way, from the figure, overall impacts (overall level and trend changes) post-crisis and post-QE are calculated as the *difference* between the U.S. observed and counterfactual values post-crisis and post-QE, respectively. In other words, although the overall impacts represent difference-in-difference changes, from the graph, they are measured *directly* as the *difference* between the values on the solid and dashed orange lines since the dashed orange lines are already based on the difference between the solid and dashed blue lines.

An alternative way to represent **Figure 20** is to construct the U.S. post-crisis and post-QE counterfactuals *independently of* the observed changes in Canada post-crisis and post-QE, respectively. The resulting graph would be **Figure 20X** below.

**Figure 20X**

*An Alternative Representation of Figure 20*

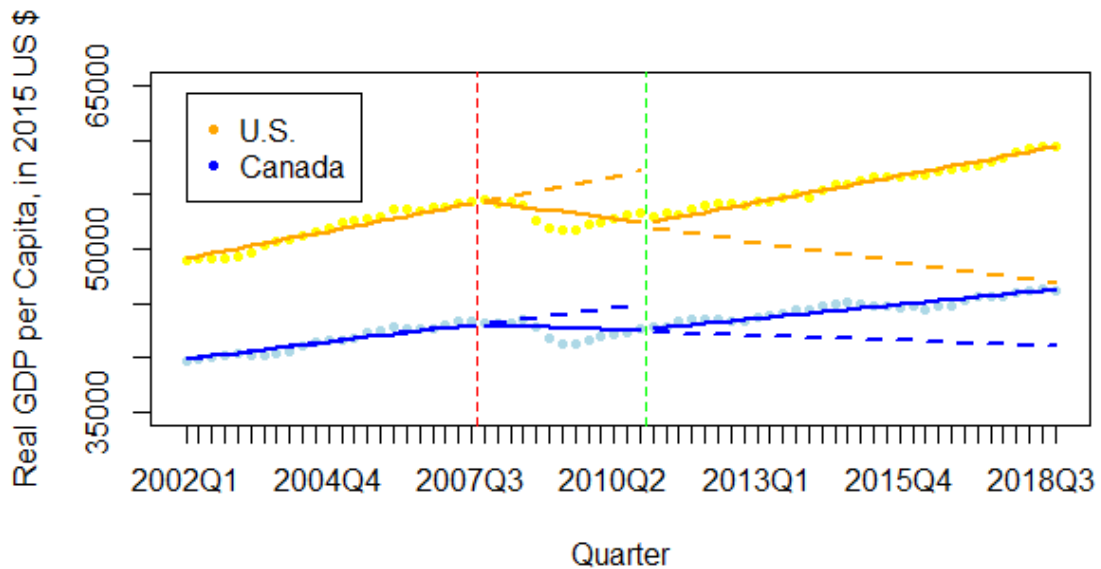

*Note*. The figure is based on the author's calculations using data from the Organization for Economic Co-operation and Development (2022a, 2022b). The data are also accessible from Kamkoum (2023a). R codes for the figure are found in **Appendix 14** and can be freely downloaded from Kamkoum (2023a).

This way, from **Figure 20X**, overall impacts post-crisis and post-QE are calculated as the *difference-in-differences* between the U.S. and Canada observed and counterfactual values post-crisis and post-QE, respectively. That is, from **Figure 20X**, overall impacts are measured *indirectly* as the *difference-in-differences* between the values on the solid and dashed orange and blue lines.

**Appendix 14: R Codes for All the Figures and Tables (Model Estimation Results) in This Study**

The R codes for all the figures and tables (model estimation results) in this paper are found below. The codes are adapted from *Policy Analysis Using Interrupted Time Series* [MOOC], by M. Law, 2015, edX. Copyright 2015 by edX.

The data used in the study and the R codes can be freely downloaded from *Data and Code for "The Federal Reserve's Response to the Global Financial Crisis and Its Long-Term Impact: An Interrupted Time-Series Natural Experimental Analysis,"* by A. C. Kamkoum, 2023, The Open Science Framework (OSF), Center for Open Science (https://doi.org/10.17605/osf.io/t7ezj). CC BY 4.0.

```
# SETTING UP ---------------------------------------------------

#**Downloading the data**
```

```
# Download the data for free from https://doi.org/10.17605/osf.io/t7ezj
# You could download either the data_qe.csv or data_qe.RData file
```

#**Setting the working directory**

```
# Set the working directory to be the computer folder that contains the data_qe.csv or data_qe.RData file
```

#**Loading required libraries**

```
# Load the following libraries
library(car)
library(nlme)
library(lmtest)
```

#**Reading the data**

```
# Read the data_qe.csv file and assign it to data_qe
data_qe <- read.csv("data_qe.csv")

# Or directly open the data_qe.RData file with RStudio
```

# FIGURE 16 --------------------------------------------------------

#**Codes for Panel A of Figure 16**

```
# Plot the time series for U.S. Real GDP per Capita
```

```r
plot(data_qe$time[1:68], data_qe$gdp[1:68],
     ylab = "Real GDP per Capita, in 2015 US $",
     ylim = c(35000, 65000),
     xlab = "Quarter",
     type = "l",
     col = "orange",
     xaxt = "n")

# Add the time series for Canada Real GDP per Capita to the plot
points(data_qe$time[69:136], data_qe$gdp[69:136],
       type = 'l',
       col = "blue")

# Add x-axis quarter labels
axis(1, at = 1:68, labels = data_qe$yearqtr[1:68])

# Add the points for the figure
points(data_qe$time[1:68], data_qe$gdp[1:68],
       col = "orange",
       pch = 20)

points(data_qe$time[69:136], data_qe$gdp[69:136],
       col = "blue",
       pch = 20)

# Label the two interruptions representing the Global Financial Crisis and the QE
# implementation
abline(v = 23.5, lty = 2, col = "red")
abline(v = 36.5, lty = 2, col = "green")
```

```r
# Add a legend to the figure
legend(x = 1, y = 64300, legend = c("U.S.", "Canada"),
       col = c("orange", "blue"),
       pch = 20)
```

#**Codes for Panel B of Figure 16**

```r
# Plot the time series for U.S. CPI Inflation
plot(data_qe$time[1:68], data_qe$cpi[1:68],
     ylab = "CPI Inflation, Index With 2015 = 100",
     ylim = c(70, 110),
     xlab = "Quarter",
     type = "l",
     col = "orange",
     xaxt = "n")

# Add the time series for Canada CPI Inflation to the plot
points(data_qe$time[69:136], data_qe$cpi[69:136],
       type = 'l',
       col = "blue")

# Add x-axis quarter labels
axis(1, at = 1:68, labels = data_qe$yearqtr[1:68])

# Add the points for the figure
points(data_qe$time[1:68], data_qe$cpi[1:68],
       col = "orange",
       pch = 20)

points(data_qe$time[69:136], data_qe$cpi[69:136],
```

```r
    col = "blue",
    pch = 20)

# Label the two interruptions representing the Global Financial Crisis and the QE
implementation
abline(v = 23.5, lty = 2, col = "red")
abline(v = 36.5, lty = 2, col = "green")

# Add a legend to the figure
legend(x=1, y=107, legend=c("U.S.", "Canada"),
    col = c("orange", "blue"),
    pch = 20)

# TABLE 6 ---------------------------------------------------------------

#**Codes for Panel A of Table 6**

# Estimate Model 1 using the ordinary least squares (OLS) method
model_1 <- lm(gdp ~ time + crisis + crisis.time_minus_24 +
        qe + qe.time_minus_37 +
        us + us.time +
        us.crisis + us.crisis.time_minus_24 +
        us.qe + us.qe.time_minus_37,
      data = data_qe)

summary(model_1)

#**Codes for Panel B of Table 6**
```

```r
# Estimate Model 2 using the ordinary least squares (OLS) method
model_2 <- lm(cpi ~ time + crisis + crisis.time_minus_24 +
        qe + qe.time_minus_37 +
        us + us.time +
        us.crisis + us.crisis.time_minus_24 +
        us.qe + us.qe.time_minus_37,
      data = data_qe)

summary(model_2)

# FIGURE 17 -------------------------------------------------------------

#**Codes for Panel A of Figure 17**

# Check Model 1 for autocorrelation using the Durbin-Watson and Breusch-Godfrey tests
for residual autocorrelation
dwtest(model_1)
bgtest(model_1, order = 16)

#**Codes for Panel B of Figure 17**

# Set RStudio to display plots in one row and two columns
par(mfrow = c(1, 2))

# Plot the autocorrelation function (ACF) and partial-autocorrelation function (Partial-ACF) for Residuals of Model 1
acf(residuals(model_1))
```

```r
acf(residuals(model_1), type = 'partial')

# FIGURE 18 ------------------------------------------------------------

#**Codes for Panel A of Figure 18**

# Check Model 2 for autocorrelation using the Durbin-Watson and Breusch-Godfrey tests
for residual autocorrelation
dwtest(model_2)
bgtest(model_2, order = 16)

#**Codes for Panel B of Figure 18**

# Set RStudio to display plots in one row and two columns
par(mfrow = c(1, 2))

# Plot the autocorrelation function (ACF) and partial-autocorrelation function (Partial-
ACF) for Residuals of Model 2
acf(residuals(model_2))
acf(residuals(model_2), type = 'partial')

# TABLE 7 ----------------------------------------------------------------

#**Codes for Panel A of Table 7**

# Estimate AR(13) Model 1 using the generalized least squares (GLS) method
```

```r
model_1_p13 <- gls(gdp ~ time + crisis + crisis.time_minus_24 +
          qe + qe.time_minus_37 +
          us + us.time +
          us.crisis + us.crisis.time_minus_24 +
          us.qe + us.qe.time_minus_37,
       data = data_qe,
       correlation = corARMA(p=13, q=0, form = ~time|us),
       method = "ML")

summary(model_1_p13)

#**Codes for Panel B of Table 7**

# Estimate AR(10) Model 2 using the generalized least squares (GLS) method
model_2_p10 <- gls(cpi ~ time + crisis + crisis.time_minus_24 +
          qe + qe.time_minus_37 +
          us + us.time +
          us.crisis + us.crisis.time_minus_24 +
          us.qe + us.qe.time_minus_37,
       data = data_qe,
       correlation = corARMA(p=10, q=0, form = ~time|us),
       method = "ML")

summary(model_2_p10)

# TABLE 8 -------------------------------------------------------------

#**Codes for Panel A of Table 8**
```

```
# Check whether adding more AR and MA terms to AR(13) Model 1 increases its
goodness of fit
model_1_p13q1 <- update(model_1_p13,
              correlation = corARMA(p=13, q=1, form = ~time|us))
anova(model_1_p13, model_1_p13q1)

model_1_p14 <- update(model_1_p13,
              correlation = corARMA(p=14, q=0, form = ~time|us))
anova(model_1_p13, model_1_p14)
```

#**Codes for Panel B of Table 8**

```
# Check whether adding more AR and MA terms to AR(10) Model 2 increases its
goodness of fit
model_2_p10q1 <- update(model_2_p10,
               correlation = corARMA(p=10, q=1, form = ~time|us))
anova(model_2_p10, model_2_p10q1)

model_2_p11 <- update(model_2_p10,
             correlation = corARMA(p=11, q=0, form = ~time|us))
anova(model_2_p10, model_2_p11)
```

# FIGURE 19 --------------------------------------------------------------

#**Codes for Panel A of Figure 19**

```
# Set RStudio to display plots in one row and one column
```

```r
par(mfrow = c(1, 1))

# Plot the time series for U.S. Real GDP per Capita
plot(data_qe$time[1:68], data_qe$gdp[1:68],
    ylab = "Real GDP per Capita, in 2015 US $",
    ylim = c(35000,65000),
    xlab = "Quarter",
    col = "yellow",
    xaxt = "n",
    pch = 20)

# Add the time series for Canada Real GDP per Capita to the plot
points(data_qe$time[69:136], data_qe$gdp[69:136],
    col = "lightblue",
    pch =20)

# Add x-axis quarter labels
axis(1, at=1:68, labels = data_qe$yearqtr[1:68])

# Label the two interruptions representing the Global Financial Crisis and the QE
implementation
abline(v = 23.5, lty = 2, col = "red")
abline(v = 36.5, lty = 2, col = "green")

# Add fitted regression lines to the three segments of the time series for U.S. Real GDP per
Capita
lines(data_qe$time[1:23], fitted(model_1_p13)[1:23],
    col = "orange", lwd = 2)
lines(data_qe$time[24:36], fitted(model_1_p13)[24:36],
    col = "orange", lwd = 2)
```

```r
lines(data_qe$time[37:68], fitted(model_1_p13)[37:68],
    col = "orange", lwd = 2)

# Add fitted regression lines to the three segments of the time series for Canada Real GDP per Capita
lines(data_qe$time[69:91], fitted(model_1_p13)[69:91],
    col = "blue", lwd = 2)
lines(data_qe$time[92:104], fitted(model_1_p13)[92:104],
    col = "blue", lwd = 2)
lines(data_qe$time[105:136], fitted(model_1_p13)[105:136],
    col = "blue", lwd = 2)

# Add a legend to the figure
legend(x = 1, y = 64400, legend = c("U.S.", "Canada"),
    col = c("orange", "blue"),
    pch = 20)

#**Codes for Panel B of Figure 19**

# Set RStudio to display plots in one row and one column
par(mfrow = c(1, 1))

# Plot the time series for U.S. CPI Inflation
plot(data_qe$time[1:68], data_qe$cpi[1:68],
    ylab = "CPI Inflation, Index With 2015 = 100",
    ylim = c(70, 110),
    xlab = "Quarter",
    col = "yellow",
    xaxt = "n",
    pch = 20)
```

```r
# Add the time series for Canada CPI Inflation
points(data_qe$time[69:136], data_qe$cpi[69:136],
    col = "lightblue",
    pch = 20)

# Add x-axis quarter labels
axis(1, at=1:68, labels = data_qe$yearqtr[1:68])

# Label the two interruptions representing the Global Financial Crisis and the QE implementation
abline(v = 23.5, lty = 2, col = "red")
abline(v = 36.5, lty = 2, col = "green")

# Add fitted regression lines to the three segments of the time series for U.S. CPI Inflation
lines(data_qe$time[1:23], fitted(model_2_p10)[1:23],
    col = "orange", lwd = 2)
lines(data_qe$time[24:36], fitted(model_2_p10)[24:36],
    col = "orange", lwd = 2)
lines(data_qe$time[37:68], fitted(model_2_p10)[37:68],
    col = "orange", lwd = 2)

# Add fitted regression lines to the three segments of the time series for Canada CPI Inflation
lines(data_qe$time[69:91], fitted(model_2_p10)[69:91],
    col = "blue", lwd = 2)
lines(data_qe$time[92:104], fitted(model_2_p10)[92:104],
    col = "blue", lwd = 2)
lines(data_qe$time[105:136], fitted(model_2_p10)[105:136],
    col = "blue", lwd = 2)
```

```r
# Add a legend to the figure
legend(x = 1, y = 107, legend = c("U.S.", "Canada"),
       col = c("orange", "blue"),
       pch = 20)

# FIGURE 20 ------------------------------------------------------------

#**Codes for Figure 20**

# Set RStudio to display plots in one row and one column
par(mfrow = c(1, 1))

# Plot the time series for U.S. Real GDP per Capita
plot(data_qe$time[1:68], data_qe$gdp[1:68],
     ylab = "Real GDP per Capita, in 2015 US $",
     ylim = c(35000, 65000),
     xlab = "Quarter",
     col = "yellow",
     xaxt = "n",
     pch = 20)

# Add the time series for Canada Real GDP per Capita to the plot
points(data_qe$time[69:136], data_qe$gdp[69:136],
       col = "lightblue",
       pch = 20)

# Add x-axis quarter labels
```

```
axis(1, at = 1:68, labels = data_qe$yearqtr[1:68])

# Label the two interruptions representing the Global Financial Crisis and the QE
implementation
abline(v = 23.5, lty = 2, col = "red")
abline(v = 36.5, lty = 2, col = "green")

# Add fitted regression lines to the three segments of the time series for U.S. Real GDP per
Capita
lines(data_qe$time[1:23], fitted(model_1_p13)[1:23],
    col = "orange", lwd = 2)
lines(data_qe$time[24:36], fitted(model_1_p13)[24:36],
    col = "orange", lwd = 2)
lines(data_qe$time[37:68], fitted(model_1_p13)[37:68],
    col = "orange", lwd = 2)

# Add fitted regression lines to the three segments of the time series for Canada Real GDP
per Capita
lines(data_qe$time[69:91], fitted(model_1_p13)[69:91],
    col = "blue", lwd = 2)
lines(data_qe$time[92:104], fitted(model_1_p13)[92:104],
    col = "blue", lwd = 2)
lines(data_qe$time[105:136], fitted(model_1_p13)[105:136],
    col = "blue", lwd = 2)

# Add the counterfactuals for the post-crisis and post-QE fitted regression lines of the time
series for Canada Real GDP per Capita
segments(1, model_1_p13$coef[1] + model_1_p13$coef[2],
     36, model_1_p13$coef[1] + model_1_p13$coef[2]*36,
     lty = 2,
```

```r
               col = "blue",
               lwd = 2)

segments(37, model_1_p13$coef[1] + model_1_p13$coef[2]*37 +
             model_1_p13$coef[3] + model_1_p13$coef[4]*14,
         68, model_1_p13$coef[1] + model_1_p13$coef[2]*68 +
             model_1_p13$coef[3]+model_1_p13$coef[4]*45,
         lty = 2,
         col = "blue",
         lwd = 2)

# Add the counterfactuals for the post-crisis and post-QE fitted regression lines of the time
series for U.S. Real GDP per Capita
segments(24, model_1_p13$coef[1] + model_1_p13$coef[2]*24 +
             model_1_p13$coef[7] + model_1_p13$coef[8]*24 +
             model_1_p13$coef[3] + model_1_p13$coef[4],
         36, model_1_p13$coef[1] + model_1_p13$coef[2]*36 +
             model_1_p13$coef[7] + model_1_p13$coef[8]*36 +
             model_1_p13$coef[3] + model_1_p13$coef[4]*13,
         lty = 2,
         col = "orange",
         lwd = 2)

segments(37, model_1_p13$coef[1] + model_1_p13$coef[2]*37 +
             model_1_p13$coef[7] + model_1_p13$coef[8]*37 +
             model_1_p13$coef[3] + model_1_p13$coef[4]*14 +
             model_1_p13$coef[9] + model_1_p13$coef[10]*14 +
             model_1_p13$coef[5] + model_1_p13$coef[6],
         68, model_1_p13$coef[1] + model_1_p13$coef[2]*68 +
             model_1_p13$coef[7] + model_1_p13$coef[8]*68 +
```

```r
        model_1_p13$coef[3] + model_1_p13$coef[4]*45 +
        model_1_p13$coef[9] + model_1_p13$coef[10]*45 +
        model_1_p13$coef[5] + model_1_p13$coef[6]*45,
      lty = 2,
      col = "orange",
      lwd = 2)

# Add a legend to the figure
legend(x = 1, y = 64400, legend = c("U.S.", "Canada"),
       col = c("orange", "blue"),
       pch = 20)

# FIGURE 20X (APPENDIX 13) ------------------------------------------

#**Codes for Figure 20X (Appendix 13)**

# Set RStudio to display plots in one row and one column
par(mfrow = c(1, 1))

# Plot the time series for U.S. Real GDP per Capita
plot(data_qe$time[1:68], data_qe$gdp[1:68],
     ylab = "Real GDP per Capita, in 2015 US $",
     ylim = c(35000, 65000),
     xlab = "Quarter",
     col = "yellow",
     xaxt = "n",
     pch = 20)
```

```r
# Add the time series for Canada Real GDP per Capita to the plot
points(data_qe$time[69:136], data_qe$gdp[69:136],
    col = "lightblue",
    pch = 20)

# Add x-axis quarter labels
axis(1, at = 1:68, labels = data_qe$yearqtr[1:68])

# Label the two interruptions representing the Global Financial Crisis and the QE implementation
abline(v = 23.5, lty = 2, col = "red")
abline(v = 36.5, lty = 2, col = "green")

# Add fitted regression lines to the three segments of the time series for U.S. Real GDP per Capita
lines(data_qe$time[1:23], fitted(model_1_p13)[1:23],
    col = "orange", lwd = 2)
lines(data_qe$time[24:36], fitted(model_1_p13)[24:36],
    col = "orange", lwd = 2)
lines(data_qe$time[37:68], fitted(model_1_p13)[37:68],
    col = "orange", lwd = 2)

# Add fitted regression lines to the three segments of the time series for Canada Real GDP per Capita
lines(data_qe$time[69:91], fitted(model_1_p13)[69:91],
    col = "blue", lwd = 2)
lines(data_qe$time[92:104], fitted(model_1_p13)[92:104],
    col = "blue", lwd = 2)
lines(data_qe$time[105:136], fitted(model_1_p13)[105:136],
    col = "blue", lwd = 2)
```

```r
# Add the counterfactuals for the post-crisis and post-QE fitted regression lines of the time
series for Canada Real GDP per Capita
segments(1, model_1_p13$coef[1] + model_1_p13$coef[2],
     36, model_1_p13$coef[1] + model_1_p13$coef[2]*36,
     lty = 2,
     col = "blue",
     lwd = 2)

segments(37, model_1_p13$coef[1] + model_1_p13$coef[2]*37 +
         model_1_p13$coef[3] + model_1_p13$coef[4]*14,
     68, model_1_p13$coef[1] + model_1_p13$coef[2]*68 +
       model_1_p13$coef[3] + model_1_p13$coef[4]*45,
     lty = 2,
     col = "blue",
     lwd = 2)

# Add the counterfactuals for the post-crisis and post-QE fitted regression lines of the time
series for U.S. Real GDP per Capita
segments(24, model_1_p13$coef[1] + model_1_p13$coef[2]*24 +
         model_1_p13$coef[7] + model_1_p13$coef[8]*24,
     36, model_1_p13$coef[1] + model_1_p13$coef[2]*36 +
       model_1_p13$coef[7] + model_1_p13$coef[8]*36,
     lty = 2,
     col = "orange",
     lwd = 2)

segments(37, model_1_p13$coef[1] + model_1_p13$coef[2]*37 +
         model_1_p13$coef[7] + model_1_p13$coef[8]*37 +
         model_1_p13$coef[3] + model_1_p13$coef[4]*14 +
```

```r
        model_1_p13$coef[9] + model_1_p13$coef[10]*14,
      68, model_1_p13$coef[1] + model_1_p13$coef[2]*68 +
        model_1_p13$coef[7] + model_1_p13$coef[8]*68 +
        model_1_p13$coef[3] + model_1_p13$coef[4]*45 +
        model_1_p13$coef[9] + model_1_p13$coef[10]*45,
      lty = 2,
      col = "orange",
      lwd = 2)

# Add a legend to the figure
legend(x = 1, y = 64400, legend = c("U.S.", "Canada"),
       col = c("orange", "blue"),
       pch = 20)

# THE END ---------------------------------------------------------

# The codes are complete
```